\documentclass[preprint2]{aastex}
\usepackage{CJK}
\usepackage{amsmath}
\usepackage{graphicx}
\usepackage{txfonts}
\usepackage{color}
\usepackage{enumitem}
\usepackage{longtable}
\newcommand{\cgs}{erg cm$^{-2}$ s$^{-1}$}
\newcommand{\N}[1]{$N_\textrm{H}$}
\shorttitle{7Ms CDF-S spectral analysis}
\shortauthors{Liu et al.}
\begin{document}
\begin{CJK*}{UTF8}{gkai}

\title{X-ray spectral analyses of AGNs from the 7Ms {\sl Chandra} Deep Field-South survey: the distribution, variability, and evolution of AGN's obscuration}

\author{Teng Liu (刘腾) \altaffilmark{1,2,3} ,Paolo Tozzi \altaffilmark{4} ,Jun-Xian Wang (王俊贤) \altaffilmark{1,2} ,William N. Brandt \altaffilmark{5,6,7} ,Cristian Vignali \altaffilmark{8,9} ,Yongquan Xue (薛永泉) \altaffilmark{1,2} ,Donald P. Schneider \altaffilmark{5,6} ,Andrea Comastri \altaffilmark{9} ,Guang Yang \altaffilmark{5,6} ,Franz E. Bauer \altaffilmark{10,11,12} ,Maurizio Paolillo \altaffilmark{13} ,Bin Luo \altaffilmark{14} ,Roberto Gilli \altaffilmark{9} ,Q. Daniel Wang \altaffilmark{3} ,Mauro Giavalisco \altaffilmark{3} ,Zhiyuan Ji \altaffilmark{3} ,David M Alexander \altaffilmark{15} ,Vincenzo Mainieri \altaffilmark{16} ,Ohad Shemmer \altaffilmark{17} ,Anton Koekemoer \altaffilmark{18} ,Guido Risaliti \altaffilmark{4}}

\altaffiltext{1}{CAS Key Laboratory for Research in Galaxies and Cosmology, Department of Astronomy, University of Science and Technology of China, Hefei 230026, China; lewtonstein@gmail.com; jxw@ustc.edu.cn}
\altaffiltext{2}{School of Astronomy and Space Science, University of Science and Technology of China, Hefei 230026, China}
\altaffiltext{3}{Astronomy Department, University of Massachusetts, Amherst, MA 01003, USA}
\altaffiltext{4}{Istituto Nazionale di Astrofisica (INAF) -- Osservatorio Astrofisico di Firenze, Largo Enrico Fermi 5, I-50125 Firenze, Italy; ptozzi@arcetri.astro.it}
\altaffiltext{5}{Department of Astronomy and Astrophysics, 525 Davey Lab, The Pennsylvania State University, University Park, PA 16802, USA}
\altaffiltext{6}{Institute for Gravitation and the Cosmos, The Pennsylvania State University, University Park, PA 16802, USA}
\altaffiltext{7}{Department of Physics, 104 Davey Lab, The Pennsylvania State University, University Park, PA 16802, USA}
\altaffiltext{8}{Dipartimento di Fisica e Astronomia, Alma Mater Studiorum, Universit\`a degli Studi di Bologna, Viale Berti Pichat 6/2, 40127 Bologna, Italy}
\altaffiltext{9}{INAF -- Osservatorio Astronomico di Bologna, Via Gobetti 93/3, 40129, Bologna, Italy}
\altaffiltext{10}{Instituto de Astrof{\'{\i}}sica and Centro de Astroingenier{\'{\i}}a, Facultad de F{\'{i}}sica, Pontificia Universidad Cat{\'{o}}lica de Chile, Casilla 306, Santiago 22, Chile} 
\altaffiltext{11}{Millennium Institute of Astrophysics (MAS), Nuncio Monse{\~{n}}or S{\'{o}}tero Sanz 100, Providencia, Santiago, Chile} 
\altaffiltext{12}{Space Science Institute, 4750 Walnut Street, Suite 205, Boulder, Colorado 80301} 
\altaffiltext{13}{Dip. di Fisica,  Università di Napoli Federico II, C.U. di Monte Sant'Angelo,  Via Cintia ed. 6, 80126 Naples, Italy}
\altaffiltext{14}{School of Astronomy \& Space Science, Nanjing University, Nanjing 210093, China}
\altaffiltext{15}{Centre for Extragalactic Astronomy, Department of Physics, Durham University, South Road, Durham, DH1 3LE, UK}
\altaffiltext{16}{European Southern Observatory, Karl-Schwarzschild-Str. 2, 85748, Garching bei M\"unchen, Germany}
\altaffiltext{17}{Department of Physics, University of North Texas, Denton, TX 76203, USA}
\altaffiltext{18}{Space Telescope Science Institute, 3700 San Martin Drive, Baltimore, MD 21218, USA}

\begin{abstract}
We present a detailed spectral analysis of the brightest Active Galactic Nuclei (AGN) identified in the 7Ms {\sl Chandra} Deep Field South (CDF-S) survey over a time span of 16 years.
Using a model of an intrinsically absorbed power-law plus reflection, with possible soft excess and narrow Fe K$\alpha$ line, we perform a systematic X-ray spectral analysis, both on the total 7Ms exposure and in four different periods with lengths of 2--21 months.
With this approach, we not only present the power-law slopes, column densities \N{H}, observed fluxes, and absorption-corrected 2-10~keV luminosities $L_X$ for our sample of AGNs, but also identify significant spectral variabilities among them on time scales of years.
We find that the \N{H} variabilities can be ascribed to two different types of mechanisms, either flux-driven or flux-independent.
We also find that the correlation between the narrow Fe line EW and \N{H} can be well explained by the continuum suppression with increasing \N{H}.
Accounting for the sample incompleteness and bias, we measure the intrinsic distribution of \N{H} for the CDF-S AGN population and present re-selected subsamples which are complete with respect to \N{H}.
The \N{H}-complete subsamples enable us to decouple the dependences of \N{H} on $L_X$ and on redshift.
Combining our data with that from C-COSMOS, we confirm the anti-correlation between the average \N{H} and $L_X$ of AGN, and find a significant increase of the AGN obscured fraction with redshift at any luminosity.
The obscured fraction can be described as $f_{obscured}\thickapprox 0.42\ (1+z)^{0.60}$.
\end{abstract}


\section{INTRODUCTION}
\label{Section:intro}

Active Galactic Nuclei (AGN) are important for understanding the formation and evolution of galaxies.  It is now well established that the large majority of galaxies experience periods of nuclear activity, as witnessed by the ubiquitous presence of supermassive black holes (SMBHs) in their bulges \citep[e.g.,][]{2013Kormendy}.
A privileged observational window to select and characterize AGN is the 0.5-10~keV X-ray band, which has become particularly effective thanks to the advent of revolutionary X-ray facilities in the last 16 years, such as {\sl Chandra} and {\sl XMM-Newton}.
Despite the fact that only 5-10\% of the total nuclear emission emerges in the X-ray band, the relative strength of X-ray to other band (optical, infrared, radio) emission in AGN is much higher than that in stars.
This trait allows one to identify AGN out to very high redshift in deep, high-resolution surveys.
At least to first order, the majority of AGN spectra can be well described by an intrinsic power-law undergoing photoelectric absorption and Compton scattering by line-of-sight obscuring material, an unabsorbed power-law produced by scattering from surrounding ionized material, and a reflection component from surrounding cold material.
These features make X-ray spectral analysis a powerful tool to measure the accretion properties and the surrounding environment of SMBHs.
Therefore, tracing the X-ray evolution of AGN across cosmic epochs is crucial to reconstruct 
the cosmic history of accretion 
onto SMBH and the properties of the host galaxy at the same time.  

In this framework, significant results have been obtained thanks to a number of high-sensitivity, large and medium sky coverage X-ray surveys such as C-COSMOS \citep{Elvis2009,Lanzuisi2013}, XMM-COSMOS \citep{Hasinger2007}, COSMOS-Legacy \citep{Civano2016,Marchesi2016}, CDF-S \citep[][]{Giacconi2002,Luo2008,Xue2011,Luo2017}, CDF-N \citep{Brandt2001,Alexander2003,Xue2016}, Extended CDF-S \citep{Lehmer2005,Virani2006,Xue2016}, AEGIS-X \citep{Laird2009}, XMM-LSS \citep{Pierre2007,Pierre2016}, and XMM survey of CDF-S \citep[][]{Comastri2011,Ranalli2013}.
Among this set, the CDF-S survey which recently reached a cumulative exposure time of 7 Ms represents the deepest observation of the X-ray sky obtained as of today and in the foreseeable future \citep{Luo2017}.
Despite its small solid angle (484 arcmin$^2$), the CDF-S is the only survey which enables the characterization of low-luminosity and high-redshift X-ray sources.  

The 7Ms CDF-S data have been collected across the entire lifespan of the {\sl Chandra} satellite (1999 -- 2016).
Several groups have already used the CDF-S data to provide systematic investigations of the X-ray properties of AGN \citep[e.g.,][]{Rosati2002,Paolillo2004,Saez2008,Raimundo2010,Luo2011,Comastri2011,Rafferty2011,Alexander2011,Young2012,Lehmer2012,Vito2013,Castello-Mor2013,Vito2016}.
Of particular interest to this work, \citet{Tozzi2006} presented the first systematic X-ray spectral analysis of the CDF-S sources on the basis of the first 1Ms exposure using traditional spectral fitting techniques.
Based on the 4Ms CDF-S data, \citet{Buchner2014} performed spectral analysis on the AGNs with a different approach. They developed a Bayesian framework for model comparison and parameter estimation with X-ray spectra, and used it to select among several different spectral models the one which best represents the data.
Other investigations focused on the spectral analysis of specific X-ray source subpopulations, such as normal galaxies \citep{2008Lehmer,Vattakunnel2012,Lehmer2016}, high-redshift AGN \citep{Vito2013}, or single sources  \citep{2002Norman}.
The CDF-S field has also been observed for 3Ms with {\sl XMM-Newton} \citep[e.g.,][]{Comastri2011,Ranalli2013}. However, we limit this work to the 7Ms {\sl Chandra} data, because the much higher spatial resolution of {\sl Chandra} compared with {\sl XMM-Newton} is essential in resolving high-redshift sources, identifying multi-band counterparts, and eliminating contamination from nearby sources; it also brings about high spectral S/N by minimizing noise in source extraction regions.

Among the most relevant parameters shaping the X-ray emission from AGN, the equivalent hydrogen column density \N{H} represents the effect of the photoelectric absorption (mostly due to the metals present in the obscuring material, implicitly assumed to have solar metallicity) and Thompson scattering on the intrinsic power-law emission.
Generally, the obscuring material is related to the pc-scale dusty torus, which produces the largest \N{H} values, or to the diffuse Interstellar Medium (ISM) in the host galaxy, which can also create \N{H} as high as $\sim 10^{22-23.5}$ cm$^{-2}$ \citep[e.g.,][]{Simcoe1997,Golding2012,Buchner2017}.
It is also found that 100 pc-scale dust filaments, which might be the nuclear fueling channels, could also be responsible for the obscuration \citep{Prieto2014}.
The presence of the obscuring material is likely related to both the fueling of AGN from the host galaxy and the AGN feedback to the host galaxy.
The geometry of the absorbing material is another relevant factor.
Particularly, the orientation along the line of sight plays a key role in the unification model of AGN \citep{Antonucci1993,Netzer2015}.
However, the observed correlation with star formation \citep[e.g.,][]{Page2004,Alexander2005,Stevens2005,Chen2015,Ellison2016} indicates that AGN obscuration can be related to a phase of the co-evolution of galaxies and their SMBHs \citep{Hopkins2006,Alexander2012}, rather than just due to an orientation effect.
Morphological studies of AGN host galaxies show that highly obscured AGNs tend to reside in galaxies undergoing dynamical compaction \citep{Chang2017} or galaxies exhibiting interaction or merger signatures \citep{Kocevski2015,Lanzuisi2015}.
For a particular Compton-thick QSO at redshift 4.75, \citet{Gilli2014} found that the heavy obscuration could be attributed to a compact starburst region.
In brief, the intrinsic obscuration of a given AGN does not have a simple and immediate physical interpretation, due to its complex origins.
A thorough understanding of the distribution of AGN obscuration and its dependence on the intrinsic (absorption-corrected) luminosity and on cosmic epoch is mandatory to understanding, at least statistically, the nature and properties of the emission mechanism, the AGN environment, the co-evolution of AGN and the host galaxy, and the synthesis of the Cosmic X-ray background \citep{Gilli2007}.

There have been several attempts to measure the \N{H} distribution of AGN in the pre-{\sl Chandra} era \citep[e.g.,][]{Maiolino1998,Risaliti1999,Bassani1999}; however, their results were severely limited by the X-ray data.
Thanks to the excellent performance of {\sl Chandra}, \citet{Tozzi2006} corrected for both incompleteness and sampling-volume effects of the 1Ms CDF-S AGN sample and recovered the intrinsic distribution of \N{H} of the CDF-S AGN population ($\log L_X\lesssim 45$, $z \lesssim 4$) with high accuracy.
They found an approximately log-normal distribution which peaks around $10^{23}$ cm$^{-2}$ with a $\sigma\sim 1.1$ dex, not including the peak at low \N{H} (below $10^{20}$ cm$^{-2}$).
Based on a local AGN sample detected by {\sl Swift-BAT} in the 15-195~keV band, which is less biased against obscured AGN, \citet{2011Burlon} presented a similar \N{H} distribution that peaks between $10^{23}$ and $10^{23.5}$ cm$^{-2}$.
Based on a mid-infrared 12$\micron$ selected local AGN sample, which is even less biased against obscured AGN than the 15-195~keV hard X-ray emission, \citet{Brightman2011a} reported a similar distribution with an obscured peak between $10^{23}$ and $10^{24}$ cm$^{-2}$.
Using a large AGN sample selected from CDF-S, AEGIS-XD, COSMOS, and XMM-XXL surveys, \citet{Buchner2015} provided intrinsic \N{H} distributions in three segregated redshift intervals between redshift 0.5 and 2.1, and found a higher fraction of sources at \N{H}$\approx 10^{23}$ cm$^{-2}$ when the redshift increases up to $>1$.
There are other investigations which presented the observed \N{H} distribution of AGN but without any correction for selection bias \citep[e.g.,][]{Castello-Mor2013,2014Brightman}.

Many works have shown that the fraction of obscured AGNs declines at high X-ray luminosity \citep[e.g.,][]{1982Lawrence,2006Treister,Hasinger2008,Brightman2011a,2011Burlon,Lusso2013,2014Brightman}.
This behavior can be explained by a decreased covering factor of the obscuring material at high luminosity \citep{Lawrence1991,Lamastra2006,Maiolino2007}, or as a result of higher intrinsic luminosities in unobscured than in obscured AGNs \citep{Lawrence2010,2011Burlon,Liu2014,Sazonov2015}.
However, other studies suggest that the relation between intrinsic absorption and luminosity is more complex, and may be non-monotonic.
Some studies indicate that in the very-low-luminosity regime, the \N{H} distribution of AGN drops with decreasing luminosity \citep{2009Elitzur,2011Burlon,Brightman2011a,Buchner2015}.
It has also been suggested that the obscured fraction rises again in the very-high-luminosity regime \citep{2014Stern,2015Assef}. 

In general, the fraction of X-ray obscured AGN has been found by several studies to rise with redshift \citep{LaFranca2005,Ballantyne2006,Tozzi2006, 2006Treister,Hasinger2008,2012Hiroi,Iwasawa2012,Ueda2014,Vito2014,2014Brightman,Buchner2015}.
However, such evolution was not found or attributed to biases in other investigations \citep{Dwelly2006,Gilli2007,Gilli2010,Lusso2013}.
The uncertainty is mainly caused by limited sample size and rough \N{H} measurement which is often based upon X-ray hardness ratio rather than spectral fitting.
In particular, the strong dependence of average \N{H} on luminosity places a large obstacle in identifying any dependence of average \N{H} on redshift, because of the strong $L$--$z$ correlation of sources in a flux-limited sample.
A sizable sample with wide dynamical ranges in luminosity and redshift, which can be split into narrow luminosity and redshift bins while maintaining good count statistics, is essential to disentangle any redshift-dependence from the luminosity-dependence.

The picture outlined here points toward a significant complexity, where different fueling mechanisms need to be invoked at different luminosities and different cosmic epochs.
In this paper we exploit the 7Ms CDF-S -- the deepest X-ray data ever obtained -- to investigate the distribution of intrinsic absorption among AGN over a wide range of redshift and luminosity.  
Besides the unprecedented X-ray survey depth, continuous multi-band follow-ups of this field allow us to perform excellent AGN classification and redshift measurement \citep{Luo2017}, which are essential in measuring the \N{H} values of AGN and their distribution across the AGN population.
We apply updated data-processing techniques to these data, and provide systematic spectral analyses of the AGNs.
A few analysis methods which have been widely used in the past few years are applied, including astrometry correction on the data, refined selection of source and background extraction regions, a more elaborate spectral stacking method, and a more accurate spectral fitting statistic. 
The lengthy time interval (16 years) of the 7Ms exposure provides us long-term averaged properties of the sources.
To measure the AGN obscuration more accurately, we include variability in the spectral fitting strategy, which includes not only the variation of the intrinsic luminosity, but also the change in the obscuration on time scales of a few years \citep[e.g.,][]{Risaliti2002,Yang2016}.
On shorter timescales, most of our sources have insufficient statistics to measure \N{H} accurately.
Our final aim is to characterize the intrinsic distribution and evolution of AGN obscuration based on the systematic spectral analyses.

Throughout this Paper, we adopt the WMAP cosmology, with $\rm \Omega_{m}$= 0.272, $\rm \Omega_{\Lambda}$ = 0.728 and $H_{0}$ = 70.4 km $\rm s^{-1}$ $\rm Mpc^{-1}$ \citep{Komatsu2011}.
All of the X-ray fluxes and luminosities quoted throughout this paper have been corrected for the Galactic absorption, which has a column density of $8.8 \times 10^{19}$ cm$^{-2}$ \citep{Stark1992} in the CDF-S field.

\section{DATA PROCESSING}
\label{section:datareduction}

\subsection{CDF-S Observations}

\begin{figure*}[htbp]
\epsscale{1.8}
\begin{center}
\plotone{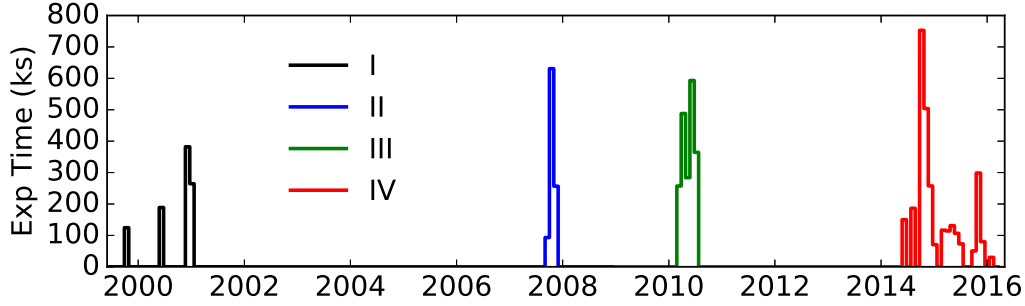}
\caption{Color coded histogram of exposure time of the four periods of CDF-S observations as listed in Table~\ref{table:obsid}. The bin size is 30 days. The approximate total exposure time of each period in sequence is 1Ms, 1Ms, 2Ms, and 3Ms.}
\label{fig:obsdates}
\end{center}
\end{figure*}

\begin{figure*}[htbp]
\epsscale{1.8}
\begin{center}
\plotone{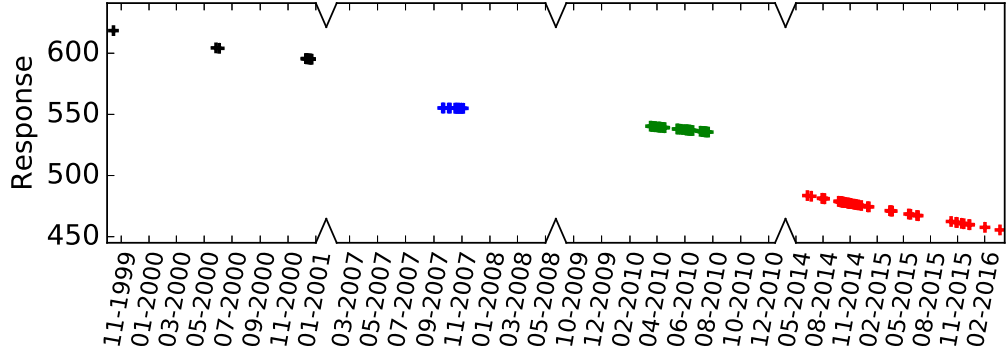}
\caption{Average aimpoint response [cm$^2$ counts/photon] of ACIS-I at 1.5~keV as a function of observation date. 
The quantum efficiency of the ACIS-I CCDs declined maximally by $26\%$.
}
\label{fig:fourperiods}
\end{center}
\end{figure*}

The 7Ms CDF-S survey is comprised of observations performed between Oct 14, 1999 and Mar 24, 2016 (UTC).
Excluding one observation compromised by telemetry saturation and other issues (ObsID 581) there are $102$ observations (observation IDs listed in Table~\ref{table:obsid}) in the dataset.
The exposures collected across 16 years can be grouped in four distinct periods, each spanning 2-21 months.
Figure~\ref{fig:obsdates} displays the distribution of the exposure and the four periods we identified.
Because of the decline of quantum efficiency of the CCD, the average response at the aimpoint \footnote{Here the average aimpoint response for each observation is defined as the average quantum efficiency [counts photon$^{-1}$], which is calculated across all the $1024\times1024$ pixels of 
CCD3, multiplied by the effective area [cm$^2$] at the aimpoint.} changes considerably across the 16 years of operation, whereas it can be considered fairly constant within a single period ($<6\%$ variation), as shown in Figure~\ref{fig:fourperiods}.
We will consider the cumulative spectra of X-ray sources in each period, in order to mitigate the effects of AGN variability on times scales of years \citep[e.g.,][]{Paolillo2004,Vagnetti2016,Yang2016} and reduce the uncertainty of combining the time-dependent instrument calibrations. 

\begin{table*}[htbp]
\tablenum{1}
\label{table:obsid}
\begin{center}
\caption{7Ms CDF-S observations divided into four periods.}
\begin{tabular}[t]{@{}|p{2cm}p{6cm}p{4cm}p{3cm}|}
\hline
Period&Observation date&Time span&Exposure time\\
\hline
I&1999.10 -- 2000.12& 14 months & 1Ms\\
11 ObsIDs:& \multicolumn{3}{p{13cm}|}{1431-0 1431-1 441   582  2406  2405  2312  1672  2409  2313  2239}\\
\hline
II&2007.09 -- 2007.11& 2 months& 1Ms\\
12 ObsIDs:& \multicolumn{3}{p{13cm}|}{8591  9593  9718  8593  8597  8595  8592  8596  9575  9578  8594  9596}\\
\hline
III&2010.03 -- 2010.07& 4 months & 2Ms \\
31 ObsIDs:& \multicolumn{3}{p{13cm}|}{12043 12123 12044 12128 12045 12129 12135 12046 12047 12137 12138 12055 12213 12048 12049 12050 12222 12219 12051 12218 12223 12052 12220 12053 12054 12230 12231 12227 12233 12232 12234}\\
\hline
IV&2014.06 -- 2016.03& 21 months & 3Ms \\
48 ObsIDs:& \multicolumn{3}{p{13cm}|}{16183 16180 16456 16641 16457 16644 16463 17417 17416 16454 16176 16175 16178 16177 16620 16462 17535 17542 16184 16182 16181 17546 16186 16187 16188 16450 16190 16189 17556 16179 17573 17633 17634 16453 16451 16461 16191 16460 16459 17552 16455 16458 17677 18709 18719 16452 18730 16185}\\
\hline
\end{tabular}
\end{center}
\end{table*}

\subsection{Data Processing}

All the data are processed with {\sl CIAO} 4.8 using the calibration release CALDB 4.7.0.
The data are reduced using the {\sc chandra\_repro} tool.
For each observation, the absolute astrometry is refined by matching the coordinates of sources detected using {\sl wavdetect} to the 100 brightest sources of the 4Ms CDF-S catalog \citep{Xue2011} which have been aligned to VLA 1.4 GHz radio astrometric frame.
We use a simple iterative sigma-clipping routine (the CIAO task {\sc deflare}) to detect and remove background flares from the data of each CCD chip in each observation.
For the VFAINT-mode exposures (92 out of 102) we apply the standard VFAINT background cleaning to remove the ``bad'' events which are most likely associated with cosmic rays. 
This cleaning procedure could remove some real X-ray events as background in the case of bright unresolved sources.
We check such an effect on the 10 brightest sources, and find that the loss of net counts is less than $2\%$.
Finally, the exposures are combined using the {\sc flux\_obs} task to create stacked images and exposure maps in the soft (0.5--2~keV) and hard (2--7~keV) bands, respectively.  Exposure maps are computed for a monochromatic energy of 1.5 and 3.8~keV for the soft and hard bands, respectively.

\subsection{Spectra extraction}

Our spectral analysis is based on the 7Ms CDF-S point source catalog including $1008$ X-ray sources \citep{Luo2017}.
To optimize the source-extraction region, we generate an accurate PSF image at the position of each source for each exposure using the ray trace simulation tool {\sc SAOTrace}, and measure the 94\% energy-enclosed contour at 2.3~keV (the effective energy for the 0.5-7~keV band).
In the cases where the extraction regions of two or more nearby sources overlap, the enclosed energy fraction is reduced to separate the regions.
For some sources which lie inside an extended source \citep[e.g.][]{Finoguenov2015} or in a very crowded region, where the extraction regions significantly overlap, we reduce the source extraction region manually.
To prevent exceptionally large extraction regions at CCD gaps and borders where the PSF is distorted by the nonuniform local exposure map, the extraction region is confined to the 95\% energy-enclosed circle measured with the {\sc CIAO psf} task.
The loss of the source flux caused by the extraction region is recovered by applying an energy-dependent aperture correction to the spectral ancillary response files.

We define a background extraction region as an annulus around each source.
To select the inner circle, which is used to mask the source signal, we measure the 97\% energy-enclosed radius $R_{97}$ at the position of each source with the {\sc psf} tool.  
At radii larger than $R_{97}$ the {\sl Chandra} PSF is highly diffused, and a negligible amount of signal from the source falls into the annular background region surrounding the source extraction region.
Only in some cases when the source is very bright, or located far from the aim point or in very crowded region, do we have to use a larger (a factor of 1.2--2) inner radius to make sure the background annulus is free of source signal.
Moreover, we manually mask visible diffuse emission from the background measurement, consulting the extended CDF-S source catalog presented by \citet{Finoguenov2015}.
After all the source signals are masked as above, we select the outer radius for each source $i$ according to the total effective area in the source extraction region $\int A_{i,src}$.
The background-regions are not necessarily complete annuli; they could be broken by the mask of nearby sources.
The background-region size is determined by the ``backscal'' parameter, which is defined as the ratio between the total effective areas in the source region and in the background region $\int A_{i,src} / \int A_{i,bkg}$. 
Effective areas are computed for each exposure at 2.3~keV.
We chose a ``backscal'' for each source by iteration in order that the background annular region determined by this ``backscal'' includes a total of $\sim 1000$ photons in the 7Ms exposure in the total (0.5-7~keV) band.
For each source, the source and background regions vary among the observations, while the ``backscal'' remains approximately constant.
An upper limit of 30$\arcsec$ is set to the outer radius to prevent exceptionally large background regions.

Calibration files,  i.e., response matrix files (RMF) and ancillary response files (ARF), are generated for each source in each exposure.
It is often the case that a source in CDF-S is only visible after stacking multiple observations and may not have any photons (even background photon) within the extraction region for a given exposure.
This is relevant for the majority of the sources, especially for faint source with less than $102$ (total number of observations) net counts lying at the aimpoint of the FOV.
Such an exposure is discarded when extracting spectra, background and calibration files for this particular source; however, its exposure time is retained, see \S~\ref{Section:combine} for details.
Although all the AGNs are expected to be unresolved in X-ray, some sources with off-axis angles $>5\arcmin$ \footnote{Sources with off-axis angles $>5\arcmin$ have 95\% energy-enclosed PSF radii $\gtrsim 5\arcsec$.} and at least 5 photons in the source extraction region in 0.5-7~keV band are treated as extended when creating the response files, by weighting the effective area for the soft photon distributions across their large extraction region.
Although in most cases this treatment has a minor effect, it is technically more valid.
In the case of hard sources without any soft photons in a specific ObsID, the hard band counts are used as weights.
An energy-dependent aperture correction is applied to the ARF with the {\sl arfcorr} task.

\begin{figure}[htbp]
\epsscale{1}
\plotone{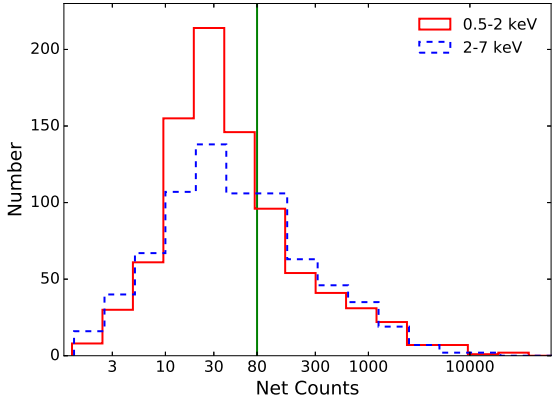}
\caption{
Distributions of Chandra net counts of the AGNs within the extraction radius in the soft and hard bands.
The green vertical line corresponds to $80$ net counts.
}
\label{fig:hist_counts}
\end{figure}

\subsection{Spectra for combined analysis}
\label{Section:combine}

The net counts of the CDF-S sources span a wide range, as shown in Figure~\ref{fig:hist_counts}.
Given the large number of individual exposures, the average number of net counts in one specific exposure is extremely low.
It is thus meaningless to perform a joint analysis keeping spectra and calibration files from all individual exposures for the bulk of the sample.
As we are interested in the spectral analysis of the largest number of sources, we choose to combine the spectra within each of the four periods and within the total 7Ms exposure, generating five sets of stacked spectra and averaged response files.

The PHA spectral files of the source and background are simply stacked using the {\sl FTOOLS} task {\sl mathpha}.
Since the ``backscal'' parameter is determined in advance and remains approximately the same in all the ObsIDs for each particular source, we set the ``backscal'' of the stacked spectra as the counts-weighted mean of the ``backscal'' of each ObsID.
The exposure time of the stacked background PHA is directly calculated by summing the exposure time of each single background spectrum.
While for the exposure time of the stacked source PHA, we have to account for the ObsIDs where there is no photon recorded in the source-extraction region; although such ObsIDs do not contribute any signal to the stacked spectrum.
The exposure time $T_j$ of such an ObsID $j$ is normalized to the mean effective area of the source and then added into the total exposure time as follows:
First, we measure $\bar{A}_j$, the mean effective area inside the extraction region of the source at the effective energy of the broad band, 2.3~keV, for each ObsID $j$.
Among the ObsIDs where signal is recorded within the source extraction region, we calculate the mean effective area of the source $\langle \bar{A}_j \rangle$.
Then, we multiply $T_j$ by $\bar{A}_j/\langle \bar{A}_j \rangle$ and add it to the total exposure time.

To compute averaged RMF and ARF, we consider only the ObsIDs where there is at least one photon within the source extraction region.
We simply use the broad-band photon counts to weight the RMF.
While for the ARF which is more variable because of the vignetting effect and the long-term degeneration of CCD quantum efficiency (see Figure~\ref{fig:fourperiods}), we use a weight of $C_j/\bar{A}_j$, where $C_j$ is the broad-band photon counts in ObsID $j$, and $\bar{A}_j$ is the mean effective area inside the extraction region at the effective energy of the broad band, 2.3~keV.
This choice leads to the most accurate average flux measurement, taking the flux variation of the AGN into account, as explained in detail in Appendix \ref{app:weights}.

\subsection{Sample selection}
\label{Section:sample}

In this work, we focus on the spectra of the AGNs among the 7Ms CDF-S main-source catalog.
As reported in \citet{Luo2017}, an AGN is selected if it satisfies any one of several criteria, including large intrinsic X-ray luminosity (with $L_{0.5-7~keV}>3\times 10^{42}$ erg s$^{-1}$), large ratio of X-ray flux to flux in other band (optical, near infrared, or radio),  hard X-ray spectrum (with an effective power-law slope $\Gamma < 1$, which is obtained without considering the intrinsic absorption of the AGN), and optical spectroscopic AGN features.

For each source, the net counts are measured from the 7Ms stacked source and background spectra.
The distributions of the soft and the hard band net counts of AGNs in the 7Ms data are shown in Figure~\ref{fig:hist_counts}.
The bulk of the spectra have less than 100 net counts, providing poor constraints on spectral parameters.
To reach meaningful characterization of the largest number of AGN, 
we set a threshold on the net counts as low as possible to select a bright subsample for spectral analyses.
To avoid possible bias induced by the low statistics, \citet{Tozzi2006} conservatively defined an X-ray bright sample suitable for spectral analysis by considering sources exceeding at least one of the following thresholds: 170 total counts, 120 soft counts, 80 hard counts, based on the first 1Ms CDF-S stacked spectra.
In this work, we select only sources with at least $80$ net counts in the hard band, which is less affected by obscuration than the soft band.
The 80 net counts threshold corresponds 
to a 2--7 keV flux of $2\times10^{-16}$
\cgs at the aimpoint of CDF-S (see Figure~\ref{fig:skycoverage}).
This threshold appears more stringent than that in \citet{Tozzi2006}; however, justified by the fact that the 7Ms background is about 7 times higher, it is actually less stringent in terms of source detection significance for the same number of net counts.
In particular for faint, off-axis sources with large extraction regions, the signal is background dominated.
In addition, at large off-axis angles where the PSF and effective exposure change dramatically, the S/N is severely reduced.
With this selection threshold, we are sampling AGNs in the luminosity range where most of the emission due to the cosmic accretion onto SMBH is produced (see \S~\ref{Section:intrinsic_distri}).
To keep our sample as large as possible, we exclude only the FOV beyond a 9.5$\arcmin$ off-axis angle from the aimpoint.
Finally, we select $269$ AGNs with at least $80$ hard-band net counts and published redshift measurements.
Besides this main sample, we select a supplementary sample from the central region within an off-axis angle of 4.5$\arcmin$ with at least $60$ net counts in the hard band, in order to fully exploit the 7Ms CDF-S data.
The supplementary sample, which contains seven AGNs, is only used in the re-selected subsamples in \S~\ref{Section:evolution}, where the $80$ net counts threshold becomes irrelevant.

\begin{figure}[htbp]
\epsscale{1}
\plotone{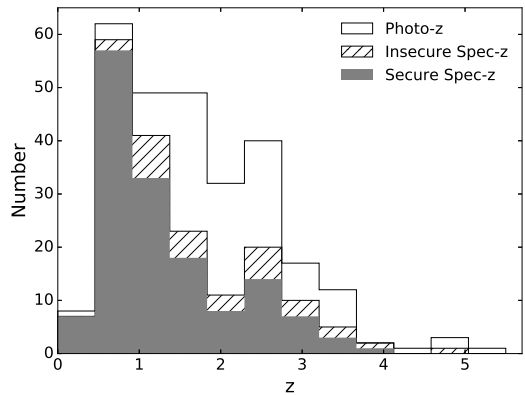}
\caption{
Redshift distribution of our sample.
}
\label{fig:hist_z}
\end{figure}

Our final sample contains $276$ AGNs, having a median redshift of 1.6 and a median number of 0.5--7~keV band net counts of 440.
The redshift measurements are collected by \citet{Luo2017} from 25 spectroscopic-$z$ catalogs and 5 photometric-$z$ catalogs. They selected preferred redshifts carefully from different catalogs and demonstrated that the photometric-$z$ measurements have a good quality by comparing the photometric-$z$ to the available spectroscopic-$z$.
Based on the detection of a narrow 6.4~keV Fe K$\alpha$ line (see \S \ref{Section:strategy}), we replace the photometric-$z$ of $5$ sources with our X-ray spectroscopic redshifts, which are considered insecure.
Finally, among all the redshift measurements, $148$ (54\%) are secure spectroscopic redshifts, $31$ (11\%) are insecure spectroscopic redshifts, and $97$ (35\%) are photometric redshifts.
As shown in Figure~\ref{fig:hist_z}, photometric measurements mostly lie at relatively high redshift.
We note that quite a number of the sources have their redshifts changed with respect to that used in \citet{Tozzi2006}. See further comparison in Section \ref{Section:comparison}.

\section{SPECTRAL ANALYSIS}

\subsection{Spectral fitting method}
\label{Section:methods}
The study of the deep X-ray sky necessarily requires the use of long exposures, often taken at different epochs, such as in the case of CDF-S.
Clearly, in spectral analyses we must take account of significant variability in AGNs, which may reflect changes not only in the intrinsic luminosity but also in the obscuration \citep[e.g.,][]{Yang2016}.
In this work, aimed at exploiting the full statistics of the deep 7Ms exposure, we group ObsIDs which are close in time into four periods and check for significant variation between periods.
To retain the energy resolution as much as possible, each stacked spectrum is grouped as mildly as possible so that each energy bin contains at least 1 photon \citep[see the Appendix in][]{Lanzuisi2013}.
We increase the grouping level (bin size) to speed up the fitting only for the brightest sources in our sample.
If a source has broad-band total counts $N_{tot}>$1000, we group its spectrum to include at least $N_{tot}/1000$+1 photons in each bin.
The low-counts regime of our spectra requires use of the {\sl C} statistic \citep{Cash1979,1989Nousek} rather than $\chi^2$.

With {\sl Xspec} v12.9.0 \citep{Arnaud1996},  we perform spectral analysis for each source following four different approaches:
\begin{description}
\item[A] Fitting the background-subtracted spectrum stacked within each period independently.
\item[B] Fitting the 7Ms stacked, background-subtracted spectrum.
\item[C] Fitting the background-subtracted spectra stacked within each period simultaneously.
\item[D] Fitting the source and background spectra stacked within each period simultaneously.
\end{description}
For a source covered by all four periods, method B deals with one spectrum, method A and C deal with four, and method D eight.
Methods A,B, and C make use of the standard {\sl C} statistic.
In model comparison, the change of the {\sl C} statistic, $\Delta C$, which follows a $\chi^2$ distribution approximately, can be used as an indicator of the confidence level of the fitting improvement.
Specifically, a model is providing a statistically significant improvement at a confidence level of $95$\% when the {\sl C} statistic is reduced by $\Delta  C > 3.84$ and  $\Delta  C > 5.99$ for one and two additional degrees of freedom (DOF), respectively.
Method D, which models both the source and the background spectra, adopts a slightly different statistic, namely, the {\sl W} statistic.
This approach mitigates a weakness of the commonly used method of fitting background-subtracted spectrum with the {\sl C} statistic, which incorrectly assumes that the background-subtracted spectrum has a Poissonian error distribution.
In this work, we use method D to obtain the final estimation of the parameters; the other three {\sl C} statistic methods are used for different purposes as described below, when the $\Delta C$ method is needed to evaluate model improvement.

\subsection{Spectral models}
\label{Section:model}
\subsubsection{Selection of our models}

\begin{figure}[htbp]
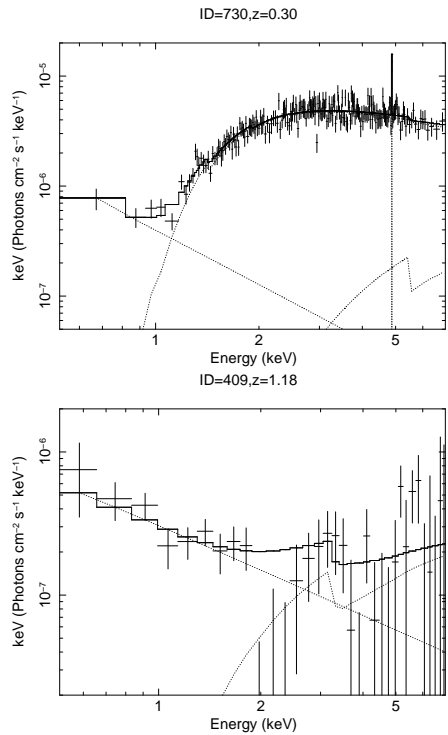

\center
\includegraphics[angle=270,width=0.8\columnwidth]{730.ps}
\includegraphics[angle=270,width=0.8\columnwidth]{409.ps}
\caption{Two examples of the spectra. Upper panel: Source 730 (with 8790 net counts in the 0.5--7~keV band) is a Compton-thin AGN fitted with the standard model, which is composed of an obscured power-law (the 
dominant
component), a narrow Fe K$\alpha$ line, a soft excess, and a {\tt zwabs*pexrav} reflection. The reflection component only contributes a small fraction of signal in the hard band.
Lower panel: Source 409 (with 440 net counts in the 0.5--7~keV band) is a Compton-thick AGN fitted with the Compton-thick model in which the relative strength of the reflection is set free. The reflection dominates the hard band emission.
Note that this Compton-thick model is only used in \S\ref{Section:Rfree} to identify Compton-thick AGN.
}
\label{fig:example}
\end{figure}

The source spectral model is {\tt wabs * (zwabs*powerlaw + zgauss + powerlaw + zwabs*pexrav*constant)}; an illustration is given in the upper panel of Figure~\ref{fig:example}.
The {\tt wabs} \citep{Wilms2000} accounts for the Galactic absorption, which is fixed at a column density of $8.8 \times 10^{19}$ cm$^{-2}$ \citep{Stark1992}. 
The model is composed of four additive components.
{\tt zwabs*powerlaw} describes the primary power-law with intrinsic obscuration, that is, the cumulative effect of the absorbing material in the circumnuclear region and possibly in the host galaxy, expressed in equivalent Hydrogen column density assuming solar metallicity.
The component {\tt zgauss} describes a gaussian emission line with a zero width to fit an unresolved 6.4~keV Fe $K\alpha$ line when present.
The second {\tt powerlaw} is used for a soft excess component, which is occasionally found in the soft band in addition to the primary power-law.
A cold reflection component is modeled with {\tt zwabs * pexrav * constant}, where the absorption is fixed to  $10^{23}$ cm$^{-2}$, as discussed later.
The four components do not always appear for each source; initially, only the primary power-law and the reflection are considered.
The emission line and the additional power-law are included only if they are statistically required, as described in \S \ref{Section:strategy}.

The absorption model {\tt zwabs} works well in the Compton-thin regime and has been widely used.
However, it considers only photoelectric absorption but not Compton scattering, which starts to be relevant at \N{H} $>$ a few $10^{23}$ cm$^{-2}$. 
In order to identify Compton-thick AGNs and measure the \N{H} of highly-obscured AGNs with more accuracy, we check how the shortage of the {\tt zwabs} model affects the results by replacing the {\tt zwabs*powerlaw} with {\tt plcabs} \citep{Yaqoob1997}, which describes X-ray transmission of an isotropic source located at the center of a uniform, spherical distribution of matter, correctly taking into account Compton scattering.

The slope and normalization of the {\tt pexrav} component are linked to those of the primary power-law, and the cut-off energy is fixed at 300~keV.
Although it has been found that the reflection strength is larger ($R\thickapprox 2.2$) in highly obscured sources ($10^{23}$--$10^{24}$ cm$^{-2}$) than in less obscured ones 
\citep[$R\lesssim 0.5$, see][]{Ricci2011},
 we fix the reflection scaling factor $R$ at $0.5$ for all the sources for simplicity.
By definition, the $R$ parameter regulates the relative strength of reflection to the primary power-law.
However, we always fix $R$ at a constant value and use the additional ``constant'' parameter to regulate the relative reflection strength, just for convenience.
In the standard model, which is Compton-thin, this ``constant'' parameter is fixed at $1$.
It is only set free and used in identifying Compton-thick sources, as shown below, where it can be large, indicating relatively strong reflection.

The X-ray reflected emission of AGN might arise from the accretion disk or the inner region of the dusty torus. Considering the realistic geometry of the torus and the AGN obscuration from torus-scale to galaxy-scale \citep{Buchner2017}, it is unlikely that all the reflected X-ray photons could leave the galaxy without any absorption as expected by the flat-surface reflection model ``pexrav''.
The X-ray reflection, if separated from the transmitted power-law as a stand-alone component, must be self-absorbed by the torus or obscured by material on a larger scale.
In a physical torus model \citep[e.g.,][]{Murphy2009,Brightman2011a} which treats the absorption, scattering, and reflection self-consistently, the ``self-absorption'' of reflection is naturally considered.
To obtain a simple rendition for a ``self-absorbed'' reflection model without adding any free parameters, we add an absorption to {\tt pexrav}.
This absorption is irrelevant to the absorption for the primary power-law, which corresponds to only the line-of-sight absorber and is highly dependent on the viewing orientation;
it corresponds to a majority of the obscuring material in the galaxy which must have a significant covering factor to the core, and is less variable among AGNs with different viewing orientations compared with the line-of-sight absorption.
According to a comparison with the MYTorus model in \S\ref{Section:torusmodels}, we set $^p$\N{H}$=10^{23}$ cm$^{-2}$ for the ``self-absorption''.
Therefore, even after adding this absorption, our reflection model still contains no free parameter.

In case of method D, the background is modeled with the {\tt cplinear} model \citep{2010Broos} plus two narrow gaussian emission lines.
The {\tt cplinear} describes the background continuum by fluxes at $10$ vertex energies.  
The vertex energies are selected dynamically between 0.5 and 7~keV, letting each segment contain the same number of photons.

The number of vertices is reduced in order to have at least 10 photons in each segment.
The gaussian lines describe the two most prominent instrument emission lines in the 0.5-7~keV band, one at 1.486~keV (Al K$\alpha$) and one between 2.1 and 2.2~keV (Au M$\alpha,\beta$).
We first fit the background spectra with this model and then fix the background parameters at the best-fit values when fitting the source and background simultaneously.

\subsubsection{Justification of our models}
\label{Section:torusmodels}

\begin{figure}[htbp]
\center
\epsscale{1}
\plotone{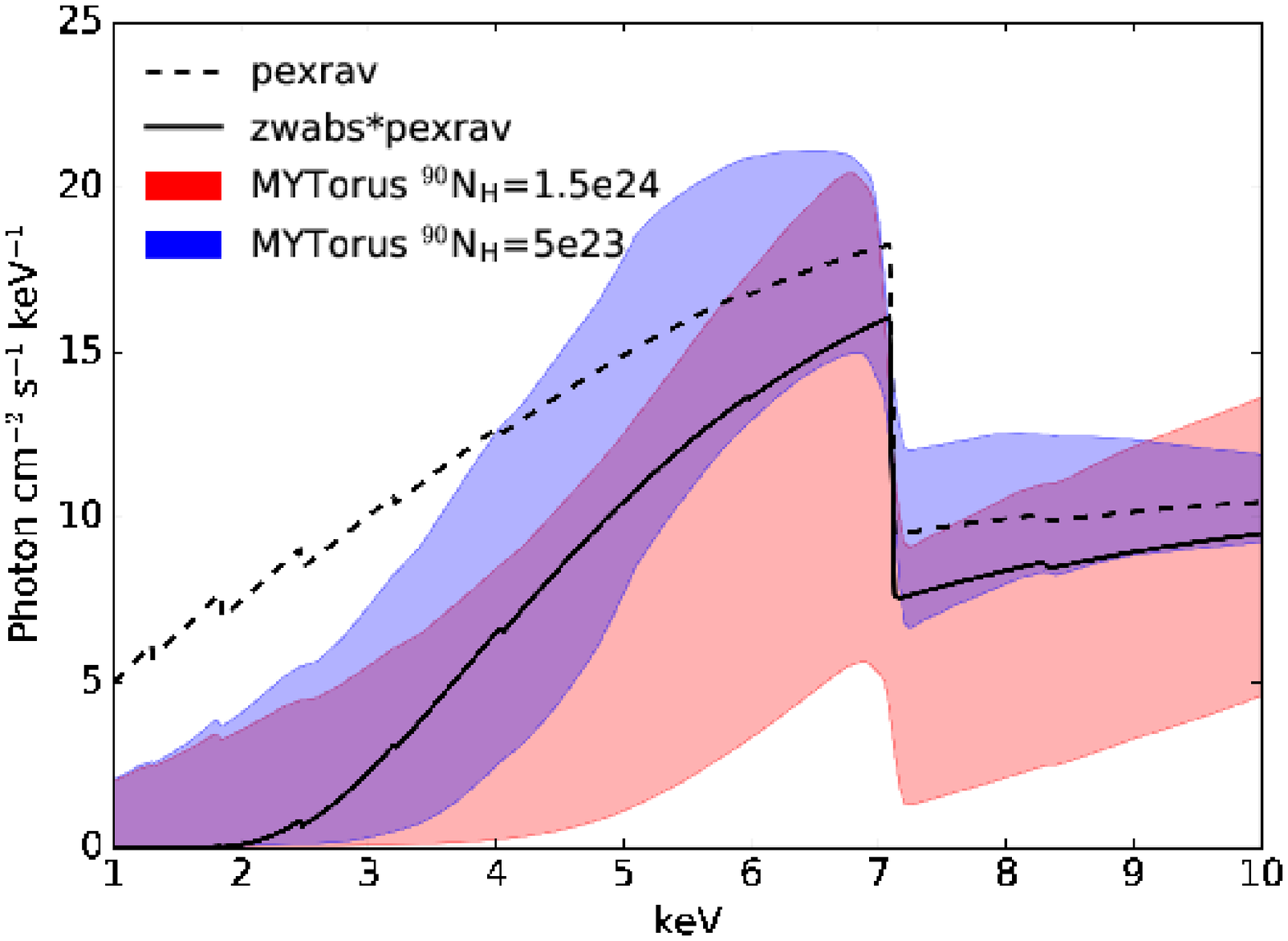}
\plotone{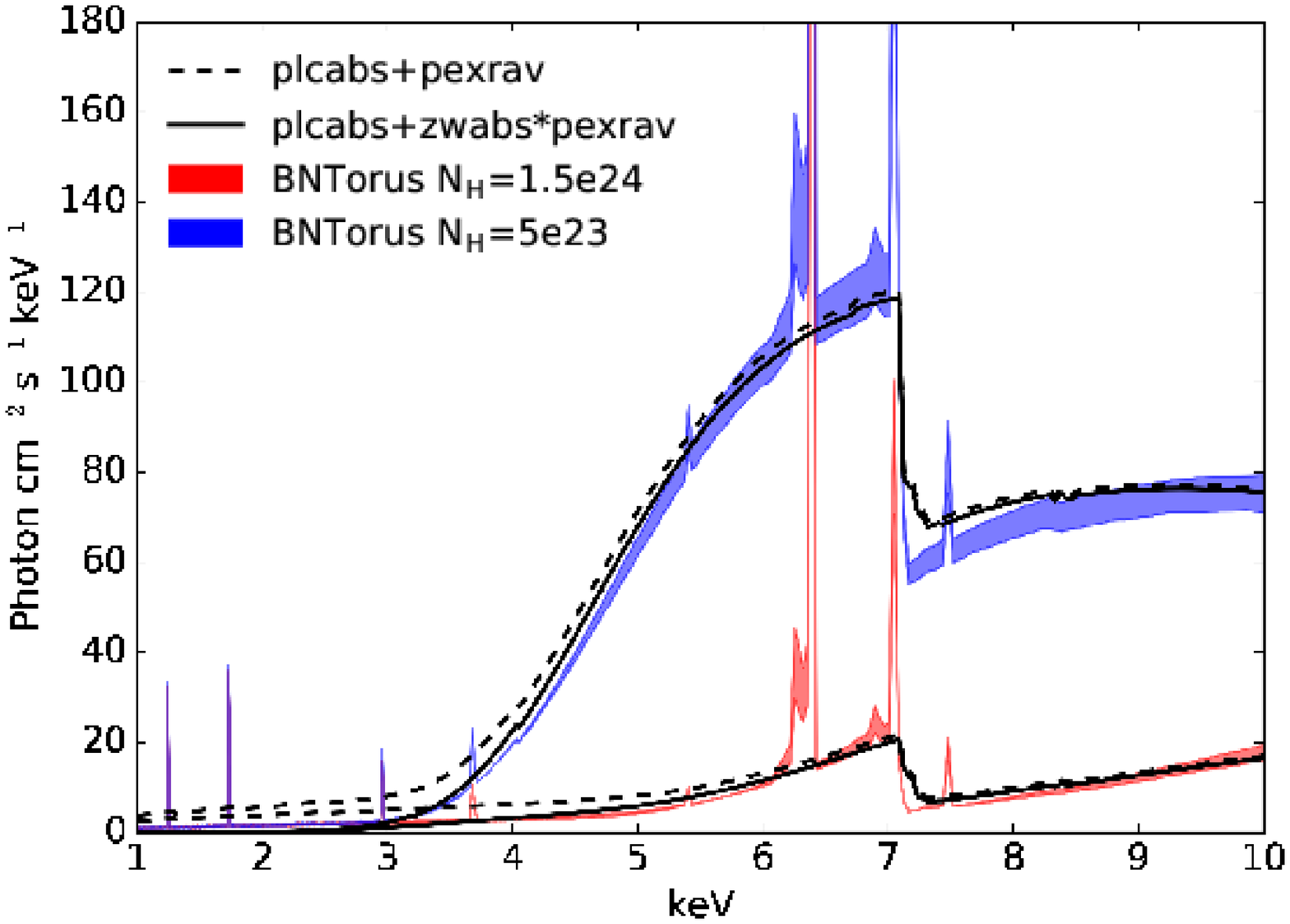}
\caption{
Upper panel: the reflection model adopted in this work ({\tt zwabs*pexrav} with an $^p$\N{H} of $10^{23}$ cm$^{-2}$, black solid line) compared with MYTorus reflection models (color filled regions) with diameter column densities of $^{90}$\N{H}$=5\times 10^{23}$ cm$^{-2}$ (blue) and $^{90}$\N{H}$=1.5\times 10^{24}$ cm$^{-2}$ (red) between inclination angles of $\theta=0^{\circ}$ (face-on) and $\theta=90^{\circ}$ (edge-on).
The unabsorbed {\tt pexrav} model is plotted with a dashed black line.
All the models are derived from the same intrinsic primary power-law and have $z=0$.
Lower panel: the model of transmitted power-law plus reflection adopted in this work ({\tt plcabs+zwabs*pexrav}, black solid lines) compared with BNTorus models (color filled regions) in two cases with \N{H}$=5\times10^{23}$ cm$^{-2}$ (blue) and \N{H}$=1.5\times10^{24}$ cm$^{-2}$ (red), respectively.
The filled ranges correspond to torus opening angles between $\theta_{torus}=30^{\circ}$ and $\theta_{torus}=60^{\circ}$.
Without the ``self-absorption'', the {\tt plcabs+pexrav} model is plotted with black dashed line.
All the models have $z=0$. They are all derived from the same intrinsic primary power-law, but the normalizations of the {\tt plcabs}-based models are multiplied by factors of 75\% and 50\% in the cases of \N{H}$=5\times10^{23}$ cm$^{-2}$ and \N{H}$=1.5\times10^{24}$ cm$^{-2}$, respectively.
}
\label{fig:torusmodel}
\end{figure}

Since the low S/N of our spectra does not allow us to constrain any parameter of the reflection component, 
we have to make proper assumptions about the model to describe a typical case of the reflected emission.
The model we adopt {\tt zwabs*pexrav} is an effective model which may not correspond to a realistic description of the torus.
Here we test its validity by comparing it with a few physical models which describe the X-ray reprocessing considering more detailed torus structures \citep[e.g.,][]{Murphy2009,Brightman2011a}.

First, we compare the spectral shape of our reflection model with that of the MYTorus model \citep{Murphy2009}, which provides the spectrum of the reflection component considering a toroidal torus structure.
In the upper panel of Figure~\ref{fig:torusmodel}, we show the MYTorus reflection in two cases with the column densities through the diameter of the torus tube $^{90}$\N{H} (not the line-of-sight \N{H}) of 5$\times 10^{23}$ cm$^{-2}$ (red lines) and 1.5$\times 10^{24}$ cm$^{-2}$ (blue) and with the inclination angles (between the line-of-sight and the symmetry axis of the torus) $\theta$ between $0^\circ$ and $90^\circ$, respectively.
Both the shape and the strength of the reflection depend significantly on $^{90}$\N{H} and $\theta$, having a large dynamical range.
It is stronger when face-on than edge-on and stronger with $^{90}$\N{H}$=5\times 10^{23}$ than $1.5\times 10^{24}$ cm$^{-2}$.
Without any self-absorption, the {\tt pexrav} model is clearly softer than the MYTorus model.
Adding the self-absorption of $^p$\N{H}$=10^{23}$ cm$^{-2}$ to {\tt pexrav}, the shape of {\tt zwabs*pexrav} is much more similar to that of MYTorus, and in both cases where $^{90}$\N{H}=5$\times 10^{23}$ cm$^{-2}$ and $^{90}$\N{H}=1.5$\times 10^{24}$ cm$^{-2}$, it lies between the face-on ($\theta=0^\circ$) and edge-on ($\theta=90^\circ$) instances of MYTorus.
Therefore, with an order-of-magnitude estimate of $^p$\N{H}$=10^{23}$ cm$^{-2}$, our reflection model can be considered as an intermediate instance of the various reflection models.
According to the MYTorus model, the reflection is weaker at higher $^{90}$\N{H} and larger $\theta$, where the line-of-sight \N{H} would be higher.
By adding the absorption to {\tt pexrav}, the strength of the {\tt pexrav} model is reduced by $32\%$ in the 2-7~keV band.
This weaker reflection setting in our model suggests that we are likely modeling the reflection in the high-\N{H} cases better than in the low-\N{H} cases.
This is helpful to our aim of measuring \N{H}, since in low-\N{H} cases where the X-ray emission is largely dominated by the primary power-law, the reflection is not as significant as in the high-\N{H} cases.

Second, to check the relative strength of the reflection to the primary power-law of our model, we compare our spectral shape of primary power-law plus reflection with that of BNTorus model \citep{Brightman2011a}, which provides the spectrum of the total transmitted and reflected emission considering a biconical torus structure.
In the lower panel of Figure~\ref{fig:torusmodel}, we compare our model with the BNTorus model in two cases with \N{H} (independent of inclination angle) of $5\times10^{23}$ cm$^{-2}$ and $1.5\times10^{24}$ cm$^{-2}$.
The BNTorus model shown in the Figure has opening angles $\theta_{torus}$ between $30^\circ$ and $60^\circ$, and is stronger with smaller $\theta_{torus}$.
We note that the BNTorus spectra are weaker than {\tt plcabs}-based models at the same \N{H}, likely because of differences in the cross-section of absorption and/or scattering and in the abundance of elements.
To compare the spectral shapes in Figure~\ref{fig:torusmodel}, the normalizations of the {\tt plcabs}-based models are multiplied by factors of 75\% and 50\% in the cases of \N{H}=$5\times10^{23}$ cm$^{-2}$ and $1\times10^{24}$ cm$^{-2}$, respectively.
In this work, we focus on how to measure \N{H} accurately and ignore this minor effect on the measurement of intrinsic luminosity, which only has a moderate impact on our results.
In both cases where \N{H}=$5\times10^{23}$ cm$^{-2}$ and $1.5\times10^{24}$ cm$^{-2}$, the spectral shape of our model is very similar to the continuum of the BNTorus model.
If the ``self-absorption'' was not added, the {\tt plcabs+pexrav} model would be softer (dashed line in the lower panel of Figure~\ref{fig:torusmodel}) than BNTorus.
It is found by \citet{LiuLi2015} that the reflection in the BNTorus model is overestimated because of a lack of torus ``self-absorption'' on the reflected emission from the inner region of the torus.
This further strengthens the necessity of adding the ``self-absorption'' to {\tt pexrav} in our model.

\subsection{Spectral fitting strategy}
\label{Section:strategy}

Before determining the final spectral fitting model for each source, we need to choose the spectral components -- whether a narrow Fe K$\alpha$ line or a soft excess component is needed, and decide whether,  in different periods, the power-law slope should be kept as a free parameter or fixed to a constant, and whether intrinsic absorption \N{H} should be linked together or left free to vary.

\subsubsection{Determining power-law slopes}
Ideally, all the parameters should be allowed to vary in each period.
However, most of the CDF-S sources have low S/N, which hamper the measurement of each single spectral parameter with sufficient accuracy.
In particular, there is a strong degeneracy between the power-law slope and the intrinsic absorption, such that in the low S/N regime, a very steep slope can be accommodated with a very high absorption level, and a very flat slope can be obtained with a severely underestimated absorption.
To avoid such a degeneracy, we link the power-law slopes of all the periods together and set it free only if the slope parameter $\Gamma$ could be well constrained, that is, the relative error 
(1$\sigma$ error divided by the best-fit value) of $\Gamma$ is lower than 10\% and the \N{H} is lower than $5\times 10^{23}$ cm$^{-2}$.
The best-fit $\Gamma$ values in these well-constrained cases have a median value of $1.8$ (see Section \ref{Section:result}) -- a typical slope of AGN found or adopted in a huge number of papers.
Therefore, in all the other cases, we fix $\Gamma$ at $1.8$.

Under this assumption, we can focus on the distribution of \N{H} of our sources, despite the disadvantage that the dispersion of \N{H} could be slightly reduced.

\subsubsection{Searching for Fe lines}
A narrow Fe K$\alpha$ line is commonly detected in AGNs, but its detection is limited by
the
quality of the spectrum.
Using fitting method B, we search for the existence of a narrow Fe K$\alpha$ line by comparing the fitting statistics with and without the line component in the model.
For simplicity, the line energy and flux are assumed to be constant in the four periods.
First, we claim a line detection when the best fit with a narrow line component at 6.4~keV has an improvement $\Delta C>3.84$, corresponding to a more than $95\%$ confidence level with $\Delta DOF=1$ (degree of freedom).
Although the $\Delta C$ method could not provide accurate probability in line detection \citep{Protassov2002}, we can still use it as a rough selection method.
Then for the sources with photometric redshifts, we search for the narrow Fe K$\alpha$ line by letting the redshift vary.
In $5$ cases when $\Delta C>9.21$, which corresponds to $99\%$ confidence with $\Delta DOF=2$, we claim a line detection and replace the photometric redshift with an X-ray spectroscopic redshift which guarantees a line energy of 6.4~keV.
The ID, old redshift, and new redshift of the $5$ sources are 98: 1.41--1.99, 646: 2.13--1.49, 733: 2.40--2.64, 940: 3.31--3.08, 958: 0.87--0.89, respectively.
For the first one (ID=98), the same X-ray spectroscopic redshift has been reported by \citet{DelMoro2014}.

Broad Fe lines are 
even harder to detect in the low S/N regime. For 29 sources which have an \N{H} below 10$^{22}$ cm$^{-2}$, a spectroscopic redshift measurement, and at least 1000 net counts in the 0.5-7~keV band, we search for the broad Fe line by setting both the line energy and width free.
This component is taken as detected if $\Delta C>11.34$, which corresponds to a 99\% confidence with $\Delta DOF=3$.
Then we fix the line width at 0. If $\Delta C<2.71$ (90\% with $\Delta DOF=1$), the line width is consistent with 0. We consider such lines as narrow Fe K$\alpha$ lines.

Eventually, we detect $50$ narrow Fe K$\alpha$ lines and $5$ broad Fe lines.
In the final model, the line component is adopted only if a line is detected.
The line energy is set free; the line width is fixed at 0 for narrow lines and set free for broad lines.

\subsubsection{Searching for the soft excess}
\label{Section:softexcess}
A soft excess is often detected in the soft X-ray band of AGN, but its origin is uncertain.
We add a secondary power-law as a phenomenological model for such a component, in order to cope with different physical origins.
The normalization of the secondary power-law component is restricted to $<10\%$ of that of the primary power-law, and its slope is constrained to be equal to or steeper than that of the primary one.
In the cases of obscured AGN, this secondary power-law could describe the power-law scattered back into the line of sight \citep[e.g.,][]{Bianchi2006,Guainazzi2007} which has the same slope but $<10\%$ flux of the primary power-law \citep{Brightman2012}.
In the cases of unobscured AGN, the soft excess could be due a blurred reflection from ionized disk \citep[e.g.,][]{Crummy2006} or warm Comptonization emission from the disk \citep[e.g.,][]{Mehdipour2011,Matt2014}.
Regardless of the physical origin, such component must have a steeper slope to rise above the primary power-law.
Meanwhile, we avoid a secondary power-law flatter than the primary one also in order to avoid severe component degeneracy problem in the fitting.
We establish the presence of a significant soft excess whenever the best fit improves by $\Delta C>5.99$ after adding the secondary power-law component ($95\%$ confidence level with $\Delta DOF=2$).
Since the soft excess can be variable on time scales of years, we search for the soft excess both in each period and in the 7Ms stacked spectrum, thus fitting methods A and B are used here.
If a soft excess is detected in one period for one source, the additional power-law component is activated for this period.
If the excess is only detectable in the 7Ms stacked spectrum, the additional power-laws are activated for all four periods, but with the same slope and scatter fraction (ratio of the power-law normalization to that of the primary power-law).

In the cases of highly obscured sources, the soft excess component is likely a scattered power-law.
Since the primary power-law is severely reduced, the soft excess could even dominate the 0.5-7~keV band, if the scattered fraction is higher than a few percent.
As described in \S~\ref{Section:Rfree}, we look for a hard excess in order to select Compton-thick sources.
For sources identified as Compton-thick in this way, for example, Source 409 (lower panel of Figure~\ref{fig:example}), we consider the power-law which dominates the soft band as a scattered component, and activate the secondary power-law components in all four periods with the same scatter fraction, so that we can measure the \N{H} and intrinsic luminosity on the basis of the hard component.
In these cases, the slope of such a secondary power-law is linked to that of the primary one.
In other cases when the \N{H} is above 5$\times 10^{23}$ cm$^{-2}$, the slopes of the primary and the secondary power-law are also linked.

In some other cases when necessary, we add a soft excess component even if it is not significantly detected, in order to ensure that the detected \N{H} variation is not due to soft excess, see \S~\ref{Section:NHvary} for details. Eventually, $85$ sources in our sample have the soft excess component activated in the final model, $27$ of them are set with a constant scatter fraction, and the others are free in one or a few of the four periods.
These sources have relatively lower redshifts, whose distribution is different from that of the whole sample at a KS-test probability of 99\%. This is because at high redshifts, the observed 0.5--7~keV band corresponds to a harder band where the soft excess is less prominent and the spectral S/N is relatively lower.

\subsubsection{Searching for \N{H} variations}
\label{Section:NHvary}
When we fit the spectra of all the periods simultaneously as in methods C and D, the primary power-law has a constant slope (either free or fixed) and an independent normalization, and by default, the intrinsic absorption has a constant \N{H}.
After the model components are selected, we check for variability of \N{H} between each pair of the four periods among $171$ sources which have at least 300 broad-band net counts in the 7 Ms exposure.
Here we use fitting method C.
First, we find the best-fit after setting \N{H} free in all the four periods. Then, for each pair of periods, we link the \N{H} parameter and measure the $\Delta C$ with respect to the best-fit obtained with all the four parameters set free.
When $\Delta C>6.64$ ($\Delta DOF=1$), \N{H} is considered to be different between the two corresponding periods at a $>99\%$ confidence level.
In some cases, since the soft excess component is detected in one period but not the other, the apparent \N{H} variation could be actually caused by the soft excess component.
In order to guarantee that the \N{H} variation is independent of the soft excess in such cases, a soft excess component, which does not improve the fit as significantly as required by our selection threshold above, is added manually into the spectral model of each period.
When no significant variability is found, the \N{H} of all the four periods are linked together.
When \N{H} is found to be significantly different between two periods, their \N{H} values are set independent of each other; for the other periods, if the best-fit \N{H} obtained when all the \N{H} were set free is between the two independent ones, it is set to the mean value of the independent ones; if the best-fit \N{H} is larger/smaller than both the independent ones, it is linked to the independent one which is closer.
Eventually, our final fitting reports more than one \N{H} for a source if its \N{H} is significantly varying, and only one average value if not.

\subsubsection{Measuring intrinsic luminosities}
At this point, the final spectral model has been set for each source, and the best-fit parameters have been obtained with using fitting method D.
In order to calculate a 7Ms averaged intrinsic luminosity, we set the flux to be constant among the four periods, and calculate the mean absorption-corrected rest-frame 2-10~keV flux of the primary power-law on the basis of the spectral modeling.
To compute the error range of this flux, we add a ``cflux'' component in front of the ``power-law'' component in the spectral model.
This component allows us to use the ``error'' task in Xspec to calculate the error of the absorption-corrected flux in the same manner as the other spectral parameters.

\subsection{Identifying Compton-thick AGNs}
\label{Section:thick}
We follow four procedures to identify Compton-thick candidates:

\subsubsection{Exceptionally strong reflection}
\label{Section:Rfree}
When \N{H}$>$ $1.5\times 10^{24}$ cm$^{-2}$, the primary power-law is severely reduced by the line-of-sight obscuration, while the reflection component, determined by the intrinsic strength of the primary power-law and the material around the core, is relatively irrespective of the line-of-sight \N{H}.
Therefore, 
Compton-thick AGN have the defining characteristic of exceptionally strong reflection
compared with the primary power-law.
In the soft band, the reflection component, which has an extremely hard spectral shape, can be easily swamped by other soft components, such as a scattered power-law; it is only prominent in the hard band.
Our first attempt to identify Compton-thick AGN is to look for an excess in the hard band, which indicates a reflection component with an exceptionally large relative strength.

Some AGNs might change their states between Compton-thick and Compton-thin. However, hampered by the low S/N, spectroscopic identification of Compton-thick AGN would be less sensitive in each single period.
Therefore, here we only use the stacked 7Ms spectra (fitting method B).
We compare our standard Compton-thin model with a Compton-thick model in order to identify sources which can be better described by the latter one.
In our standard Compton-thin model, the relative strength of the reflection to the unabsorbed primary power-law is small and fixed.
The Compton-thick model is converted from the standard model by setting the {\tt constant} in the ``{\tt zwabs*pexrav*constant}'' component free.
In this way, an exceptionally strong reflection component will manifest itself in terms of a large {\tt constant}, as illustrated with an example (Source 409) in Figure~\ref{fig:example}.
Such strong reflection,
 which could hardly be caused by the observed weak power-law, indicates that the primary power-law is hidden (obscured) and the observed power-law which dominates the soft band should be considered as a scattered component.
In order to understand the relative strength of the reflection intuitively, here we set the \N{H} in this component at $10^{22}$ cm$^{-2}$ and $R$ at $1$, which approximately describe a reflection from an infinite flat surface.
To prevent too much model flexibility which exceeds the constraining capability of the low S/N spectra, we fix the power-law slope at 1.8 and exclude the soft excess component from the Compton-thick model.
Based on such models, we find $40$ sources which are better fitted with the Compton-thick model at a $>95\%$ confidence level, which corresponds to $\Delta C >3.84$ when $\Delta DOF=1$.
Among them, we select $23$ sources whose {\tt constant} has a best-fit value $>7$ and a 90\% lower limit $>2$; since the reflection component of a Compton-thick AGN must be not only significant but also exceptionally strong.

Even though strong hard excesses are found in these sources, further checking is still needed.
In highly obscured cases (with an \N{H} of a few 10$^{23}$ cm$^{-2}$), especially in the low S/N regime, it is hard to discriminate between a highly obscured power-law and a reflection component; a hard excess might be explained by either of them.
To check this possibility, we fit the spectra using our standard Compton-thin model with the secondary power-law activated; so that the soft X-ray emission is fitted with the unobscured secondary power-law, and the hard excess is fitted with the obscured primary power-law.
This model has the same DOF as the Compton-thick model.
We find that for $13$ sources, the double power-law Compton-thin model fits the spectra better, with a best-fit \N{H} below or above $1.5\times 10^{24}$ cm$^{-2}$.
We add a scattered power-law, which has the same $\Gamma$ as the primary power-law but at most $10\%$ of the primary power-law's normalization, to the model of such sources, as mentioned in \S~\ref{Section:softexcess}. Further checking will be applied to them in the next procedure.
For the other $10$ sources, the Compton-thick model fits the spectra better. However, this does not ensure that they are Compton-thick. We note the reason they are better fitted with the Compton-thick model is that their power-law component, as opposed to the hard excess, is obscured with \N{H}$>10^{22}$ cm$^{-2}$; so that this obscured power-law cannot be well fitted with the unobscured power-law model in the Compton-thin model.
In such cases, although a strong hard excess is detected, our explanation of this excess as a reflection caused by a hidden primary power-law which is much stronger than the observed one seems not proper any more.
Because a line-of-sight absorber has been found, to imagine another independent Compton-thick absorber, which is although possible in principle, is over-explaining the data.
Alternatively, the hard excess could be caused by partial-covering obscuration or special geometry of reflecting material, which cannot be constrained with our low S/N spectra. Therefore, we do not consider such sources as Compton-thick.
By now, we have not selected any Compton-thick sources, but have done essential preparations for next procedure.

\subsubsection{Large best-fit \N{H}}
The secondary power-law in our spectral model is a flexible term of ``soft excess''.
In highly obscured cases, it can be used to describe a scattered power-law which could dominate the observed soft-band flux.
As mentioned above, when a strong hard excess is found, the secondary power-law is activated in the final model, so that the primary power-law appears as a hump in the hard-band spectrum.
This mechanism also works in some other cases where soft excess component is found in \S~\ref{Section:softexcess}.
Our second Compton-thick AGN identification procedure is based on the best-fit \N{H} obtained with the final spectral model.

Here we replace the absorption model {\tt zwabs} with the {\tt plcabs}, which gives more accurate \N{H} measurement in the high \N{H} cases, see \S \ref{Section:plcabs} for details.
In the high \N{H} cases, with a large uncertainty, the measured \N{H} often has an error range crossing the Compton-thick defining threshold of \N{H}=$1.5\times 10^{24}$ cm$^{-2}$.
For simplicity, we just select Compton-thick sources by comparing the best-fit \N{H} with $1.5\times 10^{24}$ cm$^{-2}$.
Excluding the Compton-thick candidates as selected in last section, we find $22$ Compton-thick sources with a best-fit \N{H} $> 1.5\times 10^{24}$ cm$^{-2}$.

\subsubsection{Narrow Fe K$\alpha$ line}
\label{Section:CthickFeline}
The presence of a strong narrow Fe K$\alpha$ line (EW $\gtrsim 1$~keV) is also an indicator for Compton-thick AGN \citep[e.g.,][]{Levenson2002}.
The Fe line EW is positively correlated with \N{H} \citep[e.g.,][]{Leahy1993,Bassani1999,Liu2010}, since higher line-of-sight \N{H} depresses the continuum but not the line. We will show this correlation later in \S\ref{Section:Feline}.
Generally, strong Fe K$\alpha$ lines with EW $\sim$ 1~keV are detected in highly obscured AGNs; unobscured AGNs do not display such strong lines.
For each source with a best-fit rest-frame Fe K$\alpha$ line EW $>1$~keV, if the \N{H} is $<10^{21}$ cm$^{-2}$, we consider its slightly obscured power-law emission as a scattered component rather than the primary power-law; and thus they are likely Compton-thick sources, whose expected hard excess is not detected because of the low S/N.
Besides the Compton-thick candidates found above through the continuum fitting, we find no extra ones with this method.
The low efficiency of this method is discussed in \S~\ref{Section:CthickResult}.

\subsubsection{Mid-infrared 12$\micron$ luminosity}
\begin{figure}[htbp]
\epsscale{1}
\plotone{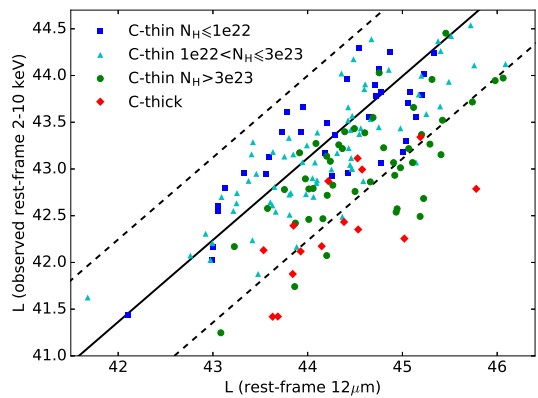}
\caption{
Scatter plot of the rest-frame 2-10~keV observed X-ray luminosity and the total rest-frame 12$\micron$ luminosity.
The sources are divided into four subsamples according to the \N{H}.
The solid line corresponds to the best-fit L$_{2-10~keV,\ intrinsic}$ -- $\log L_{12\micron}$ correlation, and the dashed lines represent its 2$\sigma$ scatter.
}
\label{fig:X_IR}
\end{figure}

Waste heat in the mid-infrared (MIR) band of AGN is an efficient tool to find deeply buried AGN \citep[e.g.,][]{Daddi2007,Fiore2008,Fiore2009,Gandhi2009,Georgantopoulos2009,Severgnini2012,Asmus2015,Stern2015,Corral2016,Isobe2016}.
We cross-correlate our sample with the GOODS-Herschel catalog \citep{Elbaz2011} with a maximum separation of 1$\arcsec$, and find 167 MIR counterparts.
These sources are provided with 8$\micron$, 24$\micron$, and 100$\micron$ fluxes, although some objects are not detected in all the bands.
Since all of our sources are AGN, we assume a power-law shape of the MIR spectrum and measure the rest-frame 12$\micron$ fluxes of these sources by simple interpolation.
This assumption is not necessarily true, as in some cases, the MIR spectrum of AGN may deviate from an ideal power-law because of a strong star burst component or Polycyclic aromatic hydrocarbon (PAH) features \citep[e.g.,][]{Ichikawa2012,Ichikawa2017,Georgantopoulos2013,DelMoro2016}. Therefore, the measured 12$\micron$ flux can be up to a few times higher than the genuine MIR emission from the AGN's torus.

Based on the sources with \N{H}$<10^{23}$ cm$^{-2}$, we perform Orthogonal Distance Regression (ODR) between the MIR luminosity and the absorption-corrected 2-10~keV luminosity $\log$L$_{2-10~keV,\ intrinsic}$ (will be presented in \S\ref{Section:result}), and find $\log$L$_{2-10~keV,\ intrinsic}$ = $\log$L$_{12\micron}$ $\times 0.88 + 4.56$, with a 1$\sigma$ scatter on $\log$L$_{2-10~keV,\ intrinsic}$ of 0.44, as shown in Figure~\ref{fig:X_IR}.
From the best-fit model of the 7Ms stacked spectra, we measure the absorbed rest-frame 2-10~keV luminosity $\log$L$_{2-10~keV,\ observed}$, and plot it with the rest-frame 12$\micron$ luminosity in Figure~\ref{fig:X_IR}.
The sources are divided into four subsamples according to the \N{H} measured with our final model. Average \N{H} is used in the cases of varying \N{H}.
Despite the large scatter, it is still visible that sources with higher \N{H} have lower observed X-ray luminosity.
If one source has $\log$L$_{2-10~keV,\ observed}$ below the 2$\sigma$ confidence interval of the above correlation, that is, 
$\log$L$_{2-10~keV,\ observed}$ $<$ $\log$L$_{12\micron}$ $\times 0.88 + 4.56 - 2\times 0.44$ (the dashed line in Figure~\ref{fig:X_IR}),
we consider it as highly obscured, but not necessarily Compton-thick.
Considering the inaccuracy of our MIR flux measurements, we are adopting a conservative selection rule here.
For X-ray sources satisfying this selection rule, intrinsic absorption with an \N{H} well above $10^{23}$ cm$^{-2}$ is expected.
If the measured \N{H} of such a source is less than $10^{22}$ cm$^{-2}$, we consider its soft-band continuum as dominated by scattered power-law and select it as a Compton-thick candidate, similar as done in \S~\ref{Section:CthickFeline}.
Besides the ones already identified above, no extra Compton-thick candidates are found with this method either.

\section{RESULTS OF SPECTRAL ANALYSES}
\subsection{Basic properties}
\label{Section:result}
We report the source net counts and observed fluxes in Table~\ref{table:flux}.
The net counts are measured directly from the source and background spectra.
The observed fluxes are calculated on the basis of our best-fit models.

We present the spectral-fitting results in Table~\ref{tab:data}, which contains:
\begin{enumerate}[nolistsep]
\item Source ID in the 7Ms CDF-S catalog \citep{Luo2017}
\item Source redshift and quality flag of the redshift \citep{Luo2017}
\item Flag of \N{H} variation.
\item intrinsic absorption \N{H}
\item primary power-law slope $\Gamma$ 
\item observed 0.5-2.0~keV net count rate
\item observed 2.0-7.0~keV net count rate
\item absorption corrected rest-frame 2-10~keV luminosity
\item sample completeness flag, which will be defined later in \S\ref{Section:completeNH}
\item existence of soft excess 
\end{enumerate}

\begin{figure}[htbp]
\epsscale{1}
\plotone{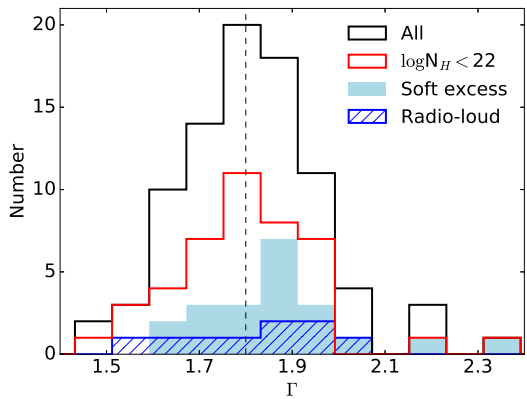}
\caption{The distribution of best-fit $\Gamma$ for the $95$ sources whose $\Gamma$ are well constrained. The vertical line shows the median value of this distribution ($\Gamma = 1.8$). The red histogram shows the subsample with \N{H}$<10^{22}$ cm$^{-2}$. The light-blue filled part corresponds to the sources which have soft excess component detected. The blue dashed part shows the $9$ radio-loud sources.}
\label{fig:Gm_distri}
\end{figure}

We are able to put good constraints on $\Gamma$ for $95$ sources (see \S\ref{Section:strategy}).
The distribution of the parameter $\Gamma$ has a median value of $1.82$ and a standard deviation of $0.15$, similar to that found by \citet{Tozzi2006} and \citet{Yang2016} for CDF-S AGNs, as shown in Figure~\ref{fig:Gm_distri}.
Having an 1$\sigma$ confidence interval of (1.80,1.83), which is measured by a bootstrap method, the median $\Gamma$ is consistent with $1.8$.
For the rest of the sources, $\Gamma$ is fixed at $1.8$ during the spectral fitting.

We check a few factors that may affect the measurement of $\Gamma$, including \N{H}, soft excess, and radio-loudness.
First, we split these sources into two subsamples with \N{H}$<10^{22}$ cm$^{-2}$ and \N{H}$\geqslant 10^{22}$ cm$^{-2}$ (see Figure~\ref{fig:Gm_distri}). They show identical $\Gamma$ distributions, with a KS-test probability as low as 30\% to be different.
Then we compare the sources with and without soft excess component detected (see Figure~\ref{fig:Gm_distri}) to make sure the addition of an secondary, soft power-law does not cause a flat $\Gamma$.
The distributions are not significantly different (KS-test probability 80\%), and the $\Gamma$ of sources with soft excess components seem slightly steeper. Therefore, we are not affected by this bias.
Since we are using a secondary power-law to describe approximately a soft excess component which might have a very different origin in different cases, we do not investigate further the properties of them.
To check the impact of radio-loud sources, we match our sources with the CDF-S 1.4~GHz radio sources \citep{Bonzini2013} within a separation of 2$\arcsec$ and find 63 counterparts; 13 of them which have a 1.4~GHz to V band flux ratio $>10^{1.4}$ are classified as radio-loud by \citet{Bonzini2013}.
We plot nine of them which have free $\Gamma$ in Figure~\ref{fig:Gm_distri}.
They show an identical $\Gamma$ distribution to the other sources, with a KS-test probability as low as 20\% to be different.

\subsection{Fe lines}
\label{Section:Feline}
\begin{figure}[htbp]
\center
\epsscale{1}
\plotone{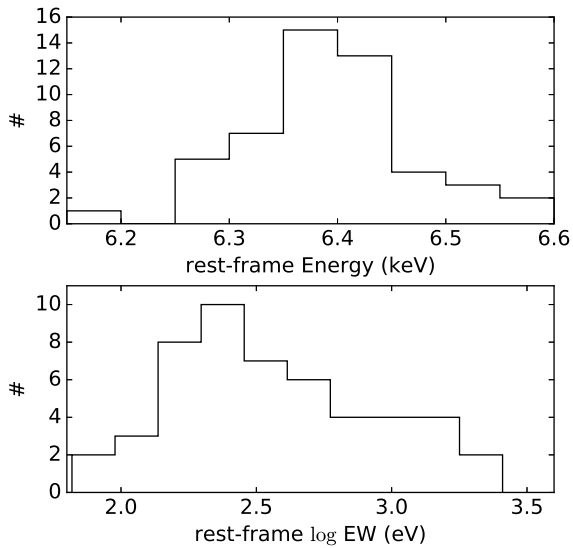}
\caption{Distributions of the central energy and EW of the detected narrow Fe K$\alpha$ lines.
The line energies are consistent with 6.4~keV and the width of the energy distribution is due to measurement uncertainties.
}
\label{fig:hist_lines}
\end{figure}

A narrow Fe K$\alpha$ line is detected in $50$ sources and a broad one is detected in $5$ sources, as listed Table~\ref{table:Fe} and Table~\ref{table:broadFe}.
Their line energies, widths (for broad line), and EWs are measured by fitting the 7Ms stacked spectra (method B).
The low S/N of the broad lines do not allow us to fit them with a physical model.
They could be relativistic broad lines from the inner region of the accretion disk, or might be a blending of multiple lines at different ionization levels.
Below we discuss the properties of the narrow lines.
The distributions of the line central energy and EW are shown in Figure~\ref{fig:hist_lines}.
Our detection procedure ensures that each line energy is consistent with 6.4~keV, which indicates an origin from cold neutral gas.
The scatter in the distribution is caused by measurement uncertainty.
The EWs of the sources span a large range from a few hundred to a few thousand eV.

\begin{figure}[htbp]
\center
\epsscale{1}
\plotone{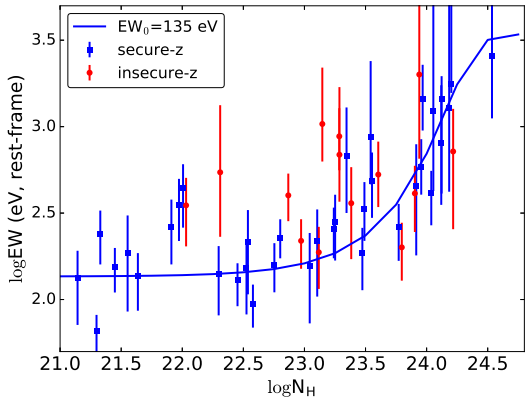}
\caption{Scatter plot of $\log$\N{H} and $\log$EW of the detected narrow Fe K$\alpha$ line.
Sources with secure and insecure redshift measurements are plotted in blue and red, respectively.
The blue line shows our model of the line with an \N{H}-independent flux and EW=135 eV when unobscured.
}
\label{fig:NH_EW}
\end{figure}

In Figure~\ref{fig:NH_EW} we show the rest-frame EW of the detected iron lines versus \N{H}.
For sources with varying \N{H}, an average value in log space $\langle\log$\N{H}$\rangle$ is used.
Considering only the sources with secure spectroscopic redshifts, a strong correlation is shown with a Spearman's rank probability $>99.999\%$.
We build a simple model to examine this correlation.
As reflection emission, the narrow Fe K$\alpha$ line originates from circumnuclear materials such as the outer region of the accretion disk, the broad line region, or the dusty torus \citep[e.g.,][]{Shu2010,Shu2011}.
It is justifiable to consider the strength of this line emission as independent of \N{H}, which is determined by the line-of-sight material.
We assume an \N{H}-independent line flux, and use our standard model to measure the flux density of the continuum at 6.4~keV at a specific \N{H}; then an EW is calculated by dividing the line flux by the continuum flux density.
Fitting this model to the sources with secure redshifts, as shown in Figure~\ref{fig:NH_EW}, the line flux can be calculated.
We find that this flux corresponds to an EW of 135 eV when unobscured, with a 1$\sigma$ scatter of 80--230 eV. It is broadly consistent with the EW measured for a local type I AGN sample by \citet{Liu2010}.
As shown in Figure~\ref{fig:NH_EW}, the data are well described by our toy model.
However, we remark that this does not mean that the Fe line or continuum strength is independent of \N{H}.
\citet{Liu2010} find that the narrow Fe K$\alpha$ line is 2.9 times weaker in Compton-thin type II AGN than in type I AGN in terms of luminosity; and \citet{Liu2014} find that the intrinsic X-ray emission (unobscured power-law) is 2.8 times weaker in Compton-thin type II AGN than in type I AGN.
Here these \N{H}-dependent effects largely cancel each other.

\subsection{\N{H} variation}
\citet{Yang2016} studied the \N{H} variations of AGN using the 6Ms CDF-S data available at the time of their analysis.
Similarly, they divide the CDF-S observations into four periods.
The first three periods are the same to ours, and the fourth period is supplemented with the last 1Ms of observations in this work.
They found $11$ sources with reliable \N{H} variations.
In this work, we use the 7Ms CDF-S data which are somewhat different, adopt a slightly different selection threshold, and search among a larger population of sources.
For $39$ sources, we find the \N{H} to be variable among $63$ pairs of periods. 
Three sources (ID 898, 328, 252) found by \citet{Yang2016} are not considered as \N{H} variable in this work.
The first one is because of low significance, and the last two are considered as having variable soft excess components.

\begin{figure}[htbp]
\center
\epsscale{1}
\plotone{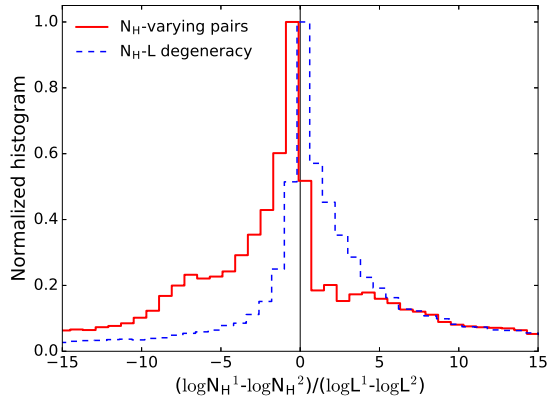}
\caption{
Resampled $k$ distributions, calculated between each pair of \N{H} variable spectra (red) and each pair of random values of the same spectrum (blue). The histograms are normalized to the peak value.
}
\label{fig:k_nH_Gm_L}
\end{figure}

The physical origin of the obscuration variability is uncertain.
The obscuration variability might be driven by the flux variability of the central engine, in the sense that a stronger illumination from the central engine could depress the line-of-sight obscuration through ionizing or blowing away the obscuring material; or the obscuration variability might be independent of the flux variability, if it is caused by obscuring clouds moving across the line-of-sight.
To study the correlation between obscuration variability and flux variability, we define a factor $k$ between each pair of \N{H} variable spectra 
$k =$  ($\log$\N{H}$^1$-$\log$\N{H}$^2$)/($\log$L$^1$-$\log$L$^2$), where $L$ is the 2--10 keV intrinsic luminosity.
A positive $k$ indicates a positive correlation and a negative $k$ signifies a reverse trend of variation.
To take parameter uncertainties into account, we use the {\sl Xspec} ``simpars'' command to generate $1000$ sets of parameters for each of the concerned spectra, and calculate the $k$ factors using them, making a smoothed $k$ distribution.
Considering the degeneracy between $\Gamma$ and \N{H}, to avoid the possible effect of unnoticed $\Gamma$ variation, we exclude the sources whose $\Gamma$ are not well-constrained and thus fixed at 1.8 from the analysis here; and for the remaining $42$ pairs of spectra of $24$ sources, we let $\Gamma$ vary freely among each period when generating the random sets of parameters.

Besides calculating the $k$ between each pair of \N{H} variable spectra, we also calculate this value between the resampled values (random parameters generated with ``simpars'') of each single spectrum, which is named $k_d$, in order to show the natural degeneracy between the parameters.
As shown in Figure~\ref{fig:k_nH_Gm_L}, the \N{H}--$L$ degeneracy (blue histogram) appears as a distribution systematically biased to $k_d>0$; the $k$ factor calculated between our \N{H}-variable pairs shows a tendency to $k<0$, in spite of the existing degeneracy effect.
The $k$ distribution has a large scatter, suggesting that the \N{H} variation could be attributed to different processes in different cases.
The distribution is likely composed of two components, one with a small scatter and the other very large.
Clearly, the small-scatter component has $k<0$. It is consistent with the case of illumination-depressed obscuration, which leads to a reverse variation trend.
In this case, a large flux variation amplitude is needed, while the obscuration variation can not be very large; therefore, $|k|$ is relatively small.
In the other case, an obscuring cloud moving across the line-of-sight could easily cause a dramatic variation of obscuration irrespective of the luminosity, resulting in a largely scattered $k$ distribution.
We remark that by excluding the sources whose $\Gamma$ are fixed at 1.8, we are biased against high-\N{H} sources (with \N{H}$>10^{23}$ cm$^{-2}$), which tend to have a fixed $\Gamma$ according to our analysis method.
Obscuring material with such high \N{H} are harder to be affected by the illumination of the central engine.
Therefore, excluding them is helpful in revealing the reverse trend which occurs more likely in the low-\N{H} cases.

\subsection{Distribution of \N{H} excluding Compton-thick sources}
\label{Section:CthickResult}
Based on the four strategies described in \S\ref{Section:thick}, we classify $22$ (8\% out of $276$) sources to be Compton-thick candidates.
As shown in Figure~\ref{fig:X_IR}, these Compton-thick candidates have significantly lower X-ray luminosities than most of the others, but are not distinctly different from some highly obscured Compton-thin sources because of the large scatter.

Although we have searched for AGN \N{H} variation among the four periods, we do not find any transition between Compton-thin and Compton-thick states.
However, this is not an evidence against the existence of such AGN in our sample.
A faint Compton-thick phase might hide in one period of a source, whose \N{H} is linked to that of other Compton-thin periods because of the low S/N.

The identification of Compton-thick sources through X-ray spectral analyses depends on detailed selection rules.
In the low S/N regime, it is hard to determine whether a highly obscured source is Compton-thick or -thin \citep[see also the comparison of Compton-thick AGN identification results from different works by][]{Castello-Mor2013}.
We note that some Compton-thick candidates identified by previous spectral analyses are classified as Compton-thin in this work, including source
375 \citep[][named BzK8608 therein]{Feruglio2011},
551 \citep[][CXOCDFSJ033229.8-275106]{Comastri2011},
328 \citep[][CXOCDFSJ033218.3-275055]{2002Norman,Comastri2011},
419 \citep[][ID=23]{DelMoro2016},
240 \citep[][XID=191]{Georgantopoulos2013}, and
867 \citep[][XID=634]{Georgantopoulos2013}.
According to our spectral analysis, these sources are found to be Compton-thin but still highly obscured, with \N{H} between $5.5\times10^{23}$--$1.1\times10^{24}$ cm$^{-2}$. Therefore, there is not any strong conflict.
Among the 14 Compton-thick candidates selected in \citet{Tozzi2006}, three are not included in this sample because of their low net counts, two are still identified as Compton-thick, and nine are classified as Compton-thin.
For these nine sources, we find that their low S/N spectra in Period I could indeed be well fitted with a reflection-dominated model. However, they are considered as Compton-thin according to our analysis of the 7Ms data.
Apparently, we are being more conservative in this work in selecting Compton-thick AGN.
The differences between our results and previous classifications are caused by both different data and different spectral models, especially the latter.
Rather than modeling Compton-thin and Compton-thick sources with pure transmitted power-law and pure reflection models, respectively, as done in \citet{Tozzi2006}, we take into account a reflection component in addition to the transmitted power-law in the Compton-thin cases, and account for a scattered power-law component besides the dominating reflection emission in the Compton-thick cases.
With such more realistic models, we can select Compton-thick candidates with higher reliabilities.

We detect neutral Fe K$\alpha$ lines in 4 out of the 22 Compton-thick sources. Although strong Fe K$\alpha$ lines are commonly detected in local bona-fide Compton-thick AGNs, for some Compton-thick AGNs, the Fe K$\alpha$ line might be weak because of high ionization state or low elemental abundance.
For example, NGC~7674, which is identified as Compton-thick by {\sl NuSTAR}, has a weak neutral Fe K$\alpha$ line of EW$\thickapprox$0.4~keV \citep{Gandhi2017}, likely because of its high bolometric luminosity (high ionization state).
Our 22 Compton-thick sources have large redshifts, 20 of them with $z>0.7$ and 6 with $z>2$.
Their bolometric luminosities might be higher than the known local Compton-thick AGNs.
It is possible that they are intrinsically weak in neutral Fe K$\alpha$ line like NGC~7674.
Even if not intrinsically weak, the line detection is severely hampered by the low S/N.
By simulation, \citet{Koss2015} found that the strong Fe K$\alpha$ line (EW$=1.11$ keV) in a Compton-thick AGN NGC~3933 ($z=0.0125$) can be detected at S/N$=3$ by the 4~Ms CDF-S only if $z<0.2$.
This explains why the detection of neutral Fe K$\alpha$ lines is inefficient in identifying Compton-thick AGNs in our high-z sample (\S\ref{Section:CthickFeline}).

\begin{figure}[htbp]
\epsscale{1}
\plotone{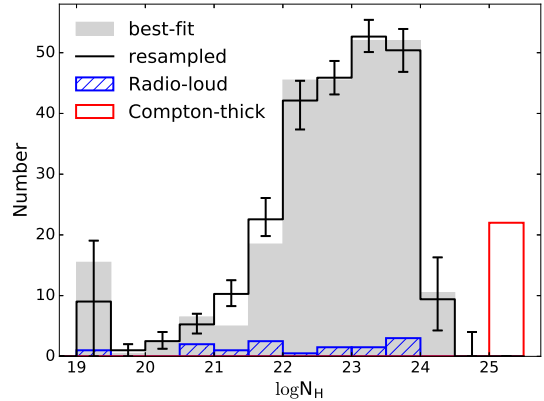}
\caption{Observed \N{H} distribution.
The gray filled histogram shows the best-fit \N{H} distribution of the Compton-thin sources.
All the sources with best-fit $\log$\N{H}$<=$19 are plotted in the leftmost bin.
Compton-thick sources are shown in the red histogram at $\log$\N{H}$>25$.
The black empty histogram shows the resampled \N{H} distribution of the Compton-thin sources, using mock values generated according to the statistical error on \N{H}.
The $1\sigma$ error bars of the histogram are obtained from the scatter of 1000 resampled histograms.
The blue dashed histogram corresponds to the $13$ radio-loud sources.
}
\label{fig:NH_distri}
\end{figure}

After classifying Compton-thick sources, we can study the \N{H} distribution of our sample.
Hereafter, since in the low obscuration regime \N{H} cannot be well constrained, we set all the sources with best-fit $\log$\N{H} below 19 at $\log$\N{H}=19.
Figure~\ref{fig:NH_distri} shows the distribution of the best-fit \N{H}.
To take into account the measurement uncertainty, which is often asymmetrical, we generate 1000 random \N{H} values for each Compton-thin spectrum, assuming two ``half-gaussian'' profiles on the lower and upper side of the best-fit value using the lower and upper 1$\sigma$ error as $\sigma$ of the profiles.
With these 1000 sets of \N{H} values, we obtain 1000 histograms, and plot the median histogram and the corresponding $1\sigma$ errors in Figure~\ref{fig:NH_distri}.
Although, the resampling approach smooths the \N{H} distribution slightly, it is still better to use the resampled values, because the \N{H} uncertainty is nonuniform among the sources.
The uncertainty is larger for sources with lower S/N (net counts), in other words, with lower luminosity or higher redshift.
It is also larger for high-$z$ sources with low \N{H}, since the observed-frame 0.5--7~keV band probes a harder rest-frame band, where the obscuration feature becomes less prominent.
It is important to take these factors into account in order to study the intrinsic distribution of \N{H} and the luminosity- and redshift-dependences of \N{H}.
By now, this distribution is simply the distribution of intrinsic absorption measured across the sample, including the statistical errors.
We correct for the selection effects to derive the intrinsic absorption distribution across the AGN population in \S~\ref{Section:NH_distr_evolu}.

In Figure~\ref{fig:NH_distri}, we also show the matched radio-loud sources from \citet{Bonzini2013}.
They show a flatter \N{H} distribution as opposed to the other sources which congregate at high \N{H}, with a KS test probability of 99\% to be different.
These radio-loud AGNs are likely a mix of mildly absorbed and strongly absorbed galaxies, similar to that found by \citet{Tundo2012}.
Constituting only $\lesssim 5\%$ of the sample, these radio-loud sources have no significant effect on the results.
We consider all the sources as a whole population in the following analyses.

\subsection{Correlation between spectral parameters}

\begin{figure}[htbp]
\epsscale{1}
\plotone{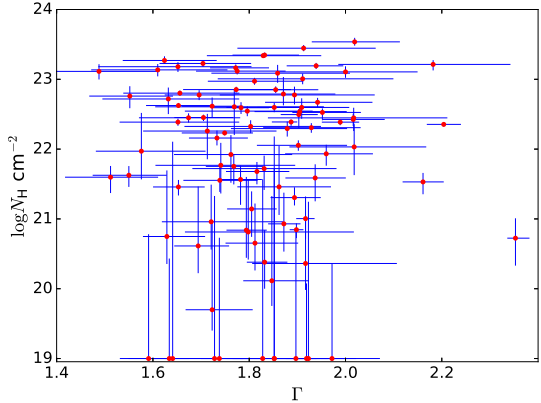}
\caption{Scatter plot of the best-fit $\Gamma$ and \N{H}. Only the sources with free $\Gamma$ are shown here.  Error bars correspond to $1\sigma$.}
\label{fig:NH_Gm}
\end{figure}

The scatter plot of $\Gamma$ and \N{H} is shown in Figure~\ref{fig:NH_Gm}.  Only sources with measured $\Gamma$ are shown.
We check for possible correlation between $\Gamma$ and \N{H}, since a degeneracy between the two parameters could show up at low S/N.  In particular, high \N{H} can be accommodated with high $\Gamma$.
We find a Spearman's rank correlation coefficient of $0.04$, which corresponds to a probability of 30\% to reject the null hypothesis that they are uncorrelated.
Therefore no correlation is found and our spectral fitting strategy is not significantly affected by this kind of bias. 

\begin{figure}[htbp]
\epsscale{1}
\plotone{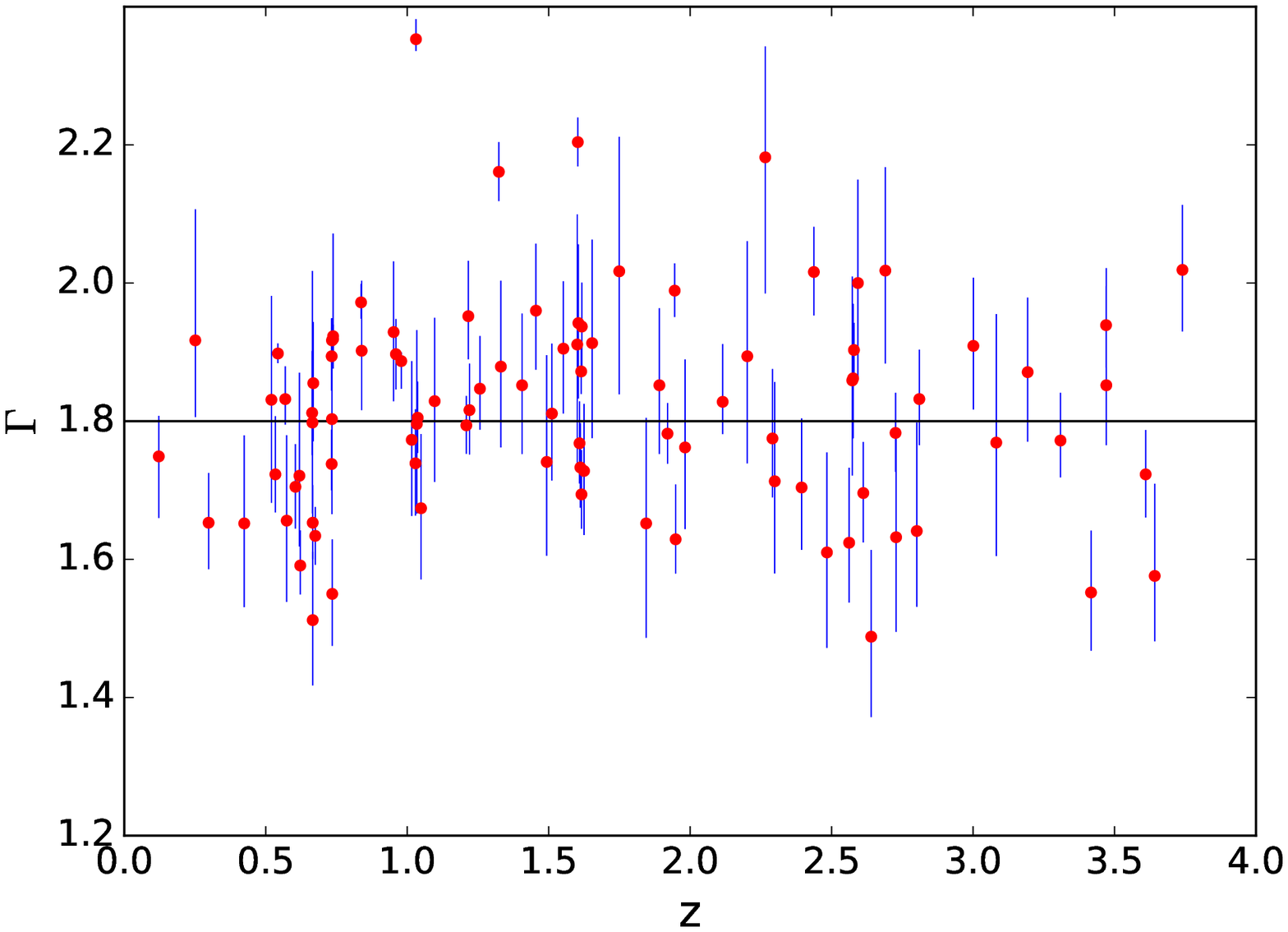}
\plotone{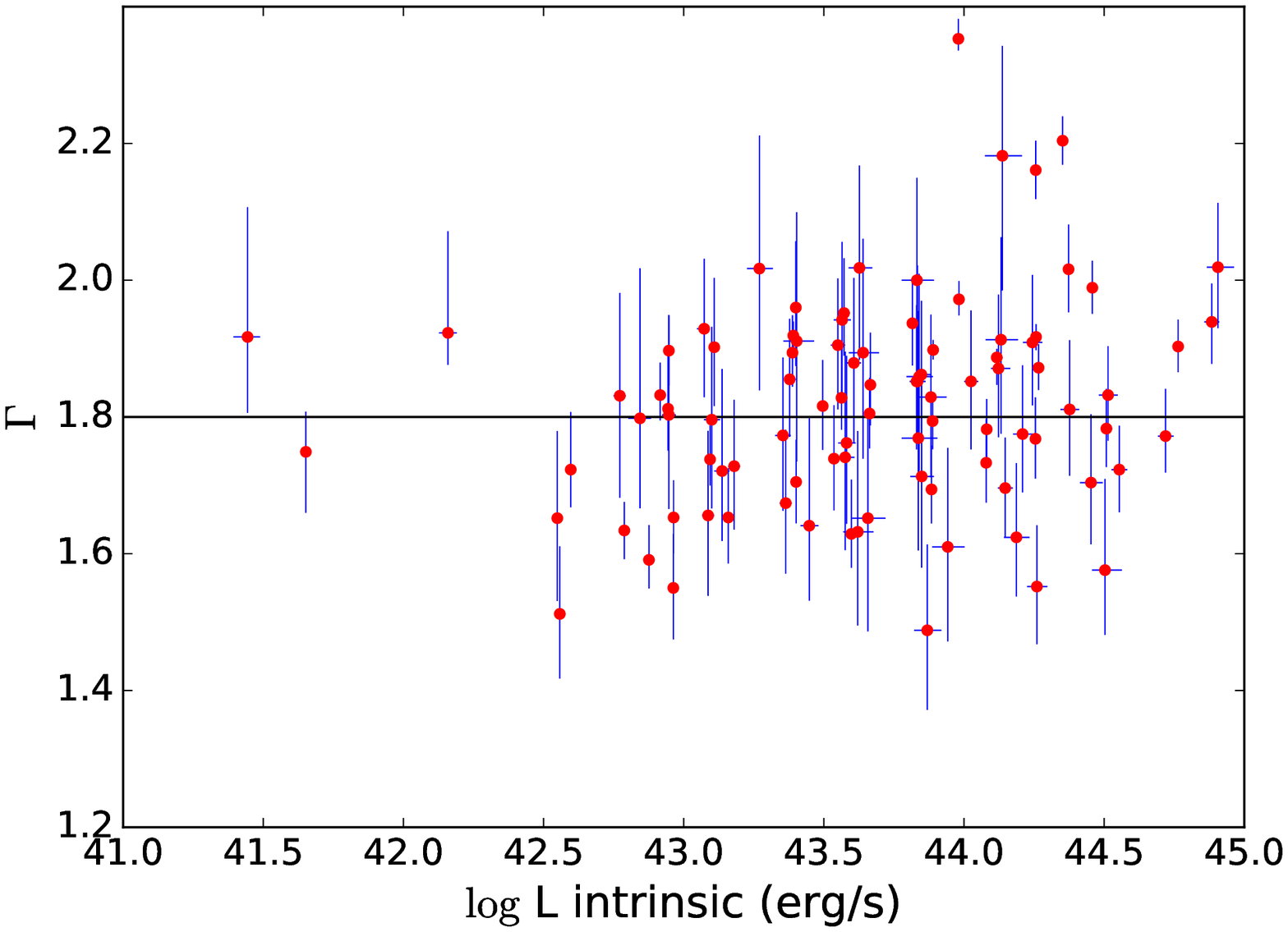}
\caption{Upper panel: scatter plot of best-fit $\Gamma$ versus redshift.
Lower panel: best-fit $\Gamma$ versus rest-frame 2--10~keV intrinsic luminosities.
Only the sources with free $\Gamma$ in the model are plotted. The line corresponds to the median $\Gamma$ of $1.8$. Error bars correspond to $1\sigma$.}
\label{fig:Gm_z_L}
\end{figure}

\begin{figure}[htbp]
\epsscale{1}
\plotone{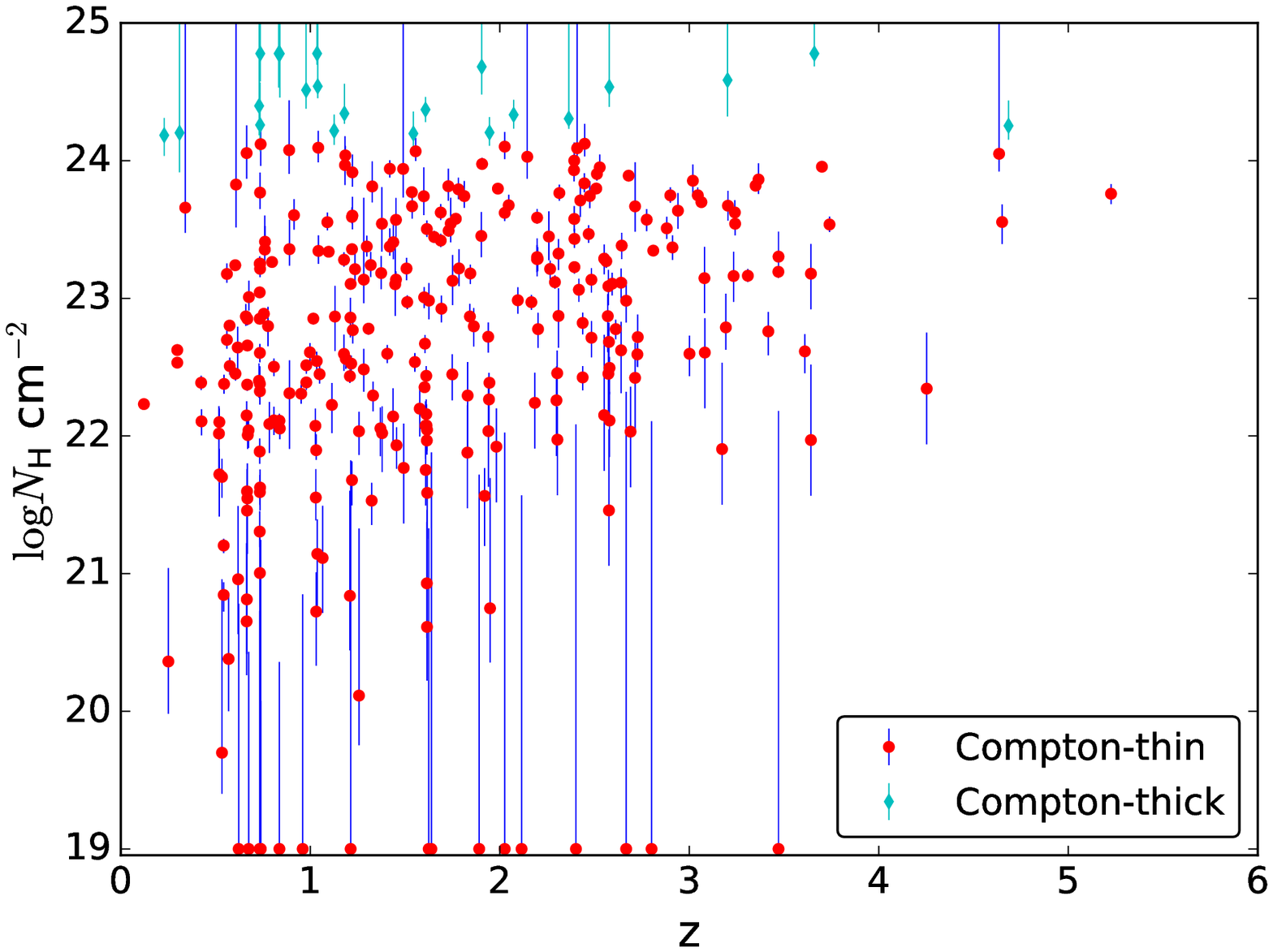}
\plotone{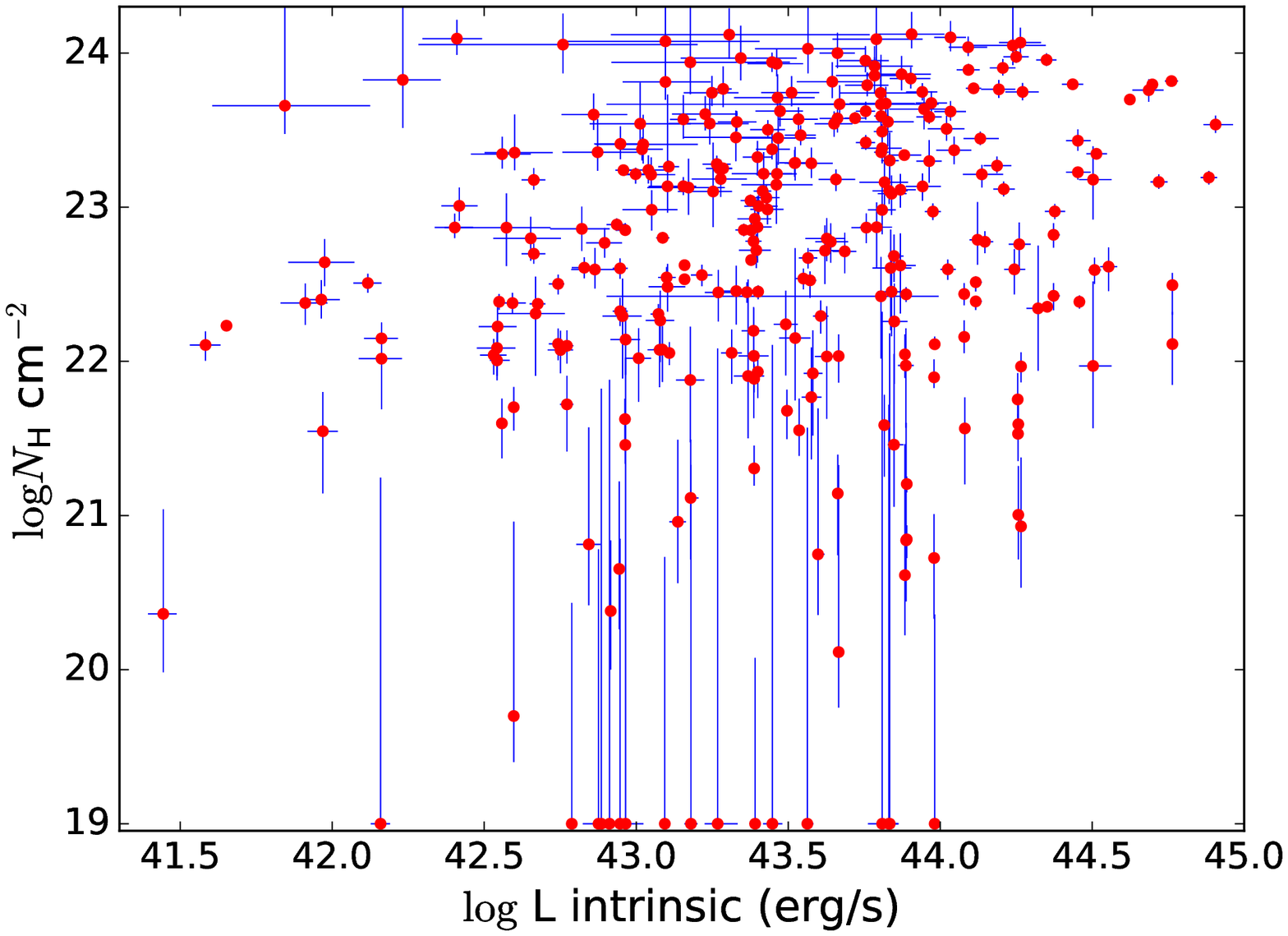}
\caption{Best-fit \N{H} versus redshift. Error bars correspond to $1\sigma$. 
Some sources with very low or very high \N{H} have very large uncertainties.
Scatter plot of the best-fit \N{H} and rest-frame 2--10~keV unabsorbed luminosities of the Compton-thin sources. Error bars correspond to $1\sigma$. }
\label{fig:NH_z_L}
\end{figure}

In Figures \ref{fig:Gm_z_L}, we show the scatter of $\Gamma$ with redshift and intrinsic luminosity.
There is no correlation between $\Gamma$ and $z$, with a Spearman's coefficient as low as $0.02$.
$\Gamma$ and $L$ show a slight correlation, with a Spearman's coefficient of $0.20$, corresponding to a probability of 95\%.
A simple linear fit results in a slope of $0.05\pm0.02$.
The $\Gamma$--$L$ correlation, if any, could be attributed to the positive correlation between $\Gamma$ and $\lambda_{Edd}$ \citep[e.g.,][]{2006Shemmer,2008Shemmer,2009Risaliti,Brightman2013,Fanali2013}, which is expected to arise from the more efficient cooling in the coronae of higher $\lambda_{Edd}$ systems.
However, considering the weak correlation and the large measurement uncertainties, we ignore this effect in this work.

In Figure \ref{fig:NH_z_L}, we show the correlation of \N{H} with redshift and intrinsic luminosity.
For the Compton-thin sources, the Spearman's correlation coefficient is $0.33$ between the best-fit \N{H} and $z$, and $0.22$ between \N{H} and intrinsic luminosity, both corresponding to a probability of $>99.9\%$.
We discussed these correlations after evaluating several selection biases in Section \ref{Section:evolution}.

\subsection{{\sl C}-statistic Versus {\sl W}-statistic}
\begin{figure}[htbp]
\plotone{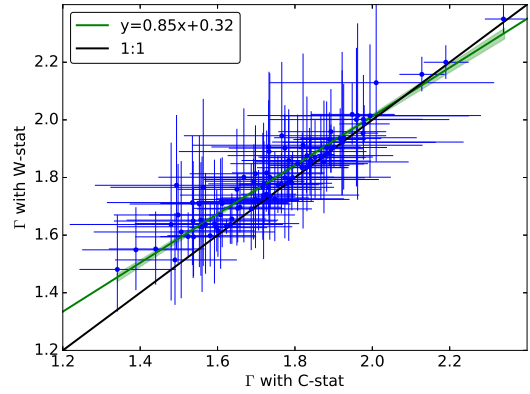}
\plotone{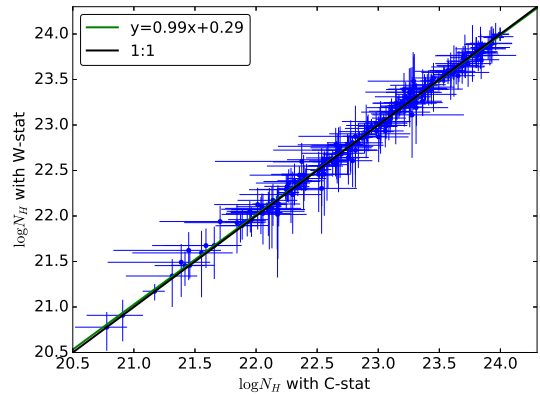}
\caption{Power-law slope and column density as obtained using spectral fitting method C (standard {\sl C} statistic) and D ({\sl W} statistic).
The best-fit lines and $1\sigma$ confidence intervals are plotted with the green lines and green shadings.
}
\label{fig:C_W}
\end{figure}

As introduced in Section \ref{Section:methods}, both methods C ({\sl C}-statistic) and D ({\sl W}-statistic) can be used to obtain the final spectral fitting results. In principle, method D uses a more accurate statistical approach.
Here we explore whether these two methods result in significant differences.
In Figure~\ref{fig:C_W}, we compare the best-fit power-law slopes and the column densities obtained using the two methods.
Only the sources with $\Gamma$ derived from fitting are involved in the $\Gamma$ comparison, and only the sources whose $\log$\N{H} are below 24 according to both methods and whose 90\% confidence intervals of $\log$\N{H} are less than 1 are involved in the \N{H} comparison.
Clearly, \N{H} is not affected by the choice of statistical method, as shown by the ODR best-fit line $y=(0.99\pm0.01)x+0.29\pm0.14$ in Figure~\ref{fig:C_W}.
However, at low values, $\Gamma$ measured using the {\sl W} statistic is typically higher ($\Gamma_W-\Gamma_C=0.09$ at $\Gamma_C=$1.4) than that measured using {\sl C} statistic. 
Meanwhile, the $<1$ slope of the $\Gamma_W$--$\Gamma_C$ best-fit line $y = (0.85\pm0.03)x + 0.32\pm0.05$ suggests that a slightly smaller scatter in the $\Gamma$ distribution of the sample will be obtained using the {\sl W}-statistic.
However, because of the low S/N of our data, the difference caused by different statistical methods is small compared to the uncertainty. Therefore, we conclude that this effect is not significant.

\subsection{{\tt zwabs} Versus {\tt plcabs}}
\label{Section:plcabs}
\begin{figure}[htbp]
\plotone{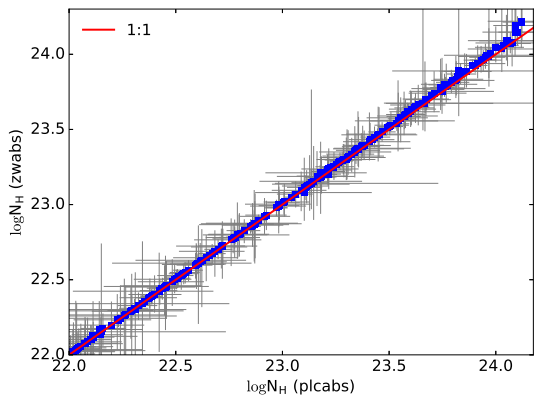}
\caption{N{H} measured using {\tt zwabs} and {\tt plcabs} with 1$\sigma$ error for the Compton-thin sources, whose \N{H} are well-constrained.
}
\label{fig:p_W}
\end{figure}

As we mentioned in \S~\ref{Section:model}, the {\tt zwabs} model considers only photoelectric absorption but not Compton scattering.
To check how this affects the results, we replace the {\tt zwabs*powerlaw} in the final model with {\tt plcabs} \citep{Yaqoob1997}, which approximately takes Compton scattering into account by modeling the X-ray transmission of an isotropic source located at the center of a uniform, spherical distribution of matter.
We set the maximum number of scattering to $3$ and the high-energy cut-off e-folding energy to 300~keV.
In Figure~\ref{fig:p_W}, we compare the \N{H} measured with this model and with {\tt zwabs}. The difference starts to appear above $10^{23}$ cm$^{-2}$, and is very small (by a factor of $\lesssim 0.3\%$ at $\log$\N{H}=24 cm$^{-2}$) compared to the uncertainty of our data.
We conclude that this shortcoming of {\tt zwabs} is negligible in this work. However, we still report the \N{H} measured with {\tt plcabs}, and use it in identifying Compton-thick AGNs.

\subsection{Comparison with previous works on X-ray spectral analyses of CDF-S sources}
\label{Section:comparison}

Several works have presented X-ray spectral analyses of CDF-S sources over the last 15 years \citep[e.g.,][]{Tozzi2006,2014Brightman,Comastri2013,Buchner2014}.
The improvement presented in this work consists in not only the much longer exposure time, but also the improving calibration of {\sl Chandra} data, the updated redshift measurements, and different spectral analysis approaches.
We compare our spectral fitting results with those obtained in \citet{Tozzi2006} and \citet{Buchner2014}, in order to understand how these different aspects affect the spectral analyses.

\begin{figure}[htbp]
\epsscale{1}
\plotone{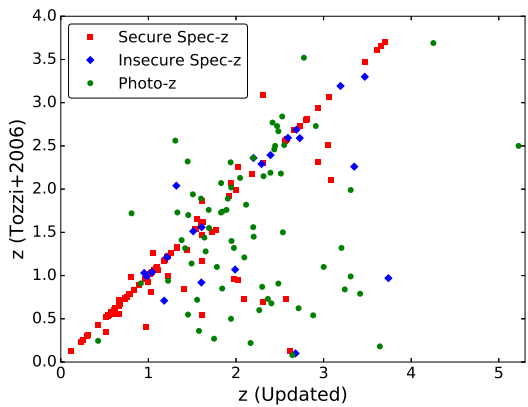}
\caption{Three types of redshifts adopted in this work, comparing with those used in \citet{Tozzi2006}.
Red square: secure spectroscopic; blue diamond: insecure spectroscopic; green circle: photometric.
}
\label{fig:compTozzi_z_z}
\end{figure}

\begin{figure}[htbp]
\epsscale{1}
\plotone{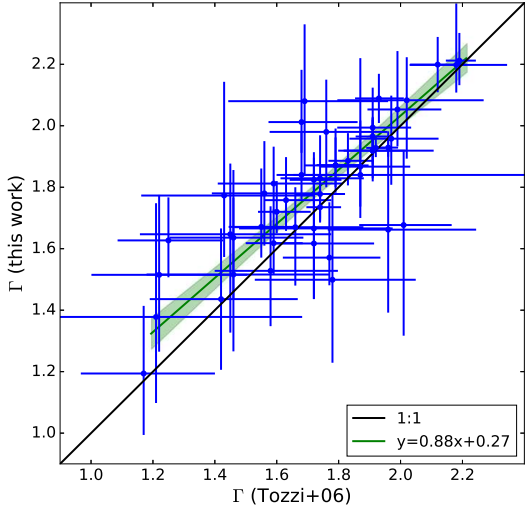}
\caption{Photon indices measured in \citet{Tozzi2006} and in this work in Period I for the sources with consistent redshifts. Sources whose photon indices are fixed at 1.8 are excluded. Errors correspond to $1\sigma$.
The linear fitting result and $1\sigma$ confidence interval are plotted with the green line and shading.}
\label{fig:compTozzi_Gm_Gm}
\end{figure}

\begin{figure}[htbp]
\epsscale{1}
\plotone{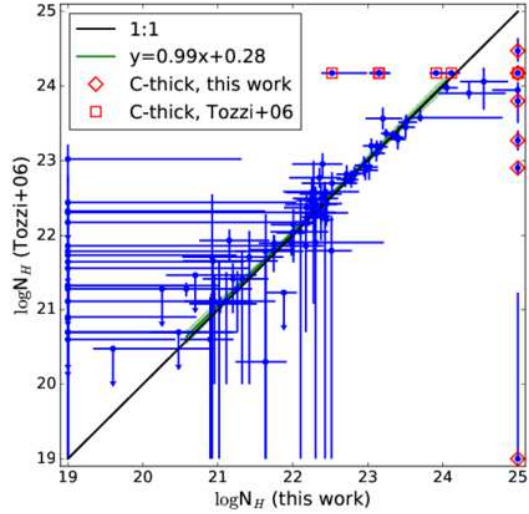}
\caption{Column densities \N{H} with $1\sigma$ errors measured in \citet{Tozzi2006} and in this work in Period I for the sources with consistent redshifts.
Sources identified as Compton-thick in \citet{Tozzi2006} and in this work are marked with red squares and red diamonds.
We set the $\log$\N{H} of our Compton-thick sources at 25.
Linear fitting result and $1\sigma$ confidence interval are plotted with green line and shade.
}
\label{fig:compTozzi_NH_NH}
\end{figure}

As shown in Figure~\ref{fig:compTozzi_z_z}, the redshifts in the CDF-S catalog were dramatically improved in the past decade, mostly thanks to the continuous multi-band follow-up and updated photometric redshift measurements, and in some cases due to the careful multi-band identification in \citet{Luo2017}.
About 50\% of the sources have a redshift different by more than $\Delta z =0.05$ with respect to the values used in \citet{Tozzi2006}, and the updated redshifts tend to be larger.
Below, we limit our comparison with the \citet{Tozzi2006} results to the sources with $\Delta z <0.05$.  We also consider the same period, which corresponds to Period I in our analyses (the first 1 Ms).
For the sources whose $\Gamma$ were fixed at 1.8 in \citet{Tozzi2006}, we fix their $\Gamma$ at 1.8 too.
For the other sources, we compare the best-fit values of $\Gamma$ obtained in the two works (see Figure~\ref{fig:compTozzi_Gm_Gm}), and get an ODR best-fit line of $y=(0.88\pm0.09)x+(0.27\pm0.16)$.
According to this line, at $\Gamma=1.4$ as obtained in \citet{Tozzi2006}, we get $\Gamma=1.5$ in this work.
The higher $\Gamma$ we get is because we use the {\sl W} statistic instead of the {\sl C} statistic and we include a reflection component in the spectral model.
In Figure~\ref{fig:compTozzi_NH_NH} we show the comparison of the best-fit values of \N{H} in the two works.
The uncertainty is very large in the cases of unobscured (\N{H}$<10^{20}$ cm$^{-2}$) low S/N sources.
For the sources which have an \N{H} between $10^{20}$ and $1.5\times 10^{24}$ cm$^{-2}$ according to both works, we perform an ODR fit.
The best-fit line $y=(0.99\pm0.05)x+(0.28\pm1.08)$ is fully consistent with 1:1.
Significant differences occur in the highly obscured regime.
They are caused by the improved spectral models used in this work.
In conclusion, despite using updated calibration, new extraction procedures for the source and the background, and different spectral models, our spectral fitting results are in broad agreement with \citet{Tozzi2006}.

\begin{figure}[htbp]
\epsscale{1}
\plotone{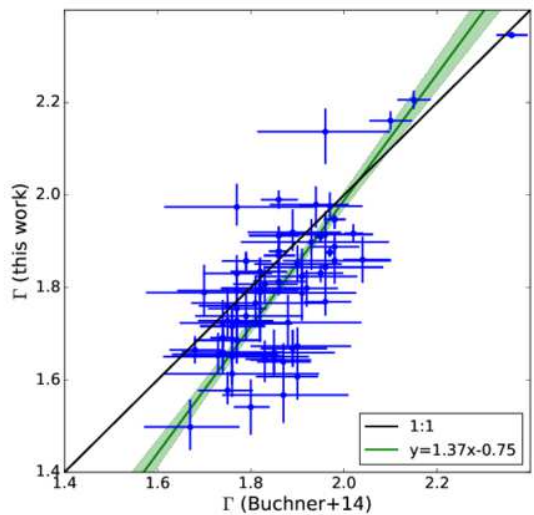}
\caption{Photon indices $\Gamma$ with $1\sigma$ errors measured in \citet{Buchner2014} on the basis of 4Ms CDF-S and in this work on the basis of 7Ms CDF-S. Sources whose photon indices are fixed at 1.8 in this work are not shown. The linear fitting result and $1\sigma$ confidence interval are plotted with the green line and shading.}
\label{fig:compBuchner_Gm_Gm}
\end{figure}

\begin{figure}[htbp]
\epsscale{1}
\plotone{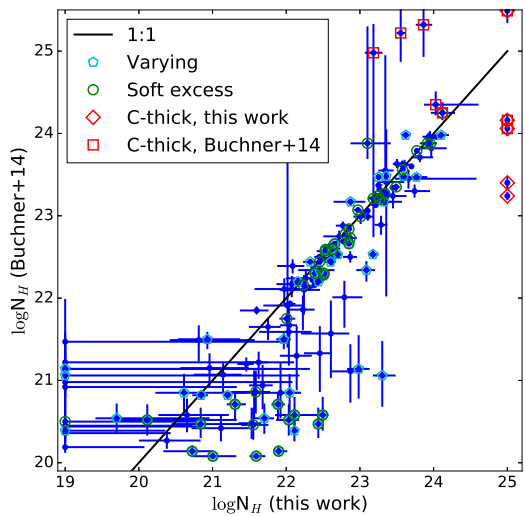}
\caption{Column densities \N{H} with $1\sigma$ errors measured in \citet{Buchner2014} and in this work.
Sources identified as Compton-thick in \citet{Buchner2014} and in this work are marked with red squares and red diamonds, respectively.
We set the $\log$\N{H} of our Compton-thick sources at 25.
If \N{H} is found to vary among different periods in this work, all the values are plotted and marked with cyan pentagons.
Sources having soft excess components, whose spectral model contains an additional power-law, are marked with green circles.
}
\label{fig:compBuchner_NH_NH}
\end{figure}

Similarly, we compare our spectral fitting results with those of \citet{Buchner2014}. Between the 4Ms CDF-S catalog they used and our updated 7Ms catalog, we still found $\geqslant 30\%$ of sources (mostly photometric) with redshifts changed by more than 0.05.
Again we compare only those with $\Delta z<0.05$.
In Figure~\ref{fig:compBuchner_Gm_Gm}, it is clearly shown that the $\Gamma$ values obtained in \citet{Buchner2014} are higher than ours.
As shown in Figure~\ref{fig:Gm_distri}, we found a median $\Gamma$ of $1.8$ in our sample.
The steeper slopes found by \citet{Buchner2014} might be caused by the prior distribution of $\Gamma$ assumed in the Bayesian method, which has a mean value of $1.95$ according to {\sl GINGA} observations by \citet{Nandra1994}, or by neglecting soft excess or setting a stronger reflection component in the spectral model.

In Figure~\ref{fig:compBuchner_NH_NH}, we compare the \N{H} obtained in both works.
Results from the two works are well consistent at around $\log$\N{H}$=23$, while at higher and lower \N{H}, there are large scatters.
In the regime of higher \N{H}, the Compton-thick AGN identifications are different for a few sources, because, as we point out in \S\ref{Section:result}, with limited information, the identification of Compton-thick AGN is highly dependent on the detailed selection method.
Meanwhile, we obtained higher \N{H} for a couple of heavily obscured Compton-thin sources comparing with \citet{Buchner2014}.
In the case of low \N{H}, the scatter is even larger.
These differences could be attributed to the different models we used and in some cases to spectral variations.
\citet{Buchner2014} used a physical torus model with additional reflection and scattering, while we use a phenomenological model in which absorption, reflection and scattering are considered but not physically connected.
\citet{Buchner2014} demonstrated that in the Compton-thin regime, the two approaches can describe the observed spectra of the full sample equally well.
However, various particular sources may prefer different model configurations, as also shown by \citet{Buchner2014}.
Particularly, we set the relative reflection strength of all the sources to a typical value, because the spectral quality does not allow us to put any constraint on it.
For highly obscured sources, whose 2--7~keV fluxes could be dominated by the reflection component, the measured \N{H} and $\Gamma$ are dependent on this configuration.
As this configuration is chosen with the aim of obtaining systematic spectral properties of the sample, it is not necessarily accurate for each particular source.
Another factor is the soft-excess component, the existence of which has a strong impact on the \N{H} measurement.
We add this component to the spectral model only if it is detectable in the spectrum.
Adding it to the soft band, the measured \N{H} on the primary power-law can be dramatically enlarged (see Figure~\ref{fig:compBuchner_NH_NH}).
In Figure~\ref{fig:compBuchner_NH_NH} we also mark sources which are found to have a varying \N{H} among the four observation periods.
Clearly, spectral variation also plays a part in the large differences of \N{H} measurements in some cases.
Considering the degeneracy between \N{H} and $\Gamma$, the difference in $\Gamma$ might be another reason for the \N{H} differences.

In conclusion, although more differences are found between this work and \citet{Buchner2014}, all these differences are understandable, which can be largely attributed to the different spectral models utilized in the two works. 
Both works apply systematic analyses to the whole sample.
But for each specific source, especially low S/N ones, different spectral-analysis strategies import different assumptions into the models, thus leading to different \N{H} measurements.

\section{PROPERTIES OF THE INTRINSIC OBSCURATION}
\label{Section:NH_distr_evolu}

In this section we dissect the sample-selection function and draw general conclusions about the properties of the distribution of intrinsic obscuration \N{H} across the AGN population in two ways.
First, using the sample selected with at least $80$ hard-band net counts, we retrieve the intrinsic \N{H} distribution which is representative of an AGN population that is well-defined according to $L$ and $z$, by correcting the sample selection bias using the known luminosity function of AGN;
Second, instead of correcting our sample for the missing part, we trim off the incomplete part without any assumption about the luminosity function and build a subsample which is complete for the distribution of the column density, in order to probe the dependence of \N{H} on the intrinsic X-ray luminosity and the cosmic epoch.
The supplementary sample is included in the second part.

\subsection{Sample selection function}
\subsubsection{Sky coverage}
\label{Section:skycoverage}

For the combined 7Ms exposure of the CDF-S field, not only does the effective area decrease as a function of the off-axis angle, but also the exposure time at off-axis angles larger than $\sim 8$ arcmin. This is because in the outskirts, abrupt variations in exposure time are due to regions imaged only in some of the 102 exposures, due to the different roll angles.
Taking into account both effective area and exposure time, we build a map of the flux limit across the field, which obviously shows low values (maximum sensitivity) in a limited region around the center, and increasingly higher values toward the edges of the field.
Using this map, we measure the sky coverage as a function of the hard flux corresponding to 80 net counts which is the selection threshold of our sample.

\begin{figure}[htbp]
\epsscale{1}
\plotone{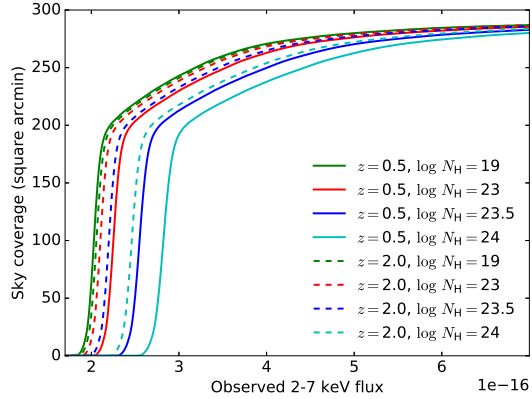}
\caption{The sky coverage as a function of the 2-7~keV observed flux (erg cm$^{-2}$ s$^{-1}$) corresponding to 80 hard-band net counts for a set of \N{H} values at redshifts of 0.5 (solid lines) and 2 (dashed lines).}
\label{fig:skycoverage}
\end{figure}

The sky coverage is relevant to measuring the \N{H} distribution because the correspondence between detected counts and emitted flux depends on the spectral shape.
In particular, intrinsic absorption can dramatically change the conversion factor between them.
In Figure~\ref{fig:skycoverage} we show the sky coverage of the CDF-S for our sample assuming our standard spectral model (absorbed power-law with $\Gamma=1.8$ plus reflection) with a set of different \N{H} and redshift values.
The sky coverage is biased against the detection of sources with high \N{H}.
This bias is less severe at high redshift, because of the inverse K-correction of highly obscured sources.
In \S\ref{Section:sample}, we choose to apply the sample-selection threshold in the hard band in order to reduce the effect of sky coverage on the \N{H} distribution.
We further correct this bias through weighting each source by the reciprocal of the sky coverage corresponding to its observed flux when computing the \N{H} distribution.

\subsubsection{\N{H}-dependent Malmquist bias}
\label{Section:Malm}
A flux-limited survey is typically biased against sources with lower luminosity at higher redshift. This effect is termed Malmquist bias.
We express the Malmquist bias by drawing a detection boundary curve in the space of luminosity and redshift, which represents the flux limit of the sample.
By the general definition, the boundary corresponding to a flux limit is determined by the survey depth, which in our case decreases with off-axis angle.
However, rather than using the flux-limited sample, our aim is to study the \N{H} distribution in a complete sample which is unbiased to the intrinsic X-ray luminosity $L_X$.
Therefore, we have to take into account the dependence of observed flux on \N{H}, and express the Malmquist bias (the boundary curve) as \N{H} dependent.

\begin{figure}[htbp]
\epsscale{1}
\plotone{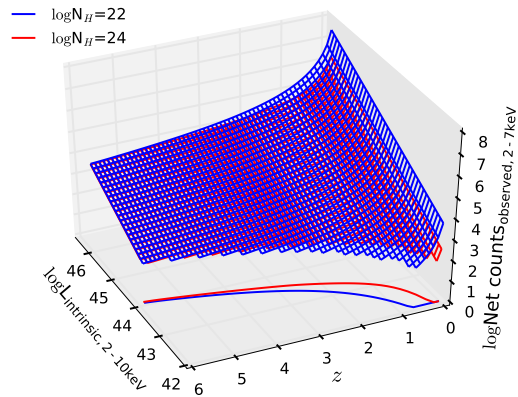}
\caption{Conversion surfaces from any point in the $L_X$--$z$ space to hard-band observed net counts, obtained at the aimpoint with a $\log$\N{H} of 22 (blue grid and line) and 24 (red grid and line). 
Only the parts above 80 net counts of the surfaces are shown.
The lines in the $L_X$--$z$ space at the base are projections of the intersections of the two surfaces with the plane where net counts = 80.
}
\label{fig:L_z_Cts}
\end{figure}

To model the \N{H}-dependent Malmquist bias, we divide the CDF-S field into 9 annular regions according to off-axis angle with steps of 1$\arcmin$ and central radii of $1-9$$\arcmin$.
The circular region within $0.5\arcmin$, which is very small in area compared to the annuli, is assigned to the first annulus.
In each annulus, at each specific $\log$\N{H} between 19 and 25 with a step of 0.5, we convert each point in the $L_X$--$z$ space into observed net counts by running {\sl Xspec} with a model spectrum and the real response files and exposure time.
A visual example of the conversion is shown in Figure~\ref{fig:L_z_Cts}.
We take the standard model which was used in our spectral fitting, that is, 
an absorbed power-law plus a scattered power-law and a reflection, with $\Gamma$=1.8, $R$=0.5, $E_{cutoff}$=300~keV.
The scattered fraction is set to 1.7\%, which is the average scattered fraction for 4~Ms CDF-S sources \citep{Brightman2012}.
At each typical off-axis angle, the averaged response files and exposure time of the sources are used.
Besides doing it in each annulus, we also carry out the whole field averaging case by taking the response files and exposure time averaged from all the sources in the field.

\begin{figure}[htbp]
\epsscale{1}
\plotone{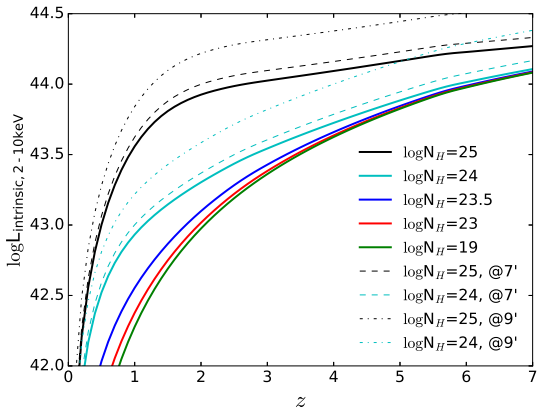}
\caption{The boundary of detectable regions in $L_X$--$z$ space with respect to different \N{H}.
Solid lines correspond to the aimpoint region (off-axis angle $<1.5\arcmin$).
Dashed lines and dash-dotted lines corresponds to off-axis angles of 7$\arcmin$ and 9$\arcmin$, respectively.
}
\label{fig:L_z_boundary}
\end{figure}

Using the conversion surface from $L_X$--$z$ to net counts, the sample-selection threshold of $80$ net counts is in turn converted back to the $L$--$z$ space, defining a detectable boundary curve -- the luminosity limit at each redshift.
In Figure~\ref{fig:L_z_boundary}, we show the detectable boundaries at a few specific \N{H} values at the aimpoint (within $1.5\arcmin$).
The boundary is less affected by \N{H} below $\log$\N{H}=23, thanks to our hard-band sample-selection threshold.
We also plot the boundaries corresponding to $\log$\N{H}=24 and 25 at off-axis angles of $7\arcmin$ (median off-axis angle of our sample) and $9\arcmin$ (maximum) in Figure~\ref{fig:L_z_boundary}, to show the variation of the survey depth.

\subsubsection{Eddington bias}
\begin{figure}[htbp]
\epsscale{1}
\plotone{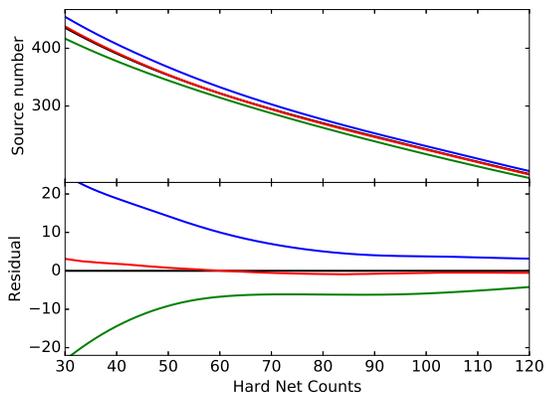}
\caption{In the upper panel, the black line, on which an almost identical red line is superimposed, shows the smoothed cumulative number counts distribution of the CDF-S 7Ms point source catalog.
Convolving the black line with the measurement error of net counts results in the blue line. The green line is the horizontal reflection of the blue line about the black line.
As a consistency check, the green line is convolved with the measurement error, giving rise to the red line, which is found to be almost identical to the black line at $>50$ net counts.
The residuals of the lines to the black line are shown in the lower panel.
}
\label{fig:Eddington}
\end{figure}
As shown in Figure~\ref{fig:hist_counts}, faint sources naturally outnumber bright ones in the CDF-S point-source catalog. Considering that the net-counts measurement could have large uncertainty in the low S/N regime, especially at large off-axis angles where the background is high, the sample-selection threshold on net counts introduces Eddington bias.
In other words, more sources are included than expected because of net-counts measurement uncertainties.
We take this effect into account with an approach as follows.
As shown in Figure~\ref{fig:Eddington}, we convolve the number-counts distribution of the 7Ms CDF-S point source catalog as a function of hard-band net counts with the measurement uncertainty. For each source the net counts is replaced by thousands of Poissonian random values following its measurement uncertainty.
The uncertainty-convolved number counts distribution is flatter, with relatively larger net counts. As shown in Figure~\ref{fig:Eddington}, we take the rightward shift of the number counts caused by the convolution as an estimation of the Eddington bias.
Shifting the number counts leftward by the same amount, we take the horizontal reflection of the uncertainty-convolved distribution about the original distribution as an approximation of the deconvolved number counts.
Such is our pseudo-deconvolving method.
To check its accuracy, we convolve the deconvolved number counts with the measurement uncertainty.
As shown in Figure~\ref{fig:Eddington}, we get a number counts which is almost identical to the original one at $>50$ net counts.

Figure~\ref{fig:Eddington} is truncated at $30$ net counts, because at lower counts, a ``faint end'' effect becomes significant, which is caused by the undetected sources which are not considered in the simulation.
However, $80$ net counts is well above the peak value of the net-counts distribution (see Figure~\ref{fig:hist_counts}), the ``faint end'' effect is negligible at $80$ net counts.

Comparing the original and the deconvolved number counts, with our selection threshold of $80$ net counts, we include $\sim10$ more sources into our sample because of Eddington bias.
Allowing the same sample size, the $80$ net counts threshold on the original number-counts distribution corresponds to $74$ net counts on the deconvolved one.
To correct the Eddington bias, we use this value as an ``effective'' threshold.
Specifically, in the correction described above in \S\ref{Section:skycoverage} and \S\ref{Section:Malm}, we use $74$ instead of the nominal net-count threshold $80$.

\subsection{Intrinsic obscuration distribution}
\label{Section:intrinsic_distri}
\begin{figure*}[htbp]
\epsscale{1}
\plotone{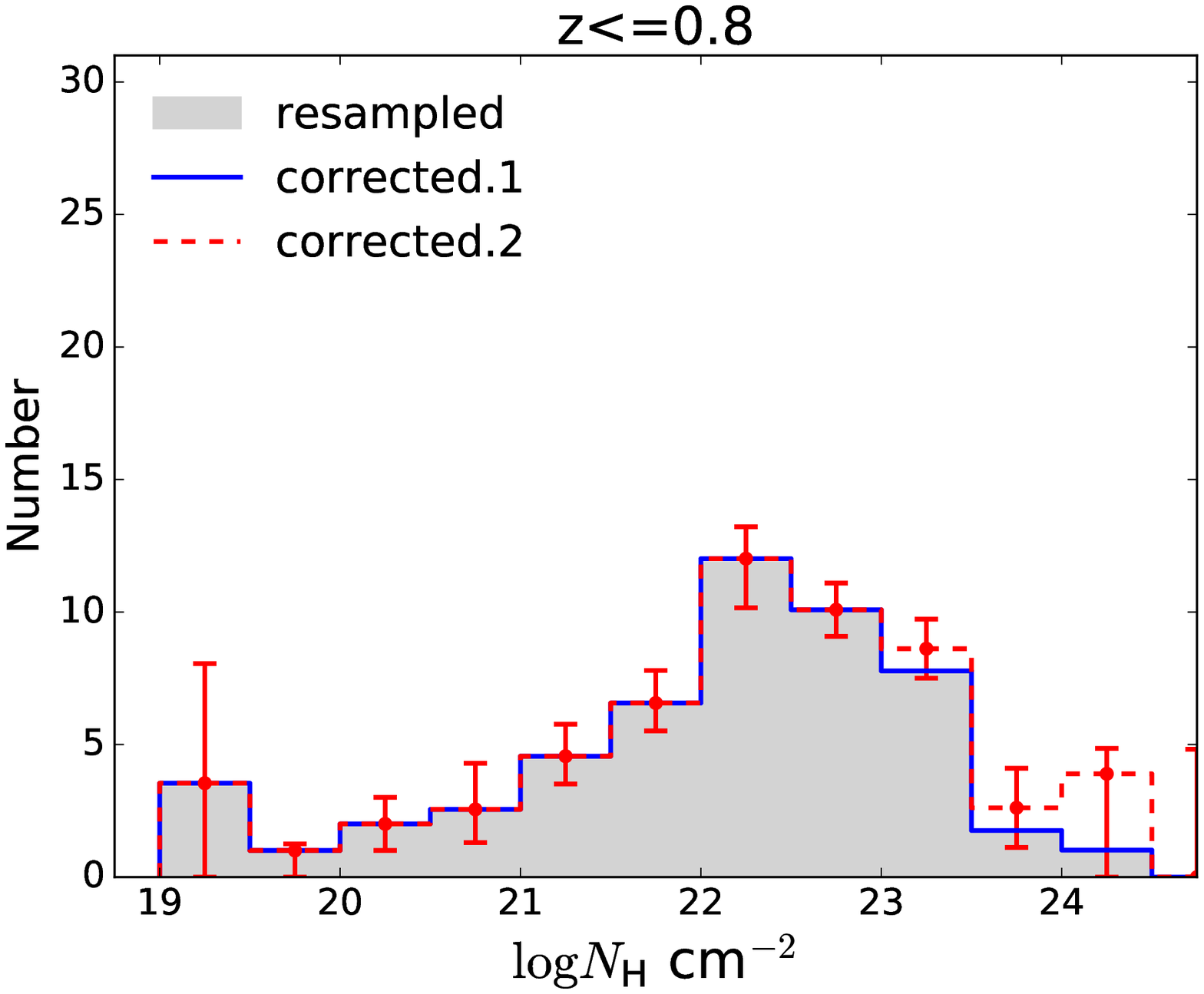}
\plotone{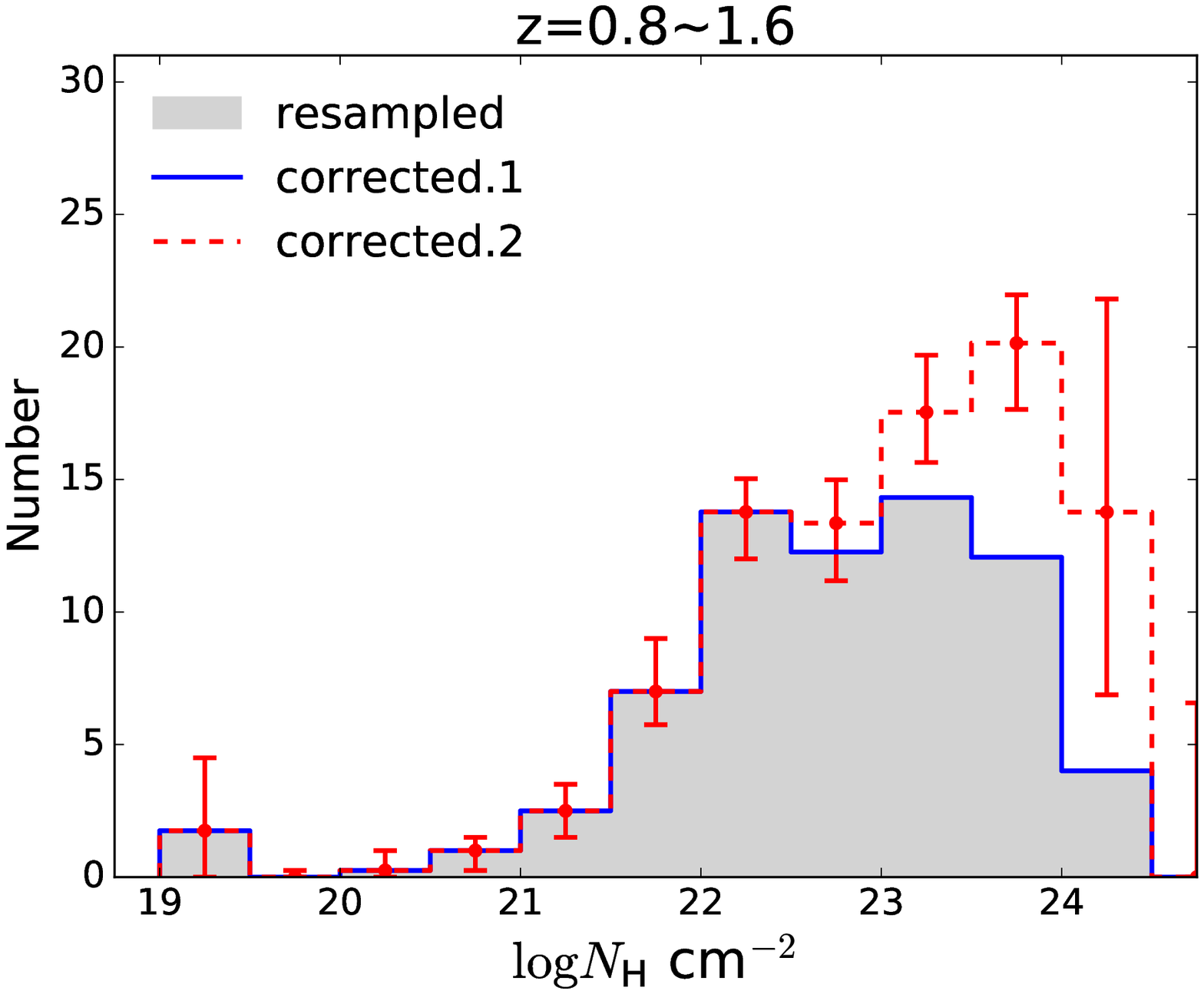}\\
\plotone{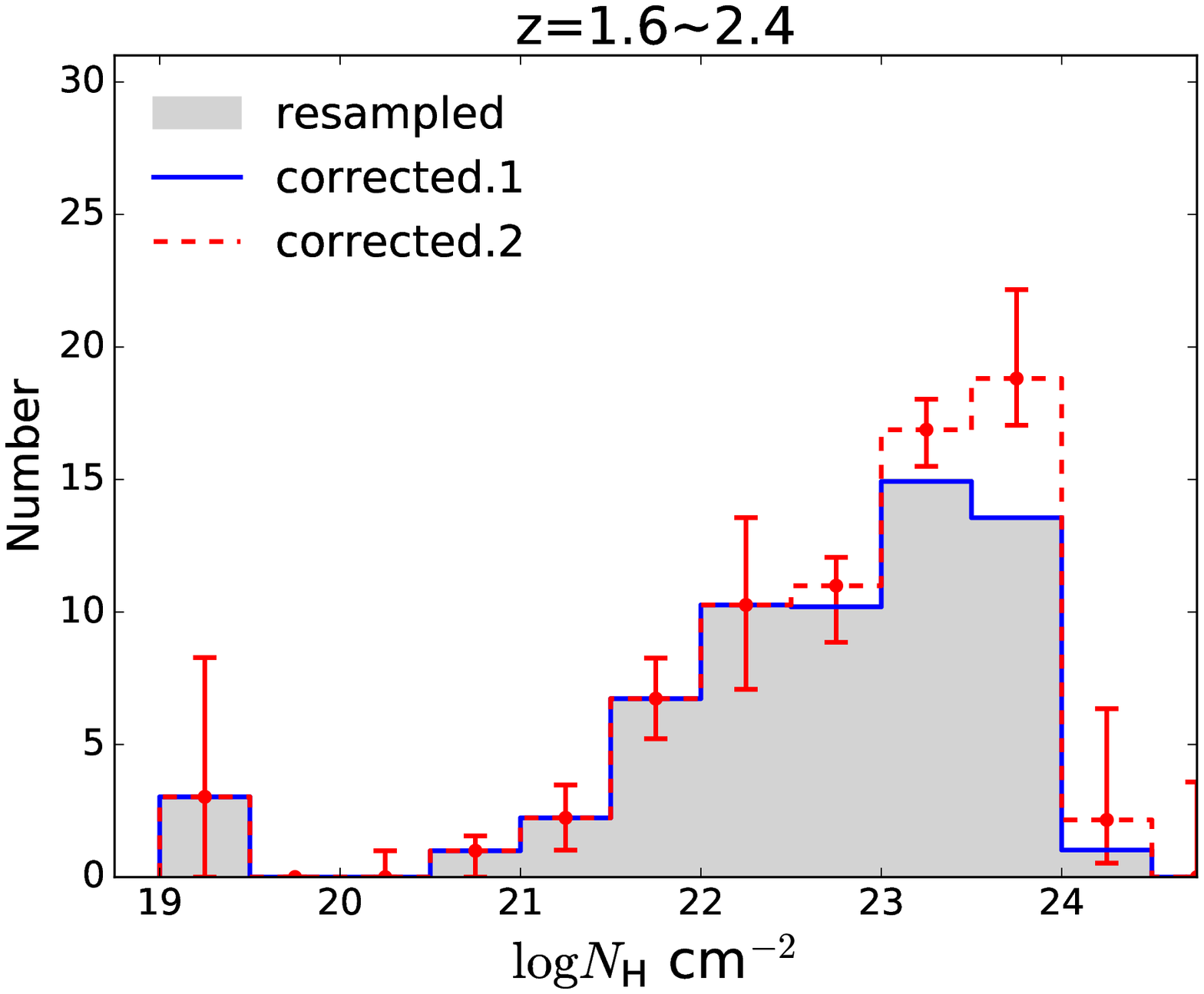}
\plotone{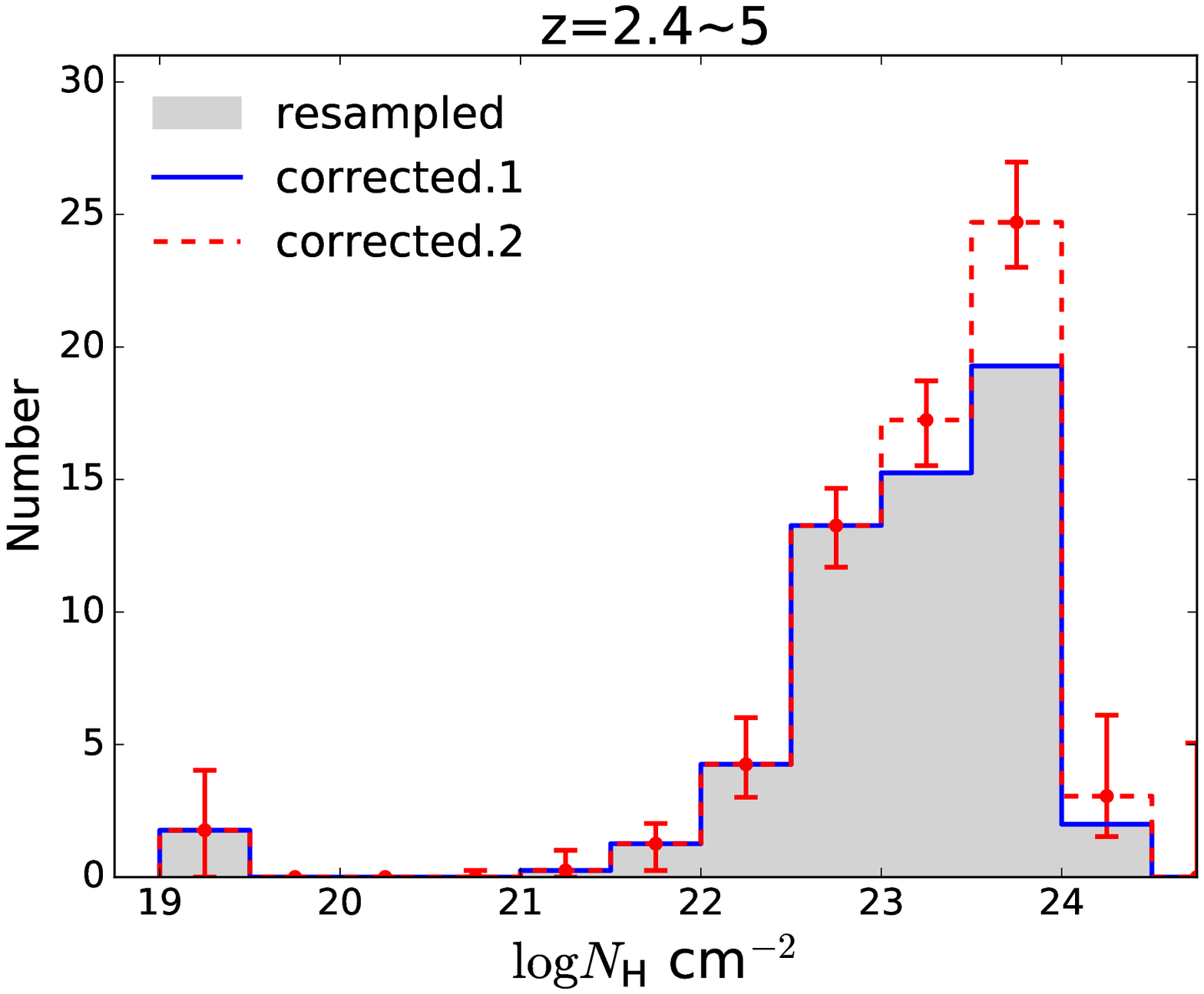}
\caption{
\N{H} distributions in the observable spaces corresponding to four redshift bins, as show in Figure~\ref{fig:L_z_boundary}.
The 90\% percentile luminosity range of the sources in the four redshift bins are 41.65$\sim$43.40, 42.57$\sim$44.12, 43.05$\sim$44.35, 43.46$\sim$44.76, respectively, in the order of arising $z$.
Grey filled histograms show the resampled \N{H} distribution of the Compton-thin sources, the same as plotted in Figure~\ref{fig:NH_distri}.
The ``corrected.1'' (blue solid line) histogram shows the \N{H} distribution after correction for the inhomogeneous survey depth, which has a minor effect.
Then after further correcting for the \N{H}-dependent Malmquist bias, the \N{H} distribution is plotted as the ``corrected.2'' (red dashed line) histogram.
The 1$\sigma$ error obtained from bootstrapping is plotted on the ``corrected.2'' histogram.
}
\label{fig:NH_intri_Z}
\end{figure*}

Correcting for the effects described above, we can retrieve the intrinsic \N{H} distribution representative of a well-defined AGN population from the observed one as follows. 
First, we correct for the sky coverage effect, that is, a source with higher \N{H} tends to have lower observed flux and thus lower sky coverage.
While calculating the re-sampled distribution of \N{H} from the spectral-fitting results, we weight each source by the reciprocal of its corresponding sky coverage.
As shown in Figure~\ref{fig:NH_intri_Z} (the ``corrected.1'' histogram), this correction has a minor effect.

\begin{figure}[htbp]
\epsscale{1}
\plotone{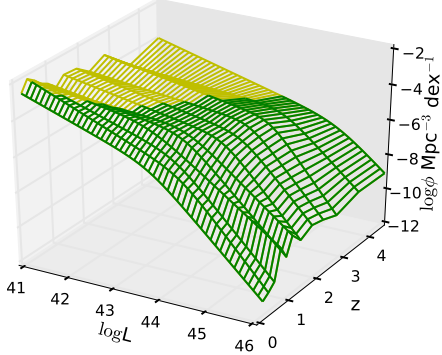}
\caption{2--10~keV intrinsic luminosity functions of AGNs in a series of redshift bins, obtained combining
the results by \citet{2015Miyaji} and \citet{Georgakakis2015}. The redshift grid values correspond to 
the central values of the redshift bins.
The green part is covered by our sample assuming that AGN have no intrinsic obscuration.
Note that this boundary corresponds to our sample selection threshold, not the source-detection threshold. The 7Ms CDF-S catalog \citep{Luo2017} extends well beyond this boundary in the $L$-$z$ space.
}
\label{fig:LF}
\end{figure}

\begin{figure}[htbp]
\epsscale{1}
\plotone{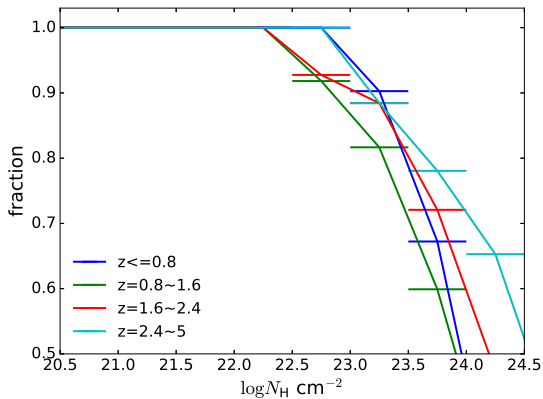}
\caption{The fraction of sources detectable within our sample with a given \N{H},
computed for the luminosity range $\log$L within 42-45, in four redshift bins.}
\label{fig:frac_NH}
\end{figure}

\begin{figure}[htbp]
\epsscale{1}
\plotone{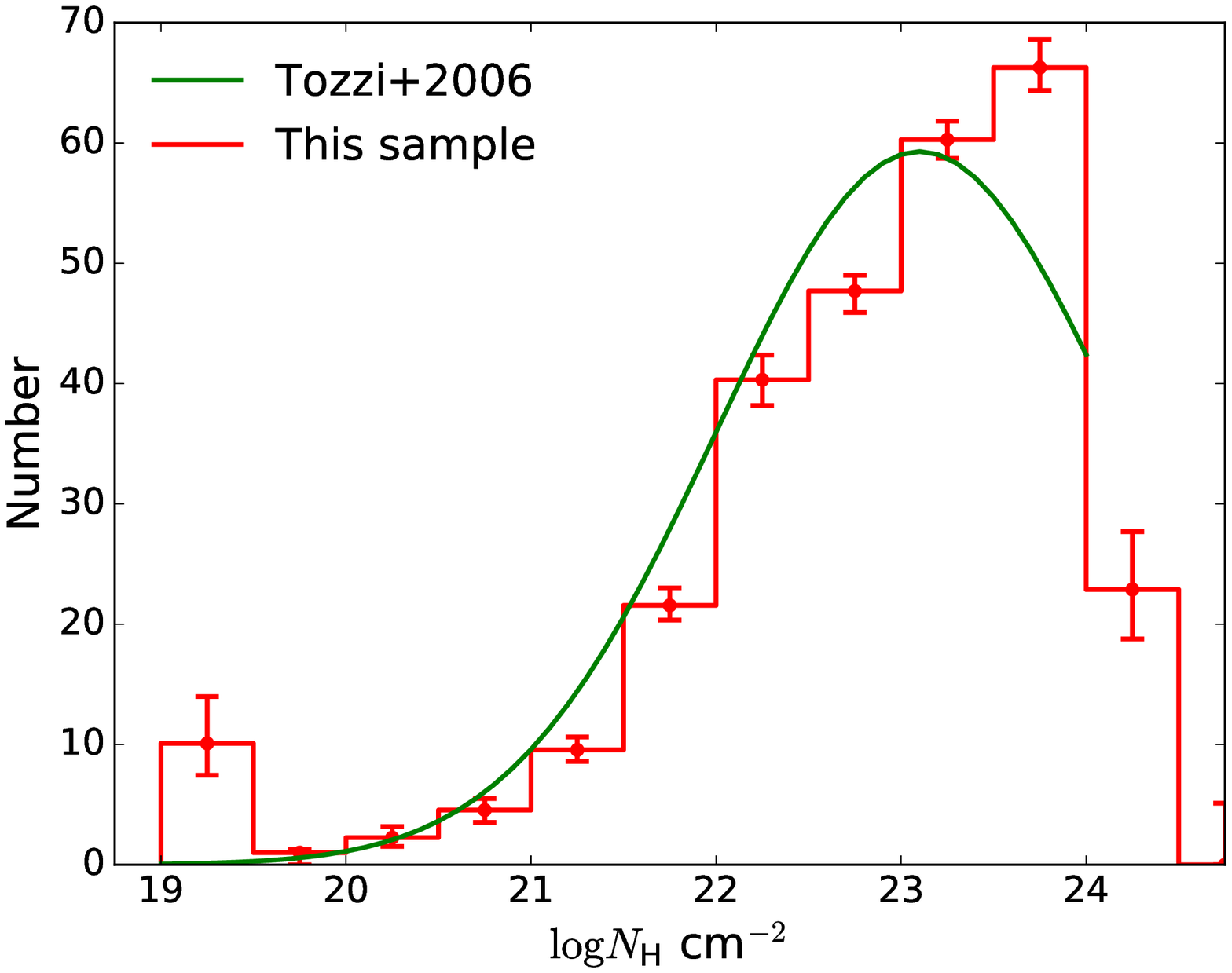}
\caption{
Intrinsic \N{H} distribution of our whole sample, measured by summing up the intrinsic \N{H} distributions in the four redshift bins ($z<5$) as shown in Figure~\ref{fig:NH_intri_Z}.
The 90\% percentile luminosity range of our sample is 42.41$\sim$44.45.
The green line which is derived from the 1Ms CDF-S by \citet{Tozzi2006}, is a log-normal distribution centered at $\log$\N{H}$=$23.1 with a $\sigma$=1.1 and normalized to the number of AGN used in this work between $\log$\N{H} of 19 and 24.
}
\label{fig:NH_intri_distri_sum}
\end{figure}

Second, we correct for the \N{H}-dependent Malmquist bias.
To do this, we need to know the distribution of AGN across the $L$--$z$ space.
\citet{2015Miyaji} modeled the AGN intrinsic X-ray luminosity function as a double power-law in a series of redshift bins from 0 to 5.8.
\citet{Georgakakis2015} also provided a double power-law luminosity function, but focusing on the high-redshift section of $3<z<5$.
We combine their works to model the AGN luminosity function in the $0<z<3$ and $3<z<5$ bins using \citet{2015Miyaji} and
\citet{Georgakakis2015}, respectively.
The combined luminosity function is shown in Figure~\ref{fig:LF}.

Concerning a specific redshift range, for each specific \N{H}, we define an observable space as the $L$--$z$ space between $\log L$=45 and the whole-field averaged completeness boundary on $L$.
As shown in Figure~\ref{fig:L_z_boundary}, a lower \N{H} corresponds to a lower detectable $L$ boundary, and thus a larger observable space.
Among a set of selected $\log$\N{H} between 19 and 25 with a step of 0.5, $\log$\N{H}=19 corresponds to the largest observable space; all the others are sub-spaces of it.
By integrating the AGN luminosity function over the observable space, we get the AGN number density in it.
For each specific \N{H}, the fraction of the AGNs in its detectable space over the AGNs in the largest, $\log$\N{H}=19 detectable space is shown in Figure~\ref{fig:frac_NH} for a few redshift bins.
Using this curve of fraction, we can correct the unobservable part below the completeness boundaries, so that sources with different \N{H} cover the same $L$--$z$ space -- the $\log$\N{H}=19 observable space.
This space is illustrated in Figure~\ref{fig:LF}, the corrected \N{H} distribution, as shown in Figure~\ref{fig:NH_intri_Z} (the ``corrected.2'' histogram), is representative of the AGN population in this space.
Clearly, the lower boundary of $L$ is well below the knee of the LF, indicating that we are sampling the typical AGNs in the universe which produce a large fraction of the cosmic accretion power.
For $\log$\N{H} below 24, the unobservable part in the $L$--$z$ space to be corrected is small compared to the observed space. Therefore, this correction is not sensitive to the selection of LF.

In the calculation, we assume the shape of the luminosity function is independent of \N{H}.
In other words, the luminosity functions of AGN with a different \N{H} have different normalizations but the same shape.
It has been found that obscured AGN fraction is dependent on intrinsic luminosity, as discussed in \S\ref{Section:intro} and \S\ref{Section:NH-2D}.
Apparently, this obscuration--luminosity correlation suggests different LFs of obscured and unobscured AGNs.
However, it does not necessarily indicate different LF shapes at different \N{H}; it could also be explained as a result of anisotropy of AGN X-ray emission \citep{Lawrence2010,2011Burlon,Liu2014,Sazonov2015}, that is, different LF normalizations rather than shapes at different \N{H}.
Therefore, the assumption of an \N{H}-independent LF shape does not conflict with our results; we do not discuss more complex corrections based on \N{H}-dependent LFs.

We remark that the two biases corrected above are independent of each other.
Although they both relate to how the \N{H} affects the observed flux, the sky-coverage effect is caused by the varying depth across the whole field, while the Malmquist bias occurs at a specific depth, which in the correction above corresponds to the average depth across the field.

As shown in Figure~\ref{fig:NH_intri_Z}, the \N{H} distribution of the Compton-thin sources shows large differences between redshift bins, because they sample AGN populations with different $L$ and different $z$.
The higher \N{H} in high-$z$ bins represents the combined dependence of \N{H} on $L$ and $z$.
Summing up the intrinsic \N{H} distributions in the four redshift bins, we plot the intrinsic \N{H} distribution of our whole sample in Figure~\ref{fig:NH_intri_distri_sum}.
It is consistent with that found by \citet{Tozzi2006} below $\log$\N{H}$<23$.
At $\log$\N{H}$>23$, we find a larger fraction of highly obscured sources with $\log$\N{H}$>23.5$.
This is because the samples are different.
In this work, the 7 times longer exposure allows spectral analyses on more low-$L$ and high-$z$ sources, which tend to have higher \N{H}, as will be shown in \S\ref{Section:evolution}.
Rather than selecting the bright sample on the basis of soft, hard, and total band emission, as done in \citet{Tozzi2006}, the sample selection in this work only makes use of the hard-band emission.
Therefore, selecting less sources with very soft spectra, our sources in this work are systematically harder and thus more obscured.
Additionally, we noticed that the updated redshift measurements are systematically higher than those used in \citet{Tozzi2006} (\S\ref{Section:comparison}), such that the degeneracy between $z$ and \N{H} in the spectral fitting also leads to higher \N{H} measurements in this work.

\subsection{Luminosity dependence and evolution}
\label{Section:evolution}
\subsubsection{Selection of subsamples unbiased with respect to \N{H}}
\label{Section:completeNH}
\begin{figure}[htbp]
\epsscale{1}
\plotone{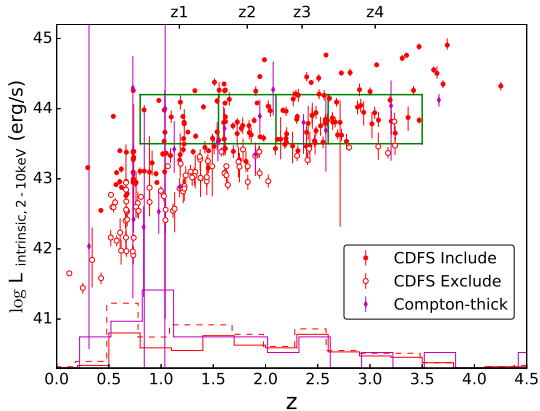}
\caption{Scatter plot of the AGN sample in the luminosity -- redshift space, with $90\%$ errors on luminosity. The $\log$\N{H}$<24$ \N{H}-complete subsample is plotted with red solid points (marked as ``Include'') and the other Compton-thin sources with red empty points (marked as ``Exclude'').
At the bottom, the normalized redshift distributions of the complete subsample (``Include'') and the whole sample (``Include''$+$``Exclude'') are shown as red solid and dashed histograms, respectively.
The Compton-thick sources are also shown with the magenta points and blue histogram.
The four green boxes, which are marked as $z1$, $z2$, $z3$, and $z4$, are the regions selected to analyze the \N{H} dependence on redshift.
Empty points are excluded from this analysis even if they fall in the selected bins.
}
\label{fig:L_z_grid}
\end{figure}

To investigate the luminosity and redshift dependence of \N{H}, we split the sample into 2D bins in the $L$-$z$ space.
This allows us to check for luminosity dependence of \N{H} at the same redshift and check for redshift dependence at the same luminosity.
Clearly two prerequisite conditions are required: 
first, wide ranges in the $L$-$z$ space must be covered by the survey; second, the sample in each bin must be complete.
As the deepest X-ray survey, the CDF-S provides an essential jigsaw piece among the X-ray surveys, which extends the coverage of X-ray surveys to the lowest luminosity and the highest redshift.
However, in some bins at low luminosity and high redshift, the flux-limited sample is incomplete with respect to \N{H}.
As shown in Figure~\ref{fig:L_z_boundary}, the completeness boundaries corresponding to different \N{H} are widely distributed in the $L$-$z$ space.
The sample will be biased against high \N{H} sources unless we apply the highest boundary, which corresponds to the highest \N{H} and largest off-axis angle, to the whole sample.
This is clearly infeasible because it excludes most of the sources in the CDF-S.
In the previous section, we handled this incompleteness with respect to \N{H} by correcting for the unobserved AGN population assuming a luminosity function.
The disadvantages of this correction are that it is dependent on the uncertain part of the luminosity function at low $L$ and high $z$, and it must be performed every time before this sample is used jointly with other X-ray surveys in the future.
Therefore, in this section we select subsamples which are complete with respect to \N{H} by trimming off the incomplete part while still maximizing the CDF-S sample.
As the flux limit is irrelevant here, we include the supplementary sample prepared in \S~\ref{Section:sample} in selecting the complete subsamples.
Such complete subsamples can be directly used in joint studies with other X-ray surveys.
Note here by ``complete'' we mean unbiased with respect to \N{H} at specific $L$ and $z$ values above the completeness boundaries, not with respect to $L$ or $z$.
The complete subsamples describe only the ``observable'' AGN population, leaving the low $L$ and high $z$ sources below the boundaries and the extremely obscured sources beyond a specific \N{H} unaccounted for.

As shown in Figure~\ref{fig:L_z_boundary}, as a function of the off-axis angle, we have determined a completeness boundary for each specific \N{H}.
Based on these boundaries, we define a ``completeness'' flag as follows.
Excluding the sources with intrinsic luminosities below the $\log$\N{H}=24 boundaries corresponding to their off-axis angles, we select sources which pass the filter of the $\log$\N{H}=24 completeness boundaries, and flag them as ``completeness'' = 24. 
Similarly, we filter our sample with the $\log$\N{H}=23.5 and $\log$\N{H}=23 boundaries, and flag the selected sources with ``completeness'' of 23.5 and 23, respectively.
A source selected by more than one filter is assigned to the highest flag, a source selected by none is flagged as 0.
Clearly, sources with a higher ``completeness'' flag represent a subsample of sources with a lower ``completeness'' flag.
Under this definition, we can simply select an \N{H}-complete subsample by requiring the ``completeness'' flag $\geqslant$ a specific \N{H} among 23, 23.5, or 24, with the aim of studying the \N{H} distribution below this specific \N{H}.
The $\log$\N{H}$<$24 complete subsample and accordingly excluded sources are shown in Figure~\ref{fig:L_z_grid}.

\subsubsection{Evolution of \N{H} with redshift}
\label{Section:NH-Z}

\begin{figure}[htbp]
\plotone{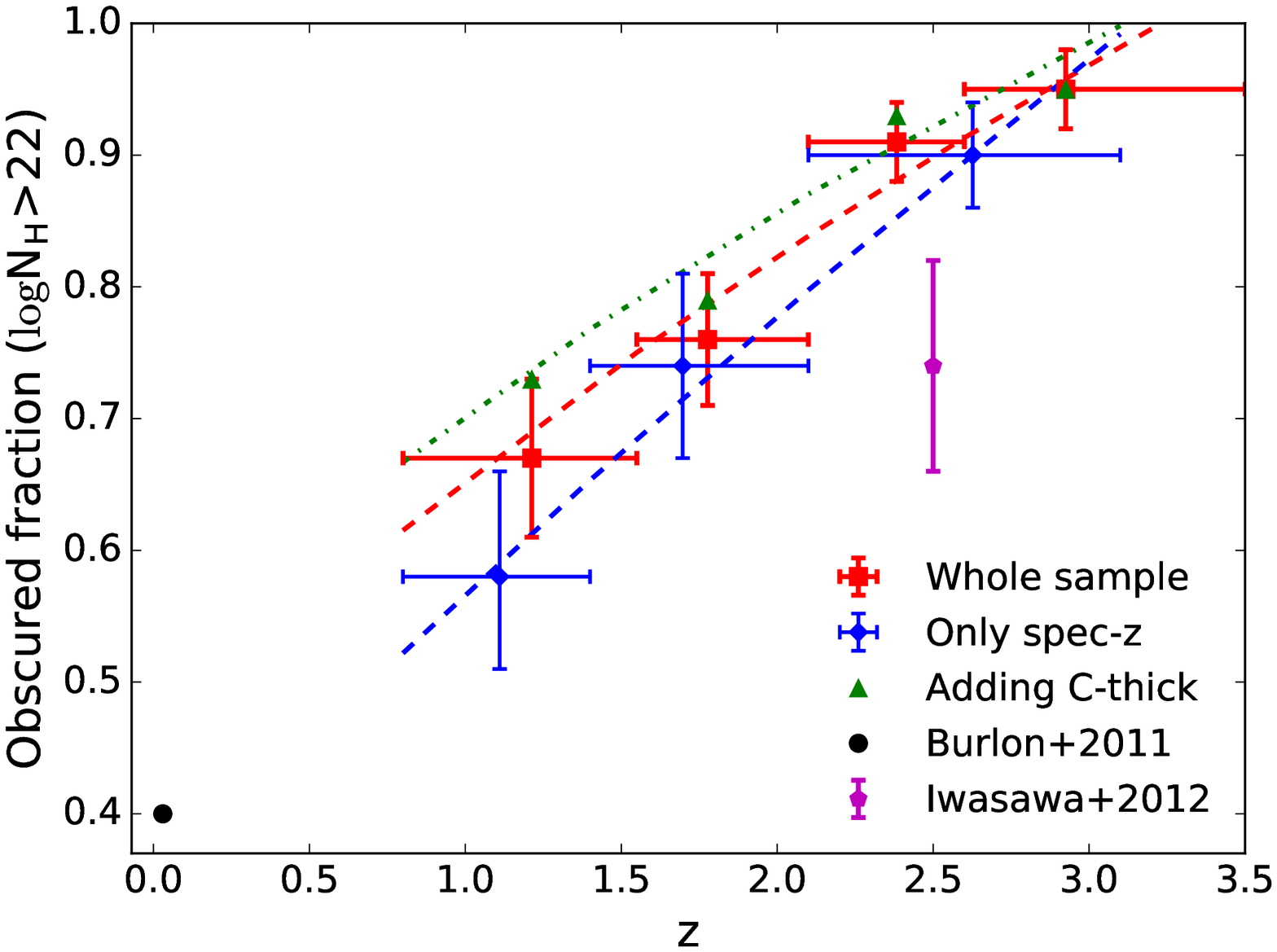}
\plotone{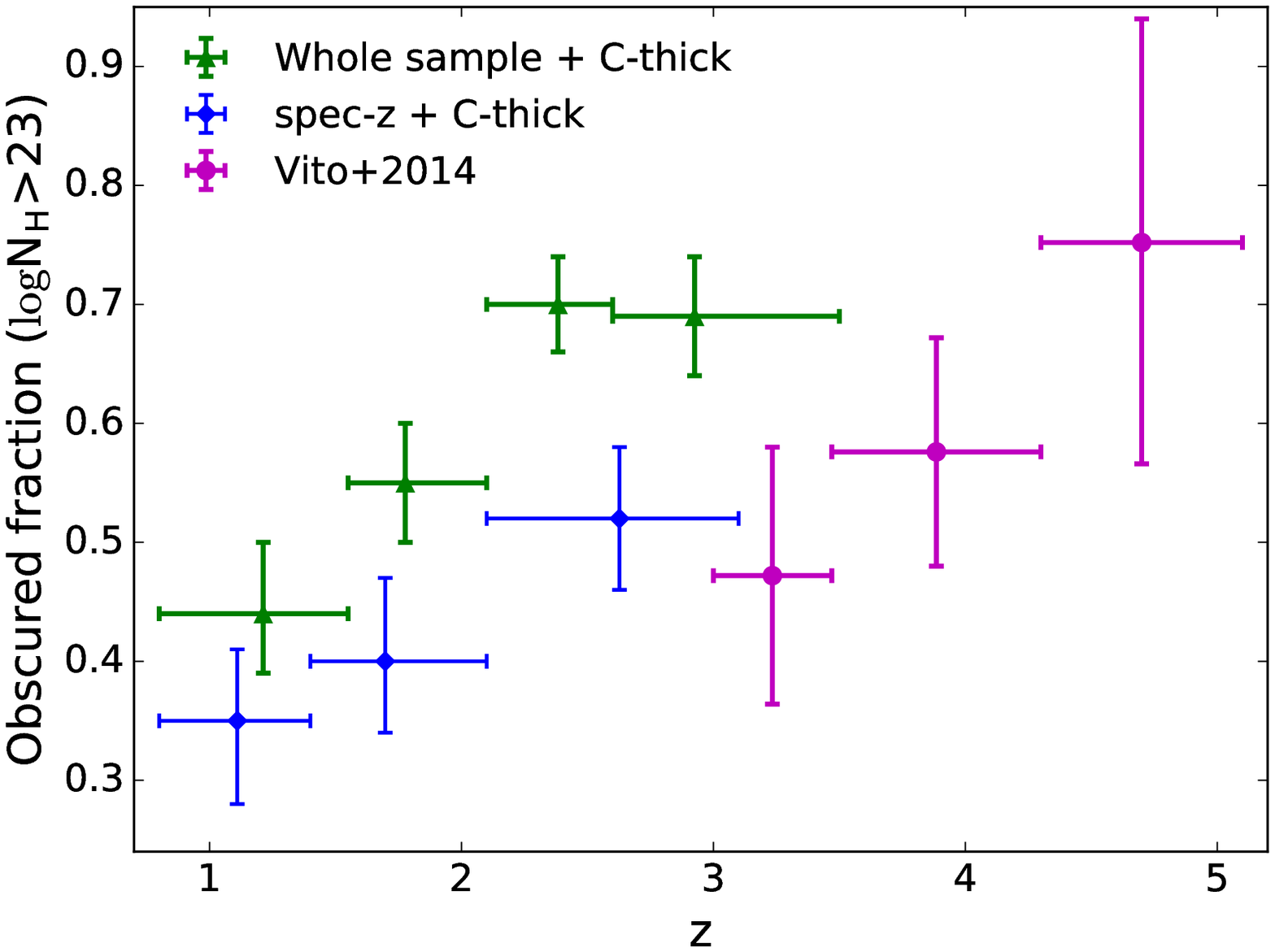}
\caption{
Upper panel: fractions of sources with $\log$\N{H}$>22$ in the four bins ($0.8<z<3.5$, $43.5<\log L<44.2$)
 as shown in Figure~\ref{fig:L_z_grid} and best-fit lines.
Results obtained from the whole ``completeness''$=24$ sample are plotted in red, the spectroscopic-z subsample in blue.
Results obtained after adding the Compton-thick sources to the whole sample are shown in green, whose errorbars are omitted for simplicity.
The \citet{2011Burlon} and \citet{Iwasawa2012} results correspond to $\log L \thickapprox 43.85$ and $\log L \thickapprox 44.2$, respectively.
Lower panel: fractions of sources with $\log$\N{H}$>23$.
Results from our sample adding the Compton-thick sources are plotted in green, from the spectroscopic-z subsample adding the Compton-thick sources are plotted in blue.
Results from the $42.85<\log L<44.5$ sample of \citet{Vito2014} are plotted in magenta.
}
\label{fig:frac_z}
\end{figure}

The unprecedented survey depth of the CDF-S provides the best opportunity to investigate the redshift-dependence of AGN obscuration.
As shown in Figure~\ref{fig:L_z_grid}, we select a narrow luminosity band between $43.5$ and $44.2$, within which our sample has a large range of redshift.
Then we select four redshift bins at grid points of $z=$ 0.8, 1.55, 2.1, 2.6, and 3.5.
For each bin, which contains a subsample unbiased with respect to $\log$\N{H}$<24$, we calculate the average $\log$\N{H} and the obscured fraction -- the fraction of \N{H} values between $10^{22}$--$10^{24}$ cm$^{-2}$ among all those below $10^{24}$ cm$^{-2}$.
We take the \N{H} measurement error into account through a bootstrapping procedure.
For each spectrum of each Compton-thin source, we have generated 1000 random \N{H} values following its \N{H} error distribution, as done in the \N{H} distribution resampling. The random $\log$\N{H} below 24 are used to measure the $68\%$ confidence range of $\langle \log$\N{H}$\rangle$ and obscured fraction.
When calculating $\langle \log$\N{H}$\rangle$, $\log$\N{H} below 19 are set to 19.
The subsample size, $\langle \log$\N{H}$\rangle$, and obscured fractions in each bin are shown in the upper panel of Table~\ref{table:NH-Z}.
Both quantities are increasing with redshift.
The obscured fractions are plotted in Figure~\ref{fig:frac_z}.
We fit the correlation between obscured fraction and redshift with $f_{24}\ =\ \beta\ (1+z)^{\alpha}$, and find best-fit values of $\beta=0.42\pm0.09$ and $\alpha=0.60\pm0.17$.

According to the result of \citet{2011Burlon}, at $\log L=43.85$ -- the average 2-10~keV luminosity of our selected luminosity range ($43.5\sim44.2$), the AGN obscured fraction at $z\sim0.03$ is about $40\%$.
As shown in the upper panel of Figure~\ref{fig:frac_z}, the \N{H} evolution from local to $z=2$ is very strong.
However, above $z=2$ the absorbed fraction is likely saturating and shows weak evolution.
It is consistent with the result of \citet{Hasinger2008}. They found that the AGN obscured fraction increases with $z$ with an $\alpha=0.62\pm0.11$, and becomes saturated at $z>2$.
\citet{Vito2014} also noted the weak evolution of \N{H} at high $z$.

\citet{Hasinger2008} found an evolution slope of $\alpha=0.48\pm0.08$ over a wide redshift range 0--3.2. Similarly, \citet{Ueda2014} found an $\alpha=0.48\pm0.05$. 
To compare with these results, we take the Compton-thick sources into the account, as done in \citet{Hasinger2008} and \citet{Ueda2014}.
We remark the shortcoming of including Compton-thick sources that the sampling of Compton-thick AGNs can be highly incomplete and the measurements of their intrinsic luminosity can be highly unreliable. 
As shown in the upper panel of Figure~\ref{fig:frac_z}, the fraction is slightly increased, and the best-fit slope of $\alpha=0.45_{-0.09}^{+0.10}$ is consistent with the previous results.
Our Compton-thin sample shows a slightly steeper slope of $0.59$ because we find less Compton-thick AGN at high $z$, as shown in Figure~\ref{fig:L_z_grid}.

In the upper panel of Figure~\ref{fig:frac_z}, we also show the result of \citet{Iwasawa2012}, who found an obscured fraction among Compton-thin AGN of $74\pm8\%$ at $z\sim2.5$ and $\log L\sim44.2$.
It is lower than the fraction measured in this work.
To compare with the result of \citet{Vito2014}, who define an obscured fraction as the fraction of sources with $\log$\N{H}$>23$ rather than $\log$\N{H}$>22$, we also calculate this fraction in our selected redshift bins.
As shown in the lower panel of Figure~\ref{fig:frac_z}, compared with our results, the fractions obtained from their $42.85<\log L<44.5$ sample also appear lower.
This deviation might be caused by different luminosities or incompleteness of their sample.
Our conservativeness in identifying Compton-thick AGN might introduce an overestimation of obscured fraction at high $z$, in the sense that a few Compton-thick AGNs might be misclassified as highly obscured Compton-thin because of large uncertainty in \N{H} measurement.
However, the observable Compton-thick sources are few, and even if misclassified as Compton-thin, they are likely considered to be highly obscured sources with $\log$\N{H}$>23$. Therefore, we consider this as a minor effect.

Considering the short duty cycle of AGN \citep[$10^7-10^8$ years, e.g.,][]{Parma2007}, the strong evolution of AGN \N{H} with $z$ is obviously associated with the evolution of galaxy. One associated factor is the high merger rate at high-$z$, in the sense that merger triggers AGN and such AGN are more obscured \citep{DiMatteo2005,Kocevski2015,Lanzuisi2015}. Another factor is the higher gas fraction at high-$z$ \citep{Carilli2013}, which not only leads to a higher \N{H} directly, but also likely leads to a longer obscured phase and a larger covering factor.

\begin{table}[htbp]
\footnotesize
\begin{center}
\tablenum{6}
\label{table:NH-Z}
\caption{Number of sources, median 0.5-7~keV net counts, average of $\log$\N{H}, and obscured fraction in each bin, obtained from the CDF-S data.}
\begin{tabular}{ccccc}
\hline
\hline
\multicolumn{5}{c}{Whole subsample}\\
\hline
Bins& z1 &z2 &z3 &z4\\
\hline
Number& 17 & 21 & 26 & 20\\
\hline
Median Cts &2224 &770 &491 &381 \\
\hline
$\langle\log$\N{H}$\rangle$& $22.36_{-0.14}^{+0.14}$&	$22.64_{-0.13}^{+0.13}$&	$22.95_{-0.10}^{+0.10}$&	$23.11_{-0.10}^{+0.10}$\\
\hline
$ObsFrac$& $0.67_{-0.06}^{+0.06}$&	$0.76_{-0.05}^{+0.05}$&	$0.91_{-0.03}^{+0.03}$&	$0.95_{-0.03}^{+0.03}$\\
\hline
\end{tabular}
\begin{tabular}{cccc}
\multicolumn{4}{c}{Spectroscopic-$z$ Only}\\
\hline
Bins& z1 &z2 &z3\\
\hline
Number& 12 & 12 & 12\\
\hline
Median Cts &2934 &1312 &476 \\
\hline
$\langle\log$\N{H}$\rangle$& $22.18_{-0.17}^{+0.17}$&	$22.60_{-0.14}^{+0.14}$&	$22.79_{-0.12}^{+0.13}$\\
\hline
$ObsFrac$& $0.58_{-0.07}^{+0.08}$&	$0.74_{-0.07}^{+0.07}$&	$0.90_{-0.04}^{+0.04}$\\
\hline
\hline
\end{tabular}
\begin{tabular}{@{}p{\columnwidth}}
\hline
Note: The obscured fraction is the fraction of \N{H} values between $10^{22}$--$10^{24}$ cm$^{-2}$ among all those below $10^{24}$ cm$^{-2}$. 
In the upper panel, the data is obtained using the CDF-S subsample which is unbiased with respect to $\log$\N{H}$<24$.
The data in the lower panel is obtained after excluding the photometric-$z$ sources.
\end{tabular}
\end{center}
\end{table}

\subsubsection{Decomposing the luminosity- and redshift-dependence of \N{H}}
\label{Section:NH-2D}

\begin{figure}[htbp]
\epsscale{1}
\plotone{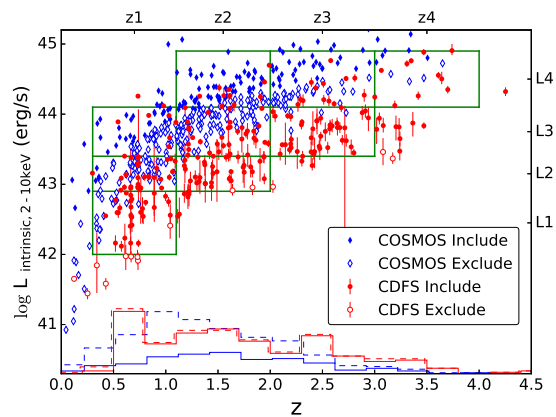}
\caption{Same as Figure~\ref{fig:L_z_grid}, but including C-COSMOS data (red for CDF-S, blue for C-COSMOS). The selected subsamples are complete for $\log$\N{H}$<$23.
}
\label{fig:L_z_grid_CM}
\end{figure}

To complement the high-luminosity regime which is not well-sampled by the CDF-S, we add AGN observed from the C-COSMOS, a wider and shallower survey \citep{Elvis2009,Lanzuisi2013,Civano2016}. 
The 2~deg$^2$ COSMOS field was observed by {\sl Chandra} for 160 ks.
\citet{Lanzuisi2013} presented the properties of AGN in this survey. We combine this sample with ours, as shown in Figure~\ref{fig:L_z_grid_CM}. As done for our sample, similar ``completeness boundaries'' are calculated for C-COSMOS, considering its 160 ks exposure time and its sample-selection threshold of 70 net counts in the broad 0.5-7~keV band.
Because of the shallow depth, C-COSMOS is severely incomplete for highly obscured AGN.
Therefore, we only make use of the $\log$\N{H}$<$23 part.
The \N{H}-complete subsamples of C-COSMOS and CDF-S for $\log$\N{H}$<23$ are shown in Figure~\ref{fig:L_z_grid_CM}.

We select a 4$\times$4 grid for the combined sample at $0.3 <z1< 1.1<z2< 2<z3< 3<z4< 4$ and $42<L1< 42.9<L2< 43.4<L3< 44.1<L4< 44.9$.
The grids are selected following several criteria:
\begin{enumerate}[nolistsep]
\item Exclude the high-$L$ low-$z$ corner because there are few sources, as limited by the total sky coverage.
\item Exclude the low-$L$ high-$z$ corner which is below the sample selection boundaries.
\item Choose a proper number of bins such that each cell contains a sufficient number of sources.
\item Make the sources evenly distributed among the cells such that each cell contains approximately the same number of sources.
\item Make the sources evenly distributed in each valid cell, avoiding an empty low-$L$ high-$z$ corner in any cell.
\end{enumerate}

\begin{figure}[htbp]
\plotone{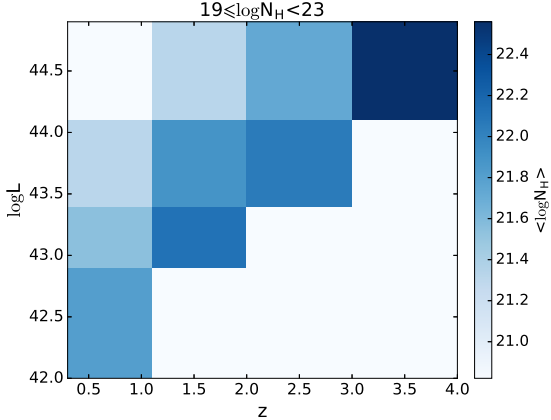}
\caption{Mean $\log$\N{H} in grid, combining CDF-S and C-COSMOS data.
Only values below 23 are considered. Values below 19 are taken as 19.
White cells are excluded.
}
\label{fig:L_z_meanNH_PC}
\end{figure}

The number of sources in each bin is listed in Table~\ref{table:meanNH_PC}.
One cell at high-$L$ low-$z$ and 6 cells at low-$L$ high-$z$ are excluded.
For the subsamples in the valid cells, the $\langle\log$\N{H}$\rangle$ below 23 are calculated similarly as above.
The obscured fraction in each bin is calculated as the fraction of the \N{H} values between $10^{22}\sim 10^{23}$ cm$^{-2}$ among all those below $10^{23}$ cm$^{-2}$.
The results are shown in Figure~\ref{fig:L_z_meanNH_PC} and Table~\ref{table:meanNH_PC}, where excluded cells are left empty.

\begin{table}[htbp]
\footnotesize
\begin{center}
\tablenum{7}
\label{table:meanNH_PC}
\caption{Number of sources, median 0.5-7~keV net counts, average of $\log$\N{H}, and obscured fraction in each bin, obtained by combining CDF-S \& C-COSMOS.
}
\begin{tabular}{ccccc}
\hline
\hline
\multicolumn{5}{c}{Whole subsample}\\
\hline
Number& $z1$ &$z2$ &$z3$ &$z4$\\
\hline
$L4$& 7& 68& 39& 13\\
$L3$& 34& 40& 43& 9\\
$L2$& 35& 33& 3& 0\\
$L1$& 29& 7& 0& 0\\
\hline
Median Cts & $z1$ &$z2$ &$z3$ &$z4$\\
\hline
$L4$ & &284 &235 &1501 \\
$L3$ &494 &654 &335 & \\
$L2$ &888 &337 & & \\
$L1$ &303 & & & \\
\hline
$\langle\log$\N{H}$\rangle$&$z1$ &$z2$ &$z3$ &$z4$\\
\hline
$L4$& &	$21.33_{-0.15}^{+0.15}$&	$21.73_{-0.21}^{+0.22}$&	$22.56_{-0.14}^{+0.12}$\\
$L3$& $21.32_{-0.17}^{+0.17}$&	$21.89_{-0.10}^{+0.11}$&	$22.06_{-0.16}^{+0.15}$& \\
$L2$& $21.55_{-0.13}^{+0.14}$&	$22.11_{-0.10}^{+0.11}$&	&	\\
$L1$& $21.81_{-0.12}^{+0.11}$&	&	&	\\
\hline
$ObsFrac$&$z1$ &$z2$ &$z3$ &$z4$\\
\hline
$L4$& &	$0.40_{-0.06}^{+0.05}$&	$0.67_{-0.07}^{+0.07}$&	$0.96_{-0.06}^{+0.04}$\\
$L3$& $0.32_{-0.07}^{+0.06}$&	$0.57_{-0.06}^{+0.05}$&	$0.76_{-0.05}^{+0.05}$&	\\
$L2$& $0.50_{-0.05}^{+0.05}$&	$0.75_{-0.06}^{+0.05}$&	&	\\
$L1$& $0.63_{-0.05}^{+0.05}$&	&	&	\\
\hline
\end{tabular}
\begin{tabular}{cccc}
\hline
\multicolumn{4}{c}{Spectroscopic-$z$ Only}\\
\hline
Number& $z1$ &$z2$ &$z3$\\
\hline
$L3$& 1& 29& 24\\
$L2$& 37& 35& 16\\
$L1$& 54& 13& 7\\
\hline
Median Cts & $z1$ &$z2$ &$z3$ \\
\hline
$L3$ & &354 &220 \\
$L2$ &518 &488 &566 \\
$L1$ &785 &228 & \\
\hline
$\langle\log$\N{H}$\rangle$&$z1$ &$z2$ &$z3$ \\
\hline
$L3$& &	$21.13_{-0.26}^{+0.25}$&	$21.31_{-0.29}^{+0.29}$\\
$L2$& $21.29_{-0.15}^{+0.15}$&	$21.79_{-0.12}^{+0.12}$&	$22.10_{-0.15}^{+0.16}$\\
$L1$& $21.65_{-0.09}^{+0.09}$&	&	\\
\hline
$ObsFrac$&$z1$ &$z2$ &$z3$ \\
\hline
$L3$& &	$0.42_{-0.08}^{+0.08}$&	$0.51_{-0.10}^{+0.09}$\\
$L2$& $0.28_{-0.06}^{+0.06}$&	$0.51_{-0.06}^{+0.06}$&	$0.72_{-0.07}^{+0.06}$\\
$L1$& $0.54_{-0.04}^{+0.04}$&	&	\\
\hline
\end{tabular}
\begin{tabular}{@{}p{\columnwidth}}
\hline
Note: The median counts, $\langle\log$\N{H}$\rangle$, and obscured fractions are given only for the valid cells.  The obscured fraction corresponds to the fraction of the \N{H} values between $10^{22}\sim 10^{23}$ cm$^{-2}$ among all those below $10^{23}$ cm$^{-2}$.
The data in the upper panel corresponds to the subsample which is unbiased with respect to $\log$\N{H}$<23$.
In the lower panel, the data is obtained after excluding the photometric-$z$ sources.
\end{tabular}
\end{center}
\end{table}

Limited by the sample size, the $\langle\log$\N{H}$\rangle$ cannot be constrained very tightly, especially at high redshift.
However, trends of luminosity- and redshift-dependences are apparent.
In each redshift bin, the mean \N{H} decreases with luminosity.
In each luminosity bin, the mean \N{H} increases with redshift.
Obscured fractions show similar luminosity- and redshift-dependences.
Limited by the luminosity range of our sample, we are not able to see any turn-over of the \N{H}--luminosity anti-correlation at low luminosities.

\subsubsection{Discussion on other systematic effects}
Our analysis so far aimed at minimizing the effects of selection bias on the correlation between \N{H}, luminosity and redshift. In this section, we discuss other systematic effects related to degeneracy between spectral parameters.

The best-fit \N{H} and intrinsic luminosity $L$ are positively correlated, in the sense that higher \N{H} gives rise to higher absorption-correction and thus higher $L$.
The degeneracy between them can be strong in low S/N and high \N{H} cases, where uncertainty of \N{H} is large.
In \S\ref{Section:NH-Z} and \S\ref{Section:NH-2D}, we have managed to attenuate this effect while comparing the \N{H} distribution in different bins in the $L$--$z$ space.
First, we consider only the uncertainty of \N{H} but not the uncertainty of $L$ while dividing the sources into bins. Second, we keep the valid cells away from sample selection boundaries (grid selection criterion 5 in \S \ref{Section:NH-2D}) to reduce boundary effect, that is, including high-$L$ high-\N{H} sources from the below-boundary region and excluding low-$L$ low-\N{H} sources from the above-boundary region because of the uncertainty of $L$ and the \N{H}-$L$ degeneracy.
The positive \N{H}--$L$ degeneracy could weaken the negative correlation we found between average \N{H} and $L$.
However, as shown in Figure~\ref{fig:L_z_meanNH_PC} and the upper panel of Table~\ref{table:meanNH_PC}, the average \N{H} declines clearly at higher $L$, indicating that the effect of the \N{H}-$L$ degeneracy is negligible.

An accurate redshift is essential in measuring both \N{H} and $L$.
In our sample, 37\% of the sources have photometric redshifts, and they lie at relatively high-$z$.
For such a photometric-$z$ source, the redshift value, which has a large uncertainty, is positively correlated with \N{H} since the absorption feature in the spectrum is considered to happen at a higher energy with a larger $z$; and $z$ is positively correlated with $L$ as a larger $z$ produces a larger luminosity distance.
The \N{H}-$z$ degeneracy enhances the \N{H} evolution we found; and the $L$-$z$ degeneracy affects the thoroughness of our decomposition of the $L$-dependence and $z$-dependence of average \N{H}.
To test the robustness of our results, we repeat the experiments with the photometric-$z$ sources excluded.

Since we have few sources in this case, instead of the 4 redshift bins show in Figure~\ref{fig:L_z_grid}, we choose 3 redshift bins at grids of $z=$ 0.8, 1.4, 2.1, and 3.1, and calculate the $\langle \log$\N{H}$\rangle$ and obscured fraction of the spectroscopic-$z$ subsample in each bin in the same manner as described above.
The results are compared with the total sample in Table~\ref{table:NH-Z} and Figure~\ref{fig:frac_z}.
We find that, compared with the results obtained from the whole sample, the $\log$\N{H}$>22$ obscured fractions are reduced slightly, while the $\log$\N{H}$>23$ obscured fractions are reduced significantly, especially at high $z$.

\begin{figure}[htbp]
\epsscale{1}
\plotone{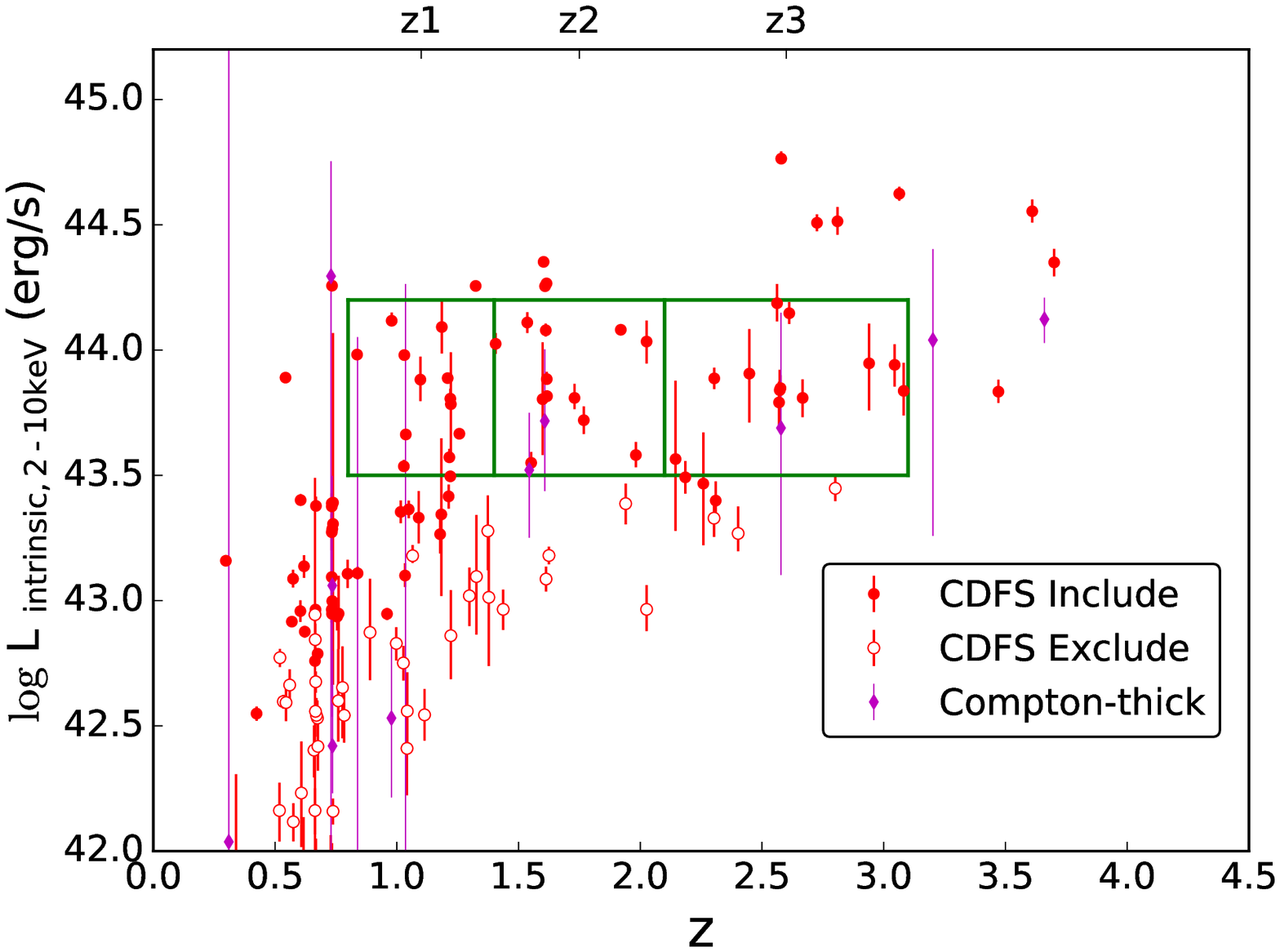}
\plotone{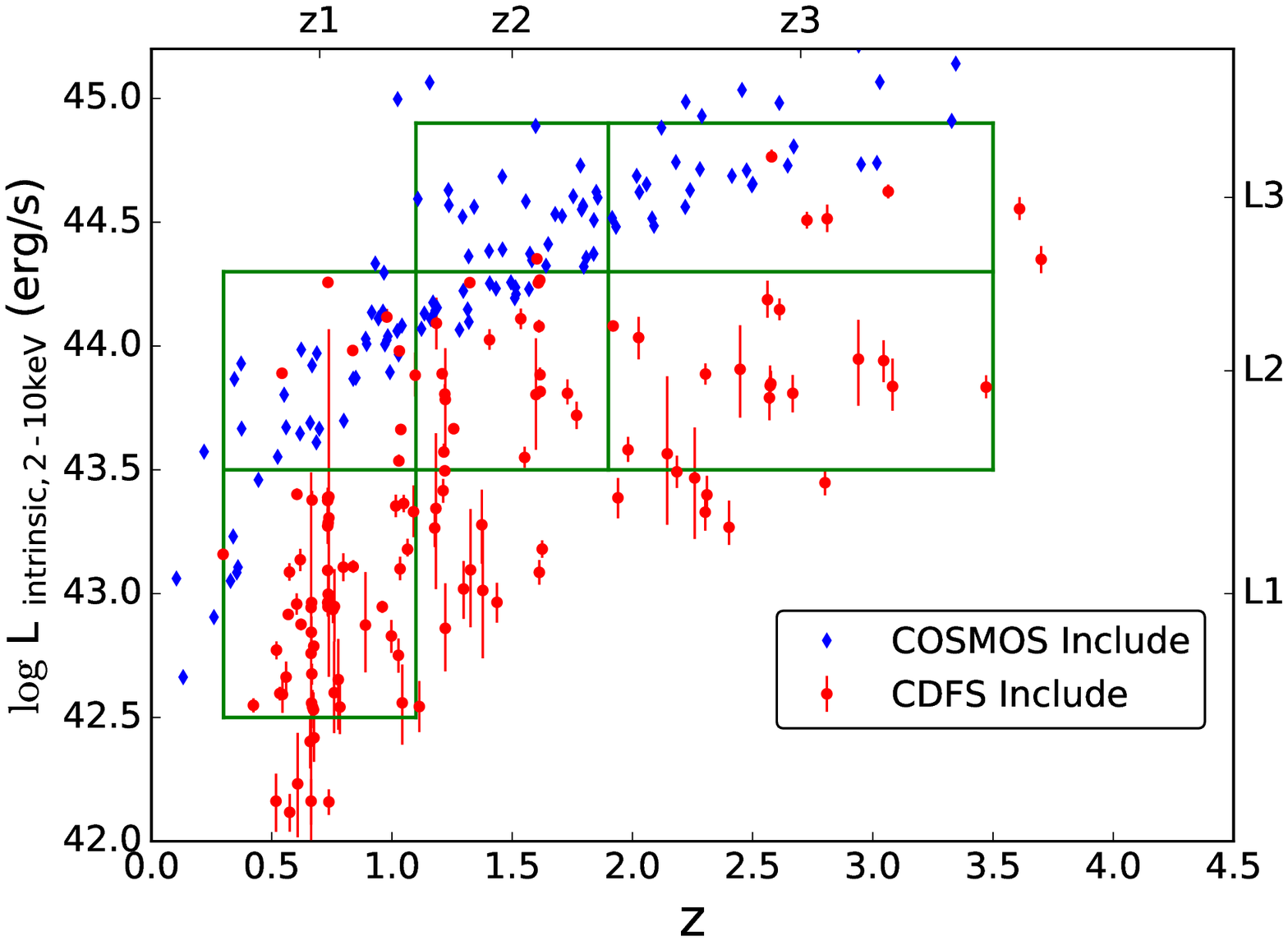}
\caption{The upper and lower panels are the same as Figure~\ref{fig:L_z_grid} and Figure~\ref{fig:L_z_grid_CM} respectively, but only include spectroscopic-$z$ sources.
}
\label{fig:L_z_grid_CM_zSpec}
\end{figure}

The 2D distribution of average \N{H} using the spectroscopic-$z$ subsample also deserves a look.
We rebin this subsample to a 3$\times$3 grid at $0.3<z1<1.1<z2<1.9<z3<3.5$ and $42.5<L1<43.5<L2<44.3<L3<44.9$, as shown in Figure~\ref{fig:L_z_grid_CM_zSpec}.
The $\langle \log$\N{H}$\rangle$ and obscured fraction are calculated in the same manner and shown in the lower panel of Table~\ref{table:meanNH_PC}.
Again, the $\langle \log$\N{H}$\rangle$ and obscured fraction are reduced for the spectroscopic-$z$ subsample, but it's still clear that the obscuration decreases with $L$ at the same $z$ and increases with $z$ at the same $L$.

The reduced fraction of obscured AGN in the spectroscopic-$z$ subsample is expected by the \N{H}-$z$ degeneracy, which leads to higher \N{H} at high-$z$ and lower \N{H} at low-$z$. However, the spectroscopic-$z$ subsample is likely biased against obscured AGN, in the sense that X-ray obscured AGNs also suffer more extinction in the optical band and are thus less likely to be spectroscopically observed.  Therefore, the difference can only be partly attributed to the \N{H}-$z$ degeneracy; the results from the spectroscopic-$z$ subsample should be considered as a conservative estimation of the \N{H} dependence on $z$.

We conclude that the luminosity- and redshift-dependence of \N{H} that we have found are robust, having all the systematics understood.

\section{SUMMARY}
Over the past 16 years, the CDF-S has been cumulatively observed for 7Ms, making it the deepest X-ray survey to date.
We present a detailed spectral analysis for a sample of the brightest AGNs in the 7Ms CDF-S.
This sample includes $276$ sources which are: 1) classified as AGN on the basis of multi-band information, 2) detected with at least 80 hard-band net counts, 3) having a redshift measurement, and 4) within an off-axis angle of 9.5$\arcmin$.
The new 7Ms CDF-S data and catalog provide not only spectra with improved S/N but also updated redshift measurements compared to those used in previous works \citep[e.g.,][]{Tozzi2006,Buchner2014}.
On the basis of these improvements, we perform a systematic X-ray spectral analysis for the bright AGN sample, putting emphasis on the properties of AGN intrinsic obscuration.
Our standard spectral model is composed of an intrinsically absorbed power-law and a 
``self-absorbed'' cold-reflection component which describes a typical reflection from the dusty torus.
A soft excess and/or a neutral Fe K$\alpha$ line are added to the model whenever the component improves the fitting significantly.
The analysis is performed both on the total 7Ms exposures and in four different periods across the 16 years.
In this way we are able to exploit the high S/N of the cumulative spectra whenever possible, and also to identify significant spectral variabilities on time scales of a few years.
The spectral fitting results are summarized as follows.

\begin{enumerate}[nolistsep]
\item We present the power-law slopes, column densities, observed fluxes, and absorption-corrected luminosities for our sample of AGNs. We also present the net counts and fluxes of each AGN in each period. For sources showing \N{H} variability, the different \N{H} in different periods are given.

\item Narrow Fe K$\alpha$ lines are found in $50$ sources and broad Fe K$\alpha$ lines are found in $5$ sources. The EWs of the narrow Fe K$\alpha$ lines show a clear correlation with the \N{H} of the sources, which is well explained by a toy model assuming a constant line flux independent of the line-of-sight \N{H}. This line flux corresponds to an EW of 135 eV when the power-law is unobscured.

\item We find \N{H} variation in $39$ sources.
By checking the correlation between the variations of \N{H} and intrinsic X-ray luminosity $L$, we show two kinds of \N{H} variations: one anti-correlated with $L$, with a relatively small amplitude, and one independent of $L$, with a large amplitude.
They can be attributed to \N{H} depression by the central engine
or to an obscuring cloud
moving across the line-of-sight, respectively.

\item On the basis of the X-ray spectral shape, Fe line EW, and relative strength of X-ray to MIR 12$\micron$ emission, we identify $22$ (8\% of the sample) Compton-thick candidates. These sources show a systematically lower X-ray to MIR ratio.
\end{enumerate}

Thanks to the well-understood selection function of our sample, we are able to measure quantitatively the sample-selection biases which impact the observed \N{H} distribution, including the sky-coverage effect which biases against sources with low observed flux, the \N{H}-dependent Malmquist bias which leads to different coverage in the $L_X$--$z$ space of the AGN population for different \N{H}, and the Eddington bias.
Based on thorough analyses of these effects, we are able to recover the intrinsic distribution and evolution of AGN obscuration from the observed data.
The results are summarized as follows.

\begin{enumerate}[nolistsep]
\item Correcting the sample selection biases, we recover the intrinsic distribution of \N{H} of an AGN population that is well defined by our sample selection function.
The intrinsic \N{H} distribution changes significantly between different redshift bins, because of the strong dependence of \N{H} on $L$ and $z$.
Our hard-band selected sample contains more highly obscured sources than the sample used in the 1Ms CDF-S AGN spectral analysis work by \citet{Tozzi2006}, thus showing a higher peak value of \N{H} (between $10^{23.5}$ and $10^{24}$ cm$^{-2}$) than therein.

\item As the deepest X-ray survey at present and in the foreseeable future, the 7Ms CDF-S provides the unique opportunity to study the low-$L$ -- high-$z$ AGN population.
The limitations of such a pencil-beam survey are also obvious, as it has a limited sky coverage and thus does not sample the high-$L$ AGN population well.
Therefore, it is essential to combine the 7Ms CDF-S with other wide and shallow surveys, in order to study the dependence of \N{H} on $L$ and $z$.
With this goal, we define a ``completeness'' flag for each source on the basis of the modeling of the sample-selection bias, with which one can easily trim off the ``incomplete part'' from our sample and select a subsample which is unbiased with respect to \N{H} in the range below a specific \N{H}.
Having the selection-bias eliminated, such \N{H}-complete samples can be easily used in a joint study with other or future surveys to investigate the obscuration of AGN.

\item Based on our \N{H}-complete sample, we measure the average \N{H} and obscured AGN fraction as a function of redshift in a narrow luminosity range ($43.5\sim44.2$).
We find a strong evolution of the obscured fraction with $z$, which can be expressed as $f_{24}\ =\ 0.42\pm0.09\ (1+z)^{0.60\pm0.17}$.
At $z>2$ the obscured fraction likely saturates, showing a weak evolution.
The obscured fraction measured from our sample is higher than those measured in previous works.

\item Using our \N{H}-complete sample, we are also able to disentangle the luminosity-dependence and redshift-dependence of \N{H}. Combining our data with that from the wider and shallower survey C-COSMOS \citep{Lanzuisi2013}, we measure the average \N{H} and obscured fraction in 2D $L_X$--$z$ bins with $L_X$ between $10^{42}$ and $10^{45}$ erg s$^{-1}$ and redshift up to our limit of measurement ($z\thickapprox 4$).
We find that at any redshift, the average \N{H} (or obscured fraction) decreases with intrinsic X-ray luminosity $L_X$, and at any $L_X$, it increases with redshift.
\end{enumerate}

\acknowledgments
This work is partly supported by the National Science Foundation of China (NSFC, grants No. 11403021, 11233002, \& 11421303) and the National Basic Research Program of China (973 program, grant No. 2015CB857005 \& 2015CB857004).
T.L. and Y.Q.X acknowledge support from Fundamental Research Funds for the Central Universities.
T.L., P.T. and J.X.W. received support from the ``Exchange of Researchers'' program for scientific and technological cooperation between Italy and the People's Republic of China for the years 2013-2015 (code CN13MO5).
T.L. acknowledges hospitalities by INAF-Osservatorio Astrofisico di Arcetri and by UMASS Amherst during the completion of this work.
J.X.W. and Y.Q.X. acknowledge the Strategic Priority Research Program ``The Emergence of Cosmological Structures'' of the Chinese Academy of Sciences (grant No. XDB09000000), and CAS Frontier Science Key Research Program (grant No. QYZDJ-SSW-SLH006).
J.X.W. thanks support from Chinese Top-notch Young Talents Program.
Y.Q.X. thanks support from National Thousand Young Talents program and NSFC-11473026.
B.L. acknowledges support from the NSFC-11673010 and the Ministry of Science and Technology of China grant 2016YFA0400702.
W.N.B acknowledges financial support from {\sl Chandra} X-ray Center grant GO4-15130A and the V.M. Willaman Endowment.
F.E.B. acknowledges support from CONICYT-Chile (Basal-CATA PFB-06/2007, FONDECYT Regular 1141218), the Ministry of Economy, Development, and Tourism's Millennium Science Initiative through grant IC120009, awarded to The Millennium Institute of Astrophysics, MAS.
D.M.A. thanks the Science and Technology Facilities Council (ST/L00075X/1) for support.


\appendix
\section{WEIGHT SELECTION IN SPECTRAL RESPONSE FILE COMBINATION}
\label{app:weights}
For a given source, the intrinsic flux $S_j$ (in units of $erg/cm^2/s$)  incident on the ACIS-I instrument during ObsID $j$ produces photon count $C_j(I)$ in the energy channel $I$:
\begin{equation}
C_j(I)= \int dE\, S_j(E)\, T_j\, A_j(E)\, R_j(E,I) \,,
\label{eq:cj}
\end{equation}
where $A_j(E)$ and $R_j(E,I)$ are the ancillary and matrix response files, specific to the ObsID and the position of the source in the field of view.

Stacking the spectra from all the ObsID in a period, we have the stacked spectrum $C_{tot}$ and the total exposure time $T_{tot}$:
\begin{align*}
C_{tot}(E) &\equiv \sum_j C_j(E)\,\\
T_{tot} &\equiv \sum_j T_j\,.
\end{align*}
To perform spectral analyses on the stacked spectrum, we will use a relation of the same kind as equation \ref{eq:cj}:
\begin{equation}
C_{tot}(I)= \int dE\, \tilde  S(E)\, T_{tot}\, \tilde A(E)\, \tilde R(E,I)\,,
\label{eq:ctot}
\end{equation}
Based on the two expressions of $C_{tot}$:
\begin{equation}
\sum_j \int dE\, S_j(E)\, T_j\, A_j(E)\, R_j(E,I) \, =  \int dE\, \tilde S(E)\, T_{tot}\, \tilde A(E)\, \tilde R(E,I)\,,
\label{eq:twoexp}
\end{equation}
we need to find proper expressions for $\tilde A(E)$ and $\tilde R(E,I)$ in order that the average flux we measured from the stacked spectrum is correct, as in this form:
$$\tilde S(E) = \frac{\sum_j S_j(E)\, T_j}{T_{tot}}\,.$$

The RMF $R(E,I)$ affects the spectrum $C$ in each channel. But as a normalized response matrix, its impact on the broad band integrated flux is secondary compared to that of ARF, which directly affects the flux measurement in terms of effective area.
Moreover, ARF is variable because of both the significant degradation of CCD quantum efficiency over the past 16 years (as shown in Figure~\ref{fig:fourperiods}) and the vignetting effect at positions off-axis from the aimpoint; while RMF is relatively constant.
Therefore, we can consider RMF of a source as roughly constant and simplify it from Equation \ref{eq:twoexp}.
To average the RMF from each ObsID within a period, we simply use the broad-band photon counts as weights.
While for ARF, which is of more significance in flux calibration, we derive the optimized weight from Equation \ref{eq:twoexp}, in order to achieve the most accurate definition of average flux.
The solution is straightforward in the case of constant flux $S(E)$. But with the aim of studying AGNs which are highly variable on timescales of months, we take $S(E)$ as time-dependent. Thus besides time and position dependence of the ACIS-I effective area, we also keep track of source variability in the weight.

After simplifying $R(E,I)$, the left side of Equation \ref{eq:twoexp} can be written as:
\begin{align*}
\sum_j \int & dE\, S_j(E)\, T_j\, A_j(E)\\
=& \int dE\, \sum_j S_j(E)\, T_j\, A_j(E)\\
=& \int dE\, \frac{\sum_j S_j(E)\,T_j}{\sum_j T_j}\, \frac{\sum_j T_j}{1}\, \frac{\sum_j S_j(E)\, T_j\, A_j(E)}{\sum_j S_j(E)\,T_j}\\
=& \int dE\, \tilde  S(E)\, T_{tot}\, \frac{\sum_j S_j(E)\, T_j\, A_j(E)}{\sum_j S_j(E)\,T_j}
\end{align*}
By a direct comparison with the right side, we find:
$$ \tilde A(E)\,=\,\frac{\sum_j S_j(E)\, T_j\, A_j(E)}{\sum_j S_j(E)\,T_j}\,.$$
Obviously, the optimized weight is $S_j(E)\,T_j$.

Consider the broad-band photon counts $C_j$, the energy-averaged flux $S_j$, and the mean effective area $\bar{A}_j$ at the effective energy of the broad band, 2.3~keV, we have $C_j \simeq S_j \times T_j \times \bar{A}_j $. 
Thus we can use $C_j /\bar{A}_j$ as an approximation of $S_j(E)T_j$.

\end{CJK*}
\bibliography{CDFS_ref}

\newpage
\begin{deluxetable}{llllll}
\centering
\tablenum{2}
\tablewidth{0pt}
\tabletypesize{\footnotesize}
\tablecaption{\label{table:flux}
Counts and fluxes of each source in each period.
}
\tablehead{
\colhead{ID}&
\colhead{Period}&
\colhead{Soft Cts}&
\colhead{Hard Cts}&
\colhead{Soft Flux}&
\colhead{Hard Flux}
}
\startdata
20& I& 33$\pm$12&6$\pm$13& $2.96_{-1.04}^{+0.693}$& $5.09_{-1.7}^{+1.66}$\\
20& II& 102$\pm$13&50$\pm$14& $21.6_{-3.83}^{+1.25}$& $37.1_{-6.07}^{+6}$\\
20& III& 153$\pm$17&78$\pm$23& $11.2_{-1.93}^{+0.565}$& $19.2_{-3.03}^{+3.03}$\\
20& IV& 102$\pm$15&102$\pm$22& $10.1_{-1.94}^{+0.726}$& $17.3_{-2.78}^{+2.79}$\\
21& I& 235$\pm$16&60$\pm$10& $25.8_{-1.9}^{+2.02}$& $35.7_{-3.09}^{+3.09}$\\
21& II& 4$\pm$3&0$\pm$5& $6.46_{-3.59}^{+3.61}$& $8.96_{-5.07}^{+5}$\\
21& III& 61$\pm$11&24$\pm$13& $5.47_{-1.04}^{+1.06}$& $7.59_{-1.48}^{+1.46}$\\
21& IV& 55$\pm$11&63$\pm$15& $5.89_{-1.25}^{+1.27}$& $8.16_{-1.7}^{+1.7}$\\
22& I& 78$\pm$12&0$\pm$13& $6.79_{-1.74}^{+0.678}$& $9.71_{-1.88}^{+1.86}$\\
22& II& 25$\pm$10&25$\pm$15& $5.91_{-2}^{+1.1}$& $8.6_{-2.23}^{+2.25}$\\
\enddata
\tablecomments{
Column 1: source id \citep{Luo2017}.
Column 2: period.
Column 3: 0.5-2 keV net counts.
Column 4: 2-7 keV net counts.
Column 5: 0.5-2 keV observed flux in $10^{-16}$ erg/cm$^2$/s.
Column 6: 2-7 keV observed flux in $10^{-16}$ erg/cm$^2$/s.
Table 2 is published in its entirety in the electronic 
edition of the {\it Astrophysical Journal}.  A portion is shown here 
for guidance regarding its form and content.}

\end{deluxetable}
\newpage
\begin{deluxetable}{lcccllllcl}
\centering
\tablenum{3}
\tablewidth{0pt}
\tabletypesize{\footnotesize}
\tablecaption{\label{tab:data} 
Spectral properties.
}
\tablehead{
\colhead{ID}	&\colhead{z}	&\colhead{Period}	&\colhead{\N{H}} &\colhead{$\Gamma$} &\colhead{soft rate} &\colhead{hard rate} &\colhead{L} &\colhead{C-\N{H}} &\colhead{Model}
}
\startdata
Compton-thin\\
\tableline
20&	1.3700i&	&	$1.14_{-0.694}^{+0.782}$&	1.8f&	5.9e-07& 3.2e-07&	$43.31_{-0.06}^{+0.05}$&	23.5&	\\
21&	1.0650s&	&	$0.13_{-0.13}^{+0.298}$&	1.8f&	8.1e-07& 3.1e-07&	$43.18_{-0.03}^{+0.04}$&	23.5&	\\
22&	1.9400s&	&	$1.08_{-1.08}^{+1.86}$&	1.8f&	2.4e-07& 1.3e-07&	$43.39_{-0.08}^{+0.08}$&	23.5&	\\
24&	2.3143p&	&	$58.2_{-11.5}^{+14.4}$&	1.8f&	1.2e-07& 4.4e-07&	$44.19_{-0.10}^{+0.09}$&	24&	\\
26&	2.3040s&	&	$0.94_{-0.94}^{+1.15}$&	1.8f&	7e-07& 2.7e-07&	$43.89_{-0.04}^{+0.04}$&	24&	\\
27&	2.9112p&	&	$23.4_{-7.31}^{+8.68}$&	1.8f&	1.6e-07& 2.4e-07&	$44.05_{-0.10}^{+0.09}$&	24&	\\
32&	1.3740s&	&	$15.2_{-5.92}^{+8.35}$&	1.8f&	8.3e-08& 2.4e-07&	$43.28_{-0.16}^{+0.14}$&	23.5&	1\\
31&	1.3310p&	&	$1.97_{-0.751}^{+0.845}$&	$1.88_{-0.19}^{+0.21}$&	7.9e-07& 5.7e-07&	$43.61_{-0.04}^{+0.04}$&	24&	\\
33&	1.7846p&	&	$16.5_{-6.97}^{+10.3}$&	1.8f&	8.9e-08& 1.7e-07&	$43.42_{-0.17}^{+0.15}$&	23.5&	\\
34&	2.9400s&	&	$43.3_{-18.7}^{+24.4}$&	1.8f&	1e-07& 1.6e-07&	$43.95_{-0.19}^{+0.16}$&	24&	\\
\tableline
\enddata
\tablecomments{
Column 1: source id.
Column 2: redshift. Flag 's': secure spectroscopic; 'i': insecure spectroscopic; 'p': photometry, 'x': X-ray spectroscopic-z.
Column 3: period, if \N{H} is found variable. This field is empty if \N{H} is not variable.
If more than one period is given in one line, their \N{H} are set as the same.
If one period of a source whose \N{H} is variable is not given in the table, its \N{H} is set as the average of the \N{H} of the other periods.
Column 4: \N{H} in $10^{22}$ cm$^{-2}$ with 90\% errors. Flag f'' means fixed. Note the \N{H} for Compton-thick sources are considered unreliable and not used in this work.
Column 5: $\Gamma$ with 90\% errors. Flag f'' means fixed.
Column 6,7: 0.5-2 and 2-7 keV observed net count rates, averaged among the four periods, weighted by the exposure time.
Column 8: $\log$ 2-10 keV unabsorbed luminosity with 90\% errors, in erg/s. This value is highly uncertain in the cases of extreme Compton-thick absorption, where the transmitted power-law becomes too weak.
Column 9: Completeness flag C-\N{H}. 
Use the filter C-\N{H} $\geqslant\widetilde{N_H}$'' to select a subsample which is complete with respect to \N{H} at $\log$\N{H}$<\widetilde{N_H}$, where $\widetilde{N_H}$ can be 23, 23.5, or 24.
-1 means Compton-thick. 
Column 10: period number in which soft excess component is added. 0 means soft excess is found in the 7Ms stacked spectrum. Empty means no soft excess is found.
Table 3 is published in its entirety in the electronic 
edition of the {\it Astrophysical Journal}.  A portion is shown here 
for guidance regarding its form and content.}

\end{deluxetable}
\newpage
\begin{deluxetable}{lll}
\centering
\tablenum{4}
\tablewidth{0pt}
\tabletypesize{\footnotesize}
\tablecaption{\label{table:Fe}
Narrow Fe K$\alpha$ lines.
}
\tablehead{
\colhead{ID}    &\colhead{Energy (keV)} &\colhead{EW (eV)}
}
\startdata
26	&$6.36_{-0.07}^{+0.05}$	&$354_{-135}^{+145}$\\
50	&$6.46_{-0.04}^{+0.05}$	&$444_{-184}^{+162}$\\
58	&$6.53_{-0.12}^{+0.10}$	&$879_{-320}^{+400}$\\
89	&$6.50_{-0.08}^{+0.19}$	&$132_{-61}^{+59}$\\
98	&$6.52_{-0.11}^{+0.12}$	&$200_{-72}^{+78}$\\
106	&$6.49_{-0.06}^{+0.06}$	&$241_{-81}^{+86}$\\
119	&$6.42_{-0.03}^{+0.04}$	&$228_{-55}^{+63}$\\
135	&$6.35_{-0.09}^{+0.09}$	&$411_{-165}^{+183}$\\
174	&$6.43_{-0.04}^{+0.03}$	&$1767_{-200}^{+4322}$\\
208	&$6.35_{-0.08}^{+0.10}$	&$154_{-44}^{+44}$\\
240	&$6.40_{-0.05}^{+0.04}$	&$411_{-142}^{+144}$\\
242	&$6.44_{-0.04}^{+0.04}$	&$130_{-39}^{+32}$\\
290	&$6.31_{-0.04}^{+0.05}$	&$1235_{-810}$\\
328	&$6.41_{-0.06}^{+0.06}$	&$261_{-94}^{+96}$\\
355	&$6.42_{-0.03}^{+0.02}$	&$1278_{-858}$\\
357	&$6.58_{-0.10}^{+0.08}$	&$688_{-320}^{+1005}$\\
367	&$6.33_{-0.07}^{+0.07}$	&$257_{-83}^{+83}$\\
386	&$6.42_{-0.10}^{+0.10}$	&$719_{-463}^{+549}$\\
399	&$6.27_{-0.06}^{+0.06}$	&$336_{-117}^{+142}$\\
402	&$6.36_{-0.07}^{+0.07}$	&$805_{-452}^{+930}$\\
430	&$6.41_{-0.06}^{+0.05}$	&$528_{-186}^{+291}$\\
447	&$6.35_{-0.08}^{+0.09}$	&$674_{-358}^{+617}$\\
448	&$6.16_{-0.07}^{+0.08}$	&$487_{-190}^{+222}$\\
458	&$6.58_{-0.24}^{+0.10}$	&$188_{-72}^{+76}$\\
485	&$6.39_{-0.07}^{+0.08}$	&$141_{-60}^{+63}$\\
495	&$6.46_{-0.03}^{+0.03}$	&$66_{-17}^{+16}$\\
507	&$6.36_{-0.05}^{+0.05}$	&$158_{-48}^{+53}$\\
551	&$6.53_{-0.08}^{+0.11}$	&$583_{-179}^{+263}$\\
557	&$6.27_{-0.06}^{+0.05}$	&$137_{-51}^{+49}$\\
614	&$6.27_{-0.18}^{+0.11}$	&$1441_{-1031}^{+517}$\\
621	&$6.42_{-0.22}^{+0.34}$	&$217_{-113}^{+115}$\\
638	&$6.36_{-0.12}^{+0.12}$	&$361_{-189}^{+222}$\\
643	&$6.25_{-0.04}^{+0.04}$	&$219_{-68}^{+73}$\\
646	&$6.39_{-0.06}^{+0.06}$	&$2002_{-1350}$\\
666	&$6.44_{-0.06}^{+0.06}$	&$2569_{-1453}$\\
730	&$6.32_{-0.03}^{+0.09}$	&$95_{-26}^{+27}$\\
733	&$6.37_{-0.06}^{+0.07}$	&$400_{-116}^{+135}$\\
735	&$6.45_{-0.31}^{+0.10}$	&$153_{-71}^{+66}$\\
748	&$6.30_{-0.11}^{+0.11}$	&$214_{-107}^{+115}$\\
752	&$6.42_{-0.05}^{+0.05}$	&$156_{-83}^{+87}$\\
805	&$6.35_{-0.09}^{+0.09}$	&$351_{-148}^{+155}$\\
826	&$6.34_{-0.07}^{+0.05}$	&$866_{-402}^{+1529}$\\
840	&$6.30_{-0.08}^{+0.05}$	&$186_{-73}^{+73}$\\
867	&$6.45_{-0.07}^{+0.08}$	&$451_{-271}^{+340}$\\
868	&$6.37_{-0.04}^{+0.03}$	&$1450_{-500}^{+830}$\\
940	&$6.42_{-0.08}^{+0.08}$	&$1036_{-408}^{+1159}$\\
958	&$6.40_{-0.06}^{+0.05}$	&$544_{-314}^{+786}$\\
981	&$6.32_{-0.05}^{+0.06}$	&$279_{-111}^{+125}$\\
986	&$6.37_{-0.04}^{+0.04}$	&$261_{-102}^{+117}$\\
988	&$6.36_{-0.05}^{+0.04}$	&$187_{-101}^{+120}$\\
\tableline
\enddata
\tablecomments{
Col 1: ID; Col 2: rest-frame central energy with 1$\sigma$ error; Col 3: rest-frame EW with 1$\sigma$ error.
}
\end{deluxetable}
\begin{deluxetable}{llll}
\centering
\tablenum{5}
\tablewidth{0pt}
\tabletypesize{\footnotesize}
\tablecaption{\label{table:broadFe}
Broad Fe K$\alpha$ lines.
}
\tablehead{
\colhead{ID}    &\colhead{Energy (keV)} &\colhead{$\sigma$ (keV)} &\colhead{EW (eV)}
}
\startdata
175	&$6.52_{-0.05}^{+0.05}$	&$0.20_{-0.09}^{+0.09}$	&$108_{-25}^{+26}$\\
479	&$6.61_{-0.79}^{+0.46}$	&$2.60_{-0.80}^{+1.41}$	&$703_{-386}^{+341}$\\
716	&$6.53_{-0.14}^{+0.16}$	&$0.81_{-0.25}^{+0.34}$	&$323_{-110}^{+112}$\\
856	&$6.59_{-0.12}^{+0.09}$	&$0.36_{-0.17}^{+0.33}$	&$267_{-262}^{+302}$\\
898	&$6.09_{-0.10}^{+0.08}$	&$0.39_{-0.21}^{+0.36}$	&$424_{-129}^{+131}$\\
\tableline
\enddata
\tablecomments{
Col 1: ID;
Col 2: rest-frame central energy with 1$\sigma$ error;
Col 3: rest-frame width with 1$\sigma$ error;
Col 3: rest-frame EW with 1$\sigma$ error.
}
\end{deluxetable}

\end{document}